\documentclass[twocolumn]{aastex63}

\usepackage{bm}
\usepackage{soul}

\newcommand{\alphaeuvb}{\ensuremath{\alpha_{\rm EUVB}}}

\newcommand{\HI}{\ensuremath{\mbox{\ion{H}{1}}}}

\newcommand{\HeII}{\ensuremath{\mbox{\ion{He}{2}}}}
\newcommand{\MgII}{\ensuremath{\mbox{\ion{Mg}{2}}}}

\newcommand{\OI}{\ensuremath{\mbox{\ion{O}{1}}}}
\newcommand{\OII}{\ensuremath{\mbox{\ion{O}{2}}}}
\newcommand{\OIII}{\ensuremath{\mbox{\ion{O}{3}}}}
\newcommand{\OIV}{\ensuremath{\mbox{\ion{O}{4}}}}
\newcommand{\OVI}{\ensuremath{\mbox{\ion{O}{6}}}}
\newcommand{\CII}{\ensuremath{\mbox{\ion{C}{2}}}}
\newcommand{\CIII}{\ensuremath{\mbox{\ion{C}{3}}}}
\newcommand{\SiIII}{\ensuremath{\mbox{\ion{Si}{3}}}}
\newcommand{\SiII}{\ensuremath{\mbox{\ion{Si}{2}}}}
\newcommand{\SIII}{\ensuremath{\mbox{\ion{S}{3}}}}

\newcommand{\logU}{\ensuremath{\log U}}
\newcommand{\xh}{\ensuremath{{\rm [X/H]}}}

\newcommand{\ca}{\ensuremath{{\rm [C/\alpha]}}}
\newcommand{\nH}{\ensuremath{n_{\rm H}}}

\newcommand{\lNHI}{\ensuremath{\log N_{\rm H\,I}}}
\newcommand{\NHI}{\ensuremath{N_{\rm H\,I}}}

\defcitealias{lehner2013}{{\rm L13}}
\defcitealias{lehner2018}{{\rm CCC~I}}
\defcitealias{wotta2019}{{\rm CCC~II}}
\defcitealias{lehner2019}{{\rm CCC~III}}
\defcitealias{wotta16}{{\rm W16}}
\defcitealias{haardt1996}{{\rm HM05}}
\defcitealias{haardt2012}{{\rm HM12}}
\defcitealias{KS19}{{\rm KS19}}

\shorttitle{CCC. IV. Effects of Varying Ionization Backgrounds}
\shortauthors{Gibson et al.}

\graphicspath{{./}{figures/}}

\begin{document}

\title{The COS CGM Compendium. IV. Effects of Varying Ionization Backgrounds on Metallicity Determinations in the $\bm{z < 1}$ Circumgalactic Medium} 

\correspondingauthor{Justus Gibson}
\email{justus.gibson@colorado.edu}

\author{Justus L. Gibson}
\affiliation{University of Colorado, Boulder,
2000 Colorado Ave, Boulder, CO 80305,
Boulder, CO 80305, USA}

\author[0000-0001-9158-0829]{Nicolas Lehner}
\affiliation{University of Notre Dame,
225 Nieuwland Science Hall,
Notre Dame, IN 46556, USA}

\author[0000-0002-3391-2116]{Benjamin D. Oppenheimer}
\affiliation{University of Colorado, Boulder,
2000 Colorado Ave, Boulder, CO 80305,
Boulder, CO 80305, USA}

\author[0000-0002-2591-3792]{J. Christopher Howk}
\affiliation{University of Notre Dame,
225 Nieuwland Science Hall,
Notre Dame, IN 46556, USA}

\author[0000-0001-5810-5225]{Kathy L.~Cooksey}
\affiliation{Department of Physics and Astronomy, University of Hawai`i at Hilo, HI 96720, USA}

\author[0000-0003-0724-4115]{Andrew J. Fox}
\affil{AURA for ESA, Space Telescope Science Institute, 3700 San Martin Drive, Baltimore, MD 21218, USA}

\begin{abstract}
Metallicity estimates of circumgalactic gas based on absorption line measurements typically require photoionization modeling to account for unseen ionization states. We explore the impact of uncertainties in the extreme ultraviolet background (EUVB) radiation on such metallicity determinations for the $z < 1$ circumgalactic medium (CGM).  In particular, we study how uncertainties in the power-law slope of the EUV radiation, \alphaeuvb, from active galactic nuclei affect metallicity estimates in a sample of 34 absorbers with \HI\ column densities between $15.25 <\log (\NHI / {\rm cm}^{-2}) < 17.25$ and measured metal ion column densities. We demonstrate the sensitivity of metallicity estimates to changes in the EUV power-law slope of active galactic nuclei, \alphaeuvb, at low redshift ($z<1$), showing derived absorber metallicities increase on average by $\approx 0.3$ dex as the EUV slope is hardened from $\alphaeuvb = -2.0$ to $-1.4$.  We use Markov Chain Monte Carlo sampling of photoionization models with \alphaeuvb\ as a free parameter to derive metallicities for these absorbers. The current sample of absorbers does not provide a robust constraint on the slope \alphaeuvb\ itself; we discuss how future analyses may provide stronger constraints. Marginalizing over the uncertainty in the slope of the background, we find the average uncertainties in the metallicity determinations increase from 0.08 dex to 0.14 dex when switching from a fixed EUVB slope to one that freely varies. Thus, we demonstrate that EUVB uncertainties can be included in ionization models while still allowing for robust metallicity inferences. 
\end{abstract}


\section{Introduction} \label{sec:intro}
Understanding the metallicity in the circumgalactic medium (CGM) is crucial for constraining the types of gas flows that occur within and around galaxies. It is generally accepted that accretion onto galaxies from cosmic filaments in the intergalactic medium (IGM) will bring pristine/metal-poor gas into the galaxy that can fuel future generations of star formation (e.g., \citealt{dekel03, stewart11, borthakur15, fox17}). On the other hand, outflows driven by supernovae or active galactic nuclei (AGN) can push gas enriched with metals from previous generations of stars into the CGM where it may condense back onto the galaxy or even accelerate all the way out into the IGM \citep[e.g.,][]{martin05, rubin12, hummels13, suresh15}. Studying the CGM, and notably its chemical enrichment, can therefore help illuminate how galaxies have evolved up to a given cosmic age by tracing the extent of past and current star formation activity as well as tracking where the products of stellar evolution end up.

A difficulty in characterizing the metallicity of CGM gas is that it is predominantly ionized (e.g., \citealt{lehner09,lehner2013,werk12,liang14}). Since we cannot directly constrain the column density of ionized hydrogen and sometimes important metal ions, it is impossible to {\em directly} determine the metallicity (or physical conditions) of CGM gas. Instead observers use observations of \HI\ and and available metal ion column densities to constrain models of the ionization and physical conditions within the gas. Assuming photoionization, the ionization state of the gas is determined by the ionization parameter, $U$, defined as the ratio of the density of particles (hydrogen), \nH, to the density of hydrogen-ionizing photons at the Lyman limit, $n_{\gamma}$. However, the underlying physical state of the CGM often remains under-constrained due to uncertainties in the model construction as well as limited access to metal ions due to, e.g., limited wavelength coverage, contamination, saturation, signal-to-noise (S/N) level, etc. An absorber observed in a QSO spectrum rarely has complete information on several successive ionization stages for a given metal (though see \citealt{howk09,tripp11}). This makes the determination of metallicity dependent on the details and assumptions of the ionization modeling employed. 

The majority of ionizing photons that affect the state of CGM gas are from the ultraviolet component of the diffuse extragalactic background (EGB) radiation. The EGB is the sum of the multi-wavelength (spanning the FIR up to TeV $\gamma$-rays) extragalactic light arising from the various emission sources (galaxies and quasars) modified by the attenuation the light experiences as it propagates through the universe. This radiation field evolves with cosmic time as both the radiation sources and the opacity of the IGM evolve \citep{haardt1996}. The extreme ultraviolet component of the EGB, referred to as the EUVB, and defined here as having energies between 1 to $\approx$ 1000 Rydberg, has an out-sized importance due to its dominant role in ionizing the IGM and CGM. 

Much work has gone into the development of models for the EGB radiation, which requires coherently integrating many different areas of astrophysics, including the distribution of hydrogen and helium gas throughout the universe, observations of QSO SEDs, models of the ionizing radiation produced from stellar populations in galaxies and its escape from galaxies, and the radiative transfer of this light as it propagates through an expanding universe \citep[e.g.,][]{haardt1996,haardt2001,haardt2012,FG09,KS19,fauchergiguere20}. The dominant radiation sources contributing to the EGB are galaxies and quasars, so understanding the radiation coming from these two populations is a necessary requirement. The light from these sources primarily interacts with hydrogen and helium gases (neglecting the contribution from metals and dust). Thus, accurate measurements of the \HI\ column density distribution function, $f(z,\NHI)$ and its evolution, and a model for how the radiation interacts with these discrete absorbers are main ingredients in an EGB model. 

Two sources of the emissivity, galaxies and quasars, contribute the vast majority of EUVB photons across all cosmic time. Since their redshift evolution is different, each is predicted to dominate at different epochs. We expect the radiation from quasars dominates in the present universe and up to $z\approx 2$--3, while leaking galactic radiation dominates at earlier times (see Figure~3 in \citealt{KS19}, hereafter \citetalias{KS19}). Uncertainties in the intensity of the EGB include the evolution of QSO luminosity functions and the escape of ionizing radiation from galaxies. An important and poorly constrained component of the EGB, especially at lower redshift, is the uncertainty in the shape of the QSO SED for energies above 1 Rydberg. There have been numerous studies of QSO SEDs hoping to constrain the shape of the underlying SED (e.g., \citealt{zheng97,telfer2002,scott04,shull12,stevans14,lusso2015,tilton16}), but these studies broadly disagree with one another. Most assume that the mean QSO SED is a power law with $f_\nu \propto \nu^{\alphaeuvb}$ for energies above 1 Rydberg. Interpretations of the available observations give a broad range in the power law index, \alphaeuvb, ranging from $-0.56$ to $-1.94$ \citep{khaire2017}. Most EGB models then resort to simply adopting a single value of \alphaeuvb\ within that range for their QSO SED templates. One of the earliest EGB models from \citet{haardt1996} adopted a slope of $-1.4$ motivated by the best available observations at the time. However, in their 2001 and 2012 iterations \citep{haardt2001,haardt2012} they adopt \alphaeuvb\ values of $-1.8$ and $-1.57$, respectively. A different EGB model from \citet{FG09} yielded somewhere in the middle with an adopted \alphaeuvb\ value of $-1.6$ whereas their updated model adopts a value of $-1.7$ based on stacking z $\approx$ 2.4 quasars and correcting for Ly$\alpha$ and Lyman continuum absorption \citep{fauchergiguere20}.

For studies of the CGM, this choice of EUVB slope can have a strong effect on derived results, notably at lower redshifts where QSO radiation is thought to dominate. As demonstrated by our group and others, the choice of the EUVB has a direct effect on the metallicities. For example, the \citealt{haardt2012}, hereafter \citetalias{haardt2012} radiation field, yields metallicities systematically $+0.2$--$0.5$ dex higher than the radiation field of \citealt{haardt1996}, hereafter \citetalias{haardt1996} (e.g., \citealt{lehner2013,wotta16,wotta2019,chen17}).The choice of QSO slope produces a {\it systematic}\ global shift in the abundances at $z\la 1$, i.e., it shifts the metallicity in the same way for all absorbers, though the magnitude of the shift depends on the absorber \HI\ column density, being smaller for Lyman-limit systems (LLSs) and larger for strong Ly$\alpha$ forest systems \citep[SLFSs;][]{wotta16,wotta2019}.

In this paper, we further explore the effects of the EUVB shape on absorber metallicity estimates. Instead of using one of the EGB models with a fixed EUVB slope, we adopt the varying EUVB models of \citetalias{KS19} to explore the effects from the unknown shapes of the EUVBs on metallicity determinations as well as to attempt to constrain the value of \alphaeuvb. The EGB from \citetalias{KS19} is calculated for wavelengths from the far-IR up to TeV $\gamma$-rays at $0\le z \le 15$. They use up-to-date information on galaxy star formation rates, dust attenuation in galaxies, quasar emissivity, the distribution of \HI\ in the IGM, and the escape fraction of ionizing photons from galaxies. To account for the uncertainties in the SEDs of quasars, they include a parameter \alphaeuvb\ to allow variations in the quasar SED slopes for energies between 12.4\,eV and the X-ray regime (with a specific upper energy that varies depending on \alphaeuvb\ in order to smoothly match constraints on the X-ray background intensity). The \citetalias{KS19} models are consistent with intergalactic \HI\ photoionization rates, which the \citetalias{haardt2012} EUVB does not (see \citealt{kollmeier2014,shull2015}) and with a broad range of additional constraints. \citetalias{KS19} provide EUVB fields encompassing the likely range of SED slopes from QSOs, calculating seven different EGBs for values of \alphaeuvb\ ranging from $-2.0$ to $-1.4$ (in steps of 0.1). The hardest slope, $\alphaeuvb =-1.4$, is based on QSO measurements at $z<1.5$ from \citet{stevans14}, while the softest, $\alphaeuvb =-2.0$, is set by observations at $z > 2.5$ from \citet{lusso2015} and \citet{telfer2002}. \citetalias{KS19} consider $\alphaeuvb = -1.8$ as their fiducial model, as \citet{khaire2017} showed that value to be consistent with \HeII\ Lyman-$\alpha$ optical depth measurements, though none of the slopes is excluded by available data. The slope of the EUVB SED from QSOs directly influences metallicity estimates of CGM gas; the \citetalias{KS19} models with varying \alphaeuvb\ allow us to assess how derived CGM properties depend on this slope. 

In this paper, we explore the effects of adopting different EUVB slopes from QSOs in the study of metallicities of predominantly ionized absorbers at $z<1$. We include \alphaeuvb\ as a free parameter in photoionization models describing the absorbers to marginalize over this uncertainty and provide inferences as to the true value of \alphaeuvb. We use observations of QSO absorbers obtained from the the Cosmic Origins Spectrograph (COS) on-board the \textit{Hubble Space Telescope} (\textit{HST}). Specifically, we make use of the sample of \HI-selected absorbers assembled and analyzed in the COS CGM Compendium---CCC, described in \citet[][hereafter \citetalias{lehner2018}]{lehner2018}, \citet[][hereafter \citetalias{wotta2019}]{wotta2019}, and \citet[][hereafter \citetalias{lehner2019}]{lehner2019}. We select a sub-sample of 34 absorbers with $15.25 \le \lNHI \le 17.25 $ from the CCC absorbers (column density units are in cm$^{-2}$ throughout). Following the CCC nomenclature (\citealt{lehner2018}), this range of column densities encompasses strong Ly$\alpha$ forest systems (SLFSs) with $\lNHI \in [15.0,16.2)$ and partial Lyman Limit Systems (pLLSs) with $\lNHI \in [16.2,17.2)$.  We also consider one absorber at Lyman Limit System (LLSs) column densities, $\lNHI \in [17.2,19.0)$, where absorption below 912~\AA\ becomes optically thick. The specific absorbers are selected such that the observations provide potentially robust constraints on the ionization state of the gas from metal ions. This selection emphasizes the availability of adjacent ionization states within a given element, e.g., \CII, \CIII, and, in particular, the suite of oxygen ions \OI, \OII, \OIII, and \OIV, which provides continuous information up to ionization energies of 77.4 eV. We use a Bayesian formalism combined with an MCMC algorithm developed by \citet{fumagalli16} and updated by \citet{wotta2019} to sample from the posterior probability distribution in order to constrain the properties of the CGM. We implement a new grid of Cloudy simulations using the \citetalias{KS19} EUVBs allowing us to consider variations in the EUVB slope in our photoionization models.

Our paper is organized as follows. In Section~\ref{sec:data}, we provide details on the absorption systems we model. In Section~\ref{sec:methods}, we describe the construction of our photoionization models and the adopted MCMC sampling procedure. In Section~\ref{sec:results}, we present the relationship between \xh\ and \alphaeuvb, the derived metallicities, and the inferred \alphaeuvb\ values. Finally, we discuss possible implications in Section~\ref{sec:disc} and present our conclusions in Section~\ref{sec:concl}. 

\section{Data Sample and Selection Criteria} \label{sec:data}

In this work, our goals are to investigate the effects of varying ionization backgrounds on metallicity determinations in the CGM and to determine if we can use the information from metal-ions to better constrain the slope of the EUVB. Naively, absorbers with the largest available information from metal ions (i.e., absorbers with column density information from low to intermediate/high ions, including suites of ions for the same species, such as \OI, \OII, \OIII, \OIV) would seem to provide the best set to reach our goals. A posteriori, we show in this paper that this is not necessarily the case, especially taking into account that some ions are not as constraining owing to saturation effects giving lower limits or non-detections providing upper limits.  

The sample we adopt consists of 34 absorbers taken from the 224 absorbers analyzed in \citetalias{lehner2018}, \citetalias{wotta2019}, and \citetalias{lehner2019}. For all the absorbers, we require information from either \OII\ $\lambda\lambda$832, 833, 834 and/or \OIII\ $\lambda\lambda$832, which requires the absorbers to be at redshift $z>0.4$. Moreover, we select absorbers to have some of the best constraints from a combination of the following ions: \CII, \CIII, \MgII, \OI, \OII, \OIII, \OIV, \SIII, \SiII, and \SiIII. All \HI\ and metal column densities are from \citetalias{lehner2018}.  The parent CCC sample contained absorbers spanning an \HI\ column density range of $15 < \lNHI < 19$ at $ 0.1 \la z < 1$ with varying numbers and types of detected ions (see \citetalias{lehner2018} for details on how column densities were measured).  In contrast, our sample covers redshifts $0.45\la z \la 0.88$ and \HI\ columns with  $15.25\le \lNHI \le 17.25$. Thus, our sample consists primarily of SLFSs and pLLSs with one lower column density LLS. The sample is biased toward the higher redshifts of CCC owing to the fact that we are particularly interested to have coverage of some of the EUV transitions of \OII, \OIII, and \OIV, which require redshifts $> 0.4$ to be observed in the COS G130M/G160M bandpass. Owing to our selection of the absorbers, we also note that the sample is not unbiased relative to the metallicity; in fact we will show that the metallicities are mostly distributed on the higher end of the metallicity distribution observed at these redshifts. In Table~\ref{tab:sample_props}, we summarize the sample (ordered in increasing \HI\ column density), and we refer the reader to \citetalias{lehner2018} for the metal atomic and ionic column densities.  In the subsequent figures and throughout the paper, we will use the unique ID provided in the first column of this table to refer to these absorbers. 

\begin{center}
\startlongtable
\begin{deluxetable*}{lllccc}
\tabcolsep=3pt
\tablecolumns{5}
\tablewidth{0pc}
\tablecaption{Basic Properties of our Sample of Absorbers\tablenotemark{a}\label{tab:sample_props}}
\tabletypesize{\footnotesize}
\tablehead{\colhead{ID}   & \colhead{Target}   & \colhead{Original Name}   & \colhead{$z_{\rm abs}$}   & \colhead{$\rm log \ N_{HI}$} & \colhead{$\rm Ions$\tablenotemark{b}}  \\
           \colhead{ }    & \colhead{ }        & \colhead{ }               & \colhead{ }               & \colhead{[cm$^{-2}$]}      &\colhead{ }   }
\startdata
1        & J011016.25$-$021851.0     & HB89$-$0107$-$025$-$NED05      & 0.717956       & $15.25 \pm 0.05$   & \underline{CII} \textbf{CIII} \underline{FeII} \underline{MgII} \underline{NII} \underline{OI} \underline{OII} \underline{\textbf{OIII}} \textbf{OIV} \underline{SIII} \underline{SiII} \\ 
2        & J100110.20+291137.5       & BZBJ1001+2911                  & 0.556468       & $15.38 \pm 0.03$   & \underline{CII} $\overline{\rm CIII}$ \underline{FeII} \underline{FeIII} \underline{NI} \underline{NII} $\overline{\rm NIII}$ \underline{OI} \underline{OII} $\overline{\rm OIII}$ $\overline{\rm OIV}$ \underline{SIII} \underline{SiII} \\ 
3        & J011013.14$-$021952.8     & LBQS$-$0107$-$0235             & 0.718956       & $15.61 \pm 0.02$     & \textbf{CII} $\overline{\rm CIII}$ \underline{FeII} $\overline{\rm MgII}$ $\overline{\rm NIII}$ $\overline{\rm OI}$ $\overline{\rm OII}$ \textbf{OIII} \textbf{OIV} SIII \underline{SiII}   \\ 
4        & J155048.29+400144.9       & SDSSJ155048.29+400144.9        & 0.492632       & $15.62 \pm 0.05$  & \underline{\textbf{CII}} \textbf{CIII} MgII \underline{NI} \underline{NII} \underline{NIII} \underline{OI} \underline{OII} \underline{SIII} \underline{SiII}      \\ 
5        & J044011.90$-$524818.0     & HE0439$-$5254                  & 0.615662       & $15.66 \pm 0.01$    & \textbf{CII} $\overline{\rm CIII}$ \underline{FeII} MgII \underline{OI} OII $\overline{\rm OIII}$ $\overline{\rm OIV}$ SIII \underline{SiII}    \\ 
6        & J063546.49$-$751616.8     & PKS0637$-$752                  & 0.452768       & $15.68 \pm 0.01$  &  CIII \underline{FeII} \underline{FeIII} MgII \underline{NI} \underline{OI} \underline{OII} \underline{OIII}  \underline{SIII} \underline{SiII}      \\ 
7        & J023507.38$-$040205.6     & HB89$-$0232$-$042              & 0.807754       & $15.71 \pm 0.01$  & \underline{CII} $\overline{\rm CIII}$ \underline{FeII} \underline{MgII} \underline{NI} \underline{NII} $\overline{\rm NIII}$ \underline{OI} \underline{OII} \textbf{OIII} \textbf{OIV} \underline{SIII}    \\ 
8        & J011013.14$-$021952.8     & LBQS$-$0107$-$0235             & 0.536490       & $15.76 \pm 0.02$  & \underline{CII} \textbf{CIII} \underline{FeII} \underline{FeIII} \underline{MgII} \underline{NI} \underline{NII} \underline{OI} \underline{OII} \textbf{OIV} \underline{SIII} \underline{SiII}      \\ 
9        & J112553.78+591021.6       & SBS1122+594                    & 0.557529       & $15.79 \pm 0.02$   & \textbf{CII} $\overline{\rm CIII}$ \underline{FeII} \underline{FeIII} \underline{NI} \underline{NII} \underline{OI} \underline{OII} $\overline{\rm \textbf{OIII}}$ \textbf{OIV} \underline{SIII} \underline{SiII}     \\ 
10       & J011013.14$-$021952.8     & LBQS$-$0107$-$0235             & 0.876403       & $15.91 \pm 0.02$    & CII \underline{FeII} \underline{MgII} \underline{OI} \textbf{OII} \textbf{OIII} SIII    \\ 
11       & J142859.03+322506.8       & J142859.03+322506.8            & 0.516578       & $15.99 \pm 0.01$  & \underline{CII} \underline{OII} \textbf{OIII} \textbf{OIV}       \\ 
12       & J075112.30+291938.3       & FBQS$-$0751+2919               & 0.829173       & $16.02 \pm 0.01$   & \textbf{\underline{CII}} \textbf{CIII} \underline{FeII} MgII \underline{NIII} \underline{OI} \underline{\textbf{OII}} \textbf{OIII} SIII    \\ 
13       & J141542.90+163413.7       & SDSSJ141542.90+163413.8        & 0.481677       & $16.02 \pm 0.01$    & \textbf{\underline{CII}} \textbf{CIII} \underline{OI} \textbf{OII} \textbf{OIII}    \\ 
14       & J134100.78+412314.0       & PG$-$1338+416                  & 0.621428       & $16.09 \pm 0.01$   & \textbf{CII} $\overline{\rm \textbf{CIII}}$ FeII MgII \underline{OI} OII \underline{SIII}    \\ 
15       & J080908.13+461925.6       & SDSSJ080908.13+461925.6        & 0.619130       & $16.10 \pm 0.01$  & \underline{CII} $\overline{\rm CIII}$ \underline{FeII} \underline{MgII} \underline{NII} \underline{OI} \underline{OII} $\overline{\rm OIII}$ $\overline{\rm OIV}$ \underline{SIII} \underline{SiII}      \\ 
16       & J112553.78+591021.6       & SBS1122+594                    & 0.558200       & $16.13 \pm 0.02$   & \underline{CII} \textbf{CIII} \underline{NII} \underline{NIII} \underline{OI} OII \underline{OIII} \underline{SIII} \underline{SiII}     \\ 
17       & J124511.26+335610.1       & SDSSJ124511.25+335610.1        & 0.688948       & $16.16 \pm 0.03$    & \underline{CII} $\overline{\rm CIII}$ \underline{FeII} MgII \underline{NIII} \underline{OI} \underline{OII} $\overline{\rm \textbf{OIII}}$ $\overline{\rm \textbf{OIV}}$ \underline{SIII}    \\ 
18       & J084349.47+411741.6       & SDSSJ084349.49+411741.6        & 0.533507       & $16.17 \pm 0.04$    & \underline{CII} $\overline{\rm CIII}$ \underline{FeII} MgII \underline{\textbf{NII}} \textbf{NIII} \underline{OI} \textbf{OII} $\overline{\rm OIII}$ \textbf{OIV} \underline{SIII} \underline{SiII}    \\ 
19       & J152424.58+095829.7       & PG$-$1522+101                  & 0.518500       & $16.22 \pm 0.02$   & CII CIII MgII \underline{NI} \underline{NII} \underline{OI} OII \underline{SIII}      \\ 
20       & J100535.25+013445.5       & SDSS$-$J100535.24+013445.7     & 0.837390       & $16.36 \pm 0.02$   & \underline{CII} $\overline{\rm CIII}$ \underline{FeII} \underline{MgII} \underline{OI} \underline{OII} \textbf{OIII} \textbf{OIV} \underline{SIII}      \\ 
21       & J140923.90+261820.9       & PG$-$1407+265                  & 0.682684       & $16.41 \pm 0.01$  & \textbf{\underline{CII}} \textbf{CIII} \underline{MgII} \underline{OI} \underline{\textbf{OII}} \textbf{OIII} \textbf{OIV} SIII \underline{SiII}      \\ 
22       & J063546.49$-$751616.8     & PKS0637$-$752                  & 0.468486       & $16.48 \pm 0.07$     & \textbf{CII} $\overline{\rm CIII}$ MgII \underline{NI} \underline{OI} \textbf{OII} \underline{SIII} \underline{SiII} \\ 
23       & J124511.26+335610.1       & SDSSJ124511.25+335610.1        & 0.712976       & $16.48 \pm 0.02$  & \underline{CII} \underline{FeII} MgII \underline{NIII} OIII \underline{SIII}    \\ 
24       & J134100.78+412314.0       & PG$-$1338+416                  & 0.686086       & $16.50 \pm 0.02$   & CII MgII \underline{OI} OII SIII \underline{SiII}     \\ 
25       & J100535.25+013445.5       & SDSS$-$J100535.24+013445.7     & 0.836989       & $16.52 \pm 0.02$   & \underline{CII} $\overline{\rm CIII}$ \underline{FeII} MgII \underline{OI} \textbf{OII} \textbf{OIII} \underline{SIII}   \\ 
26       & J155048.29+400144.9       & SDSSJ155048.29+400144.9        & 0.491977       & $16.53 \pm 0.02$    & \textbf{CII} $\overline{\rm CIII}$ FeII MgII \underline{NI} \underline{NII} NIII OII \underline{SIII} \underline{SiII}    \\ 
27       & J124511.26+335610.1       & SDSSJ124511.25+335610.1        & 0.556684       & $16.54 \pm 0.03$   & \underline{CII} CIII \underline{FeII} \underline{FeIII} \underline{MgII} \underline{NI} \underline{NIII} \underline{OI} \underline{OII} \underline{SIII} \underline{SiII}     \\ 
28       & J152424.58+095829.7       & PG$-$1522+101                  & 0.728885       & $16.63 \pm 0.05$   & \underline{CII} \textbf{CIII}  \underline{MgII} \underline{NIII} \underline{OI} \underline{OII} \textbf{OIII} \textbf{OIV} \underline{SIII} \underline{SiII}     \\ 
29       & J161916.54+334238.4       & SDSSJ161916.54+334238.4        & 0.470800       & $16.64 \pm 0.03$    &  \underline{FeII} MgII \underline{NI} \underline{NII} \underline{OI} \textbf{OII}  \\ 
30       & J023507.38$-$040205.6     & HB89$-$0232$-$042              & 0.738890       & $16.67 \pm 0.01$    & \textbf{CII} $\overline{\rm \textbf{CIII}}$ \underline{FeII} MgII \underline{OI} \textbf{OII} $\overline{\rm \textbf{OIII}}$ SIII \underline{SiII}    \\ 
31       & J091431.77+083742.6       & VV2006$-$J091431.8+083742      & 0.507220       & $16.70 \pm 0.03$    & \underline{CII} \underline{NIII} \underline{OI} \textbf{OII} \textbf{OIII}   \\ 
32       & J124511.26+335610.1       & SDSSJ124511.25+335610.1        & 0.689370       & $16.91 \pm 0.06$  & \textbf{CII} $\overline{\rm CIII}$ FeII MgII \underline{OI} OII $\overline{\rm OIII}$ $\overline{\rm OIV}$ SIII    \\
33       & J093603.88+320709.3       & VV2000$-$J093603.9+320709      & 0.575283       & $17.15 \pm 0.15$   & $\overline{\rm MgII}$ \underline{OI} $\overline{\rm OII}$ $\overline{\rm OIII}$ \underline{OIV} \underline{SIII}    \\ 
34       & J134447.55+554656.8       & J134447.56+554656.8            & 0.552615       & $17.25 \pm 0.15$   & \underline{OI} \textbf{OII} $\overline{\rm \textbf{OIII}}$ \underline{\textbf{OIV}}    \\ 
\enddata
 \tablenotetext{a}{The H I and redshift of the absorbers were measured in \citetalias{lehner2018}; all the metal atomic/ionic column densities used in this work are also from \citetalias{lehner2018}. The original name refers the name used in the HST archive. }
\tablenotetext{b}{Upper limits, lower limits, and detections are represented by underbars, overbars, and no bar, respectively. Bold-faced symbols indicate that the ion places strong constraints on the model.  
 }
\end{deluxetable*}

\label{tab:sample}
\end{center}

\section{Methodology} 
\label{sec:methods}

Here we describe the methods by which we determine the metallicities of our absorption systems and how we assess the effects of a varying EUVB slope. We perform two separate tasks to assess the effects that EUVB variations have on the metallicity of our absorbers. One assesses how the metallicity changes as a function of \alphaeuvb, and the other analyzes the effects that allowing \alphaeuvb\ to vary in our model have on the derived metallicities.

\subsection{Photoionization Modeling} \label{sec:cloudy}

We make the usual assumption of a uniform density slab geometry in thermal and ionization equilibrium exposed to a uniform EGB and the cosmic microwave background (CMB) appropriate for each redshift. In order to determine the metallicity of the absorbers, we generate a new grid of models that give predictions for all ion states of the relevant absorption lines given a set of physical parameters. We use the spectral synthesis code Cloudy, version 17.01 \citep{cloudy17}, to generate grids of models by varying the hydrogen number density (\nH), \HI\ column density (\NHI), metallicity (\xh), and redshift ($z$). Depending on the model, we also vary the ratio of carbon to $\alpha$-elements (\ca) and/or the slope of the EUVB  (\alphaeuvb). The parameter space explored in this work is summarized in Table~\ref{tab:mcmc_parameters}; we cover the same range of parameter space as \citetalias{wotta2019}, though with the addition of the EUVB slope using the \citetalias{KS19} EUVB models. The broad range in parameter values allows us to thoroughly explore the possible physical conditions that could lead to the observed column densities.

\begin{figure*}[tbp]
\plotone{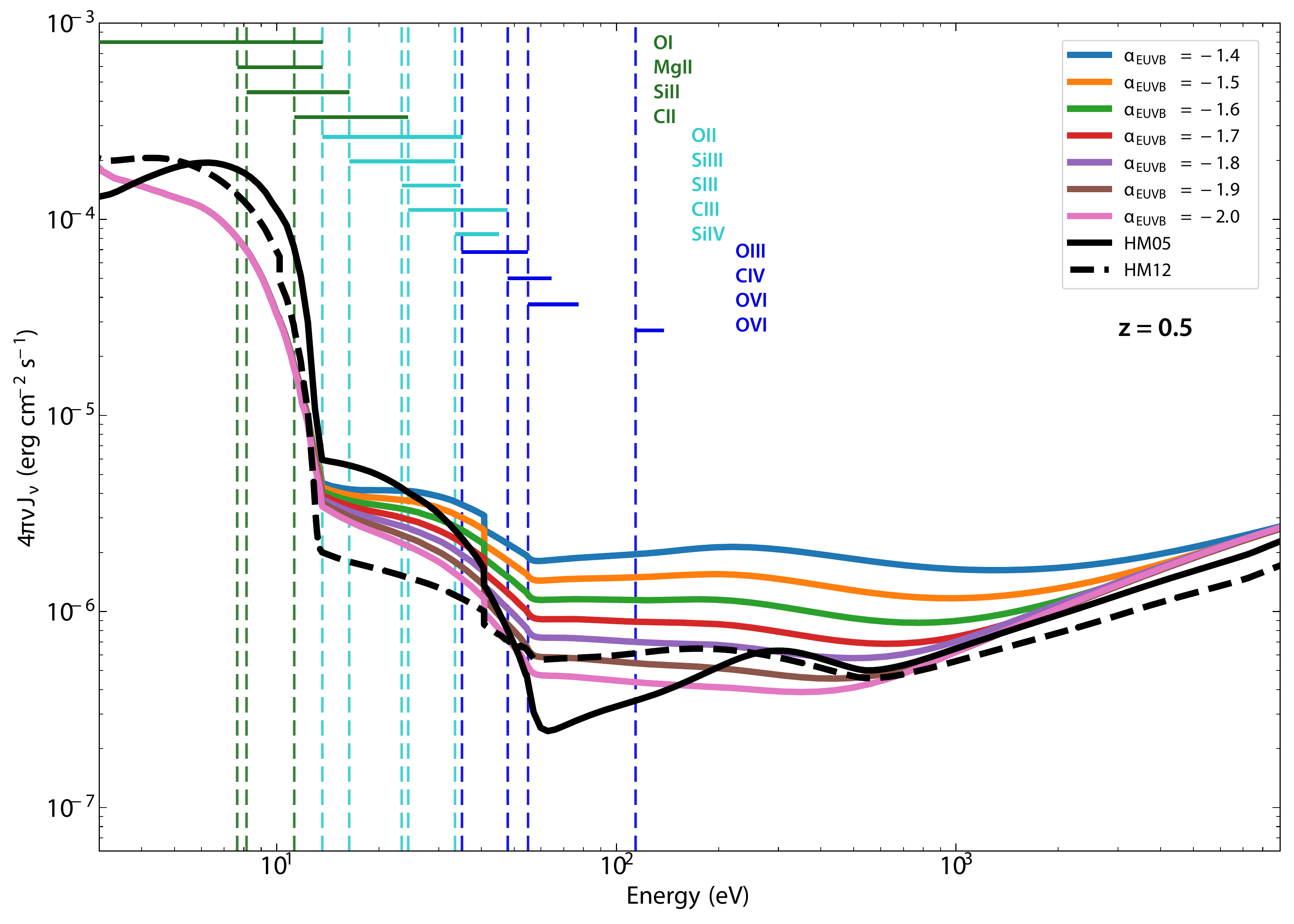}
\epsscale{1.2}
\caption{Various EUVBs used in the literature at $z = 0.5$ showing the large differences between different models. The colored lines show the \citetalias{KS19} EUVBs that we adopt here, with each color corresponding to a different \alphaeuvb\ value (see legend). \citetalias{KS19} consider $\alphaeuvb = -1.8$ to be their fiducial background. The other two backgrounds shown are \citetalias{haardt1996} (dashed black line) and \citetalias{haardt2012} (solid black line). The dashed vertical colored lines show the ionization potentials of the various ions included in our model with the addition of \OVI\ and the horizontal colored bars show the energy ranges probed. The CCC data used here measure metal species with ionization energies between 6 eV and 60 eV, although no absorber has positive detections of all ion species. Generally, the \citetalias{KS19} backgrounds are harder than either of the HM backgrounds.     
\label{fig:euvb}
}
\end{figure*}

The main difference between \citetalias{wotta2019}/\citetalias{lehner2019} and this present work is the treatment of the EUVB. We include the slope of the EUVB as a sixth variable, which allows us to explore the effects of changing EUVB slope on the ionization state of the CGM. \citetalias{KS19} calculate seven different EGBs for values of \alphaeuvb\ ranging from $-2.0$ to $-1.4$ in steps of 0.1.  We consider here these seven different values of the EUVB slope. One set of models treats the EUVB as constant and assesses how the derived metallicities vary with \alphaeuvb\ for each absorber. We also run a set of models where the slope of the \citetalias{KS19} EUVB is included as a free parameter in the modeling, which allows us us to arrive at metallicity values marginalized over \alphaeuvb. Both of these models are run once with \ca\ fixed to solar and once with \ca\ allowed to vary.

Our absorbers span a redshift range from approximately 0.4 to 0.9 and the theoretical prediction from \citetalias{KS19} is an intensity increase of a factor of 3 from the lowest redshifts in our sample to the highest for a fixed \alphaeuvb\ value. This redshift evolution is significant enough to include it in our modeling scheme and given how uncertain this value is we want to explore whether any value is favored for any absorber. We also note that none of our absorbers are at exactly the same redshift, so allowing for the variation of \alphaeuvb\ with redshift takes care of any possible variation of \alphaeuvb\ with small changes of $z$. This also permits for possible local radiation effects, which may influence the local shape of the EUVB, to be taken into account. Finally, there could be increases in the hardness of the local EUVB if the absorber is associated with a highly star-forming galaxy or with an AGN. However, this is not expected to be a likely factor which we discuss more in Section~\ref{sec:met_results_robust}.

In Figure~\ref{fig:euvb}, we show for $z = 0.5$ the seven \citetalias{KS19} EUVBs  along with the \citetalias{haardt1996} and \citetalias{haardt2012} backgrounds, both of which have been explored in previous works (see \citetalias{wotta2019,lehner2019}). At the lowest energies, the HM backgrounds are both harder than the \citetalias{KS19} backgrounds, whereas at energies between approximately 10--40 eV, the \citetalias{KS19} backgrounds are in-between \citetalias{haardt1996} (harder background) and \citetalias{haardt2012} (softer background). At energies above $\approx 40$ eV, the majority of the \citetalias{KS19} backgrounds are harder than both HM backgrounds for most energies.  

\begin{deluxetable}{lccc}
\tabcolsep=3pt
\tablecolumns{4}
\tablewidth{0pc}
\tablecaption{Cloudy Model Parameters\label{tab:mcmc_parameters}}
\tabletypesize{\footnotesize}
\tablehead{\colhead{Parameter} & \colhead{Minimum} & \colhead{Maximum} & \colhead{Step size\tablenotemark{a}} }
\startdata
    $\log \NHI$  [cm$^{-2}$]   &    12.0     &  20.0       & 0.25   \\
    $z$    &    0.0     &    5.0     &  0.25  \\
    \xh\   &    $-5.0$     &    2.5     & 0.25    \\
    $\log \nH$ [cm$^{-3}$]    &  $-4.5$       &  0.0      & 0.25  \\
   \ca\ &    $-1.0$     &   1.0      & 0.2       \\
 $\alpha_{\rm EUVB}$           & $-2.0$     &  $-1.4$       & 0.1 \\
\enddata
\tablenotetext{a}{We note that while the Cloudy grid was initially computed using the step sizes listed here, the grid was interpolated within the MCMC code.}
\end{deluxetable}
 \label{tab:cloudyparams}

\citet{lehner2013} and \citetalias{lehner2018}  show that for the majority of the systems, the low and intermediate ions have absorption profiles very similar in shape and peak optical depth as the \HI\ profiles, strongly suggesting they arise in the same gas phase. Higher ions such as \OVI\ are not used in this modeling as those likely trace a different gas phase (see e.g., \citealt{lehner2013}, \citealt{werk14}, \citealt{pointon19}), though we will explore \OVI\ in future papers (e.g., preliminary uses of the \citetalias{KS19} EUVB show for some low \NHI\ SLFSs, some \OVI\ can be photoionized in the same gas-phase as lower ions). Motivated by the alignment of low and intermediate metal ions with the \HI\ absorption profiles, we follow the methodology of \citetalias{wotta2019} and model each absorber with a single gas-phase. We note that \citet{haislmaier2021} model two of the absorption systems considered here (absorbers 19 and 24) that have a large coverage of metal ions using two to three distinct gas-phases. The metallicity differences with the single-phase CCC results and the multi-phase results of \citet{haislmaier2021} are well within 0.2 dex, implying at least in these two cases that the effects of a multi-phase medium on the derived results may not be substantial (see also \citealt{liang18}, \citealt{marra21} for similar analyses of multi-phase simulations).

\subsection{MCMC Sampling Procedure} \label{sec:mcmc}

Markov Chain Monte Carlo (MCMC) techniques provide a  rigorous way to assess uncertainties and well-informed estimates of parameter values with realistic uncertainties (see \citealt{sharma2017} for a review of MCMC in astronomy). MCMC techniques work in a Bayesian inference framework where observations are compared with model predictions by a prior probability distribution on the parameters and likelihood functions to compare the models to observations. We use publicly available codes described in \citet{fumagalli16} and \citetalias{wotta2019}, which makes use of the $\tt{emcee}$ Python module \citep{foreman2013} to carry out the MCMC sampling. We modified their codes to add the \citetalias{KS19} EGBs   and the EUVB slope as a sixth parameter.

The code uses predicted column densities from our grid of photoionization models and the observational column densities to sample the model space and determine the sets of model parameters that are most consistent with the observations. This works by defining a set of `walkers', initialized in our five-dimensional or six-dimensional model parameter space, that take a series of steps to nearby points in the space, calculating the likelihood at each step. At each new step, the likelihood is evaluated. If the likelihood is higher than the previous step, the walker will stay at the new position with some acceptance probability; if not, it will go back to the previous position. This way the ensemble of walkers will tend to converge toward the highest likelihood model and the phase-space positions of the walkers  will trace the posterior PDF in each parameter. We initialize 400 walkers and allow each walker to take 400 steps through the space, which \citetalias{wotta2019} showed to be sufficient for model convergence.  We discard the results from the first 150 steps (our burn-in) to allow the walkers time to approach higher likelihood regions of the parameter space from their initial positions. Following the methodology outlined in \citetalias{wotta2019}, we initialize the walkers randomly in the grid, which will sometimes cause some walkers to become stuck in regions of parameter space with very small likelihood. To get around this, we compute the median results from a first test run and initialize our walkers around those values in a second run. Since we use more free parameters and different radiation fields than \citetalias{wotta2019}, we have run our own convergence tests. We calculated results for two of our absorbers where the walkers and steps were increased to 2000 to check for convergence. This yielded nearly identical results to our fiducial run, consistent with a similar test done in \citetalias{wotta2019} with the \citetalias{haardt1996} background.

Following the methodology of \citetalias{wotta2019} and \citetalias{lehner2019}, the adopted likelihood function for detected metal ions is a Gaussian based on the assumption that the measured ionic column densities are Gaussian distributed with the standard deviation corresponding to the measurement error. In the case of a lower limit on a metal ion, we adopt a rescaled Q-function for the likelihood. For upper limits, we adopt a cumulative distribution function (CDF; see \citealt{fumagalli16}).

\subsubsection{Priors} \label{sec:priors}

There are up to six parameters in our models (\nH, $z$, \ca, \NHI, \xh, and \alphaeuvb) for which we need to assign with Bayesian priors. We closely follow the methodology outlined in \citetalias{wotta2019} by always adopting Gaussian priors on $\NHI$ and redshift given that these are observed quantities as well as model parameters. Flat priors are a priori assumed for $\nH$ (implying a flat prior on \logU), \xh, and for \ca\ (when this parameter is allowed to vary). \citetalias{wotta2019} only used flat priors for \nH\ for absorbers with strong observational constraints. Adopting flat priors allow our metallicity and other parameter determinations to be informed by the observational data as much as possible. 

For \alphaeuvb, we consider three different priors, though we report outcomes derived with the flat prior as our adopted results. The other two priors considered are a broad Gaussian prior and a Jeffrey's prior. Many different works favor certain values of \alphaeuvb\ over others, with fiducial values generally being around $-1.8$ to $-1.6$ (see, e.g., \citetalias{KS19,haardt1996,haardt2012} and \citealt{fauchergiguere20}). Thus, there is some basis for modeling \alphaeuvb\ with a broad Gaussian prior, whose mean is representative of the above theoretical predictions, to constrain the slope. The adoption of any prior influences the outcome of the modeling, but our purpose here is to simply explore the effect that different priors have on the inferred values of \alphaeuvb. To this end, we also run our MCMC sampling algorithm using a Gaussian prior on \alphaeuvb\ with a median of $-1.7$ and $2\sigma = 0.6$ (i.e., the full range of our EUVB slope values). Like the flat prior, a Jeffrey's prior is a non-informative prior. The central difference between a flat prior and a Jeffrey's prior is that the Jeffrey's prior is defined such that the posterior distributions for different parameterizations of some model parameter can be consistently transformed into the other parameter space. In our case, we find very little difference between the flat prior and the Jeffrey's prior. Thus, going forward, we only discuss the flat prior and the Gaussian prior results.

\subsubsection{Summary of the MCMC Inputs} \label{sec:summary}
 Gaussian priors, based on observed values, are always used for \NHI\ and redshift, and we use flat priors on the other parameters over the parameter space summarized in Table~\ref{tab:mcmc_parameters}. We constructed two sets of models for two related purposes. One set of models fixes the EUVB slope at four different values of \alphaeuvb\ spanning the full range of EUVB slopes from \citetalias{KS19}. The purpose of this model set is to test the dependence of derived metallicities on \alphaeuvb. This set of models is run once with a flat prior on \ca\ and once with \ca\ fixed to solar. Some of the absorbers do not have carbon ions detected or only have very few constraints (e.g., an upper limit on C II); for these absorbers we always keep \ca\ fixed to solar. The second set of models adds the slope of the EUVB as a free parameter to propagate the effects of the uncertainty in \alphaeuvb\ into our metallicity determinations. As in the first set of models, we run this second set of models once with \ca\ as a free parameter and once with \ca\ fixed to solar (absorbers without carbon constraints are still always modeled with fixed \ca). The results of these two models are discussed below.

\subsection{Adopted Model} \label{sec:adopted_mod}

In our exploration of \alphaeuvb\ and its effects on metallicities in the CGM, we considered several different iterations of our model, varying the inclusion and prior choice of \ca\ and \alphaeuvb\ in different ways. In order to discuss our results in a straightforward way and to allow comparison with previous works, we assemble an adopted set of models whose results---including upper and lower limits---we summarize in Table~\ref{tab:results_tab}. This set of models includes \alphaeuvb\ as a free parameter modeled with a flat prior. The flat \ca\ prior results are adopted for absorbers with good constraints on carbon and $\alpha$-elements. We adopt results with fixed \ca\ for seven absorbers with poor constraints on \ca\ as noted by the lack of \ca\ values for seven absorbers in Table~\ref{tab:results_tab}.

\section{Results} \label{sec:results}

\begin{center}
\startlongtable
\begin{deluxetable*}{lcccccc}
\tabcolsep=3pt
\tablecolumns{7}
\tablewidth{0pc}
\tablecaption{Results of Photoionization Modeling$^a$ \label{tab:results_tab}}
\tabletypesize{\footnotesize}
\tablehead{\colhead{ID}   & \colhead{$z_{\rm abs}$}   & \colhead{$\rm log \ N_{HI}$}   & \colhead{$[{\rm X/H}]$}   & \colhead{$\log U$}   & \colhead{$\rm \alpha_{EUVB}$}   & \colhead{$[{\rm C/\alpha}]$}   \\
           \colhead{ }    & \colhead{ }               & \colhead{[cm$^{-2}$]}          & \colhead{ }               & \colhead{ }          & \colhead{ }                     & \colhead{ }                   }
\startdata
1        & 0.717956       & $15.25 \pm 0.05$       & $-$1.23, $-$1.02, $-$0.81 & $-$1.72, $-$1.37, $-$1.05 & $<$$-$1.93, $<$$-$1.69, $<$$-$1.46 & $-$0.48, $-$0.23, $-$0.02  \\ 
2        & 0.556468       & $15.38 \pm 0.03$       & $-$0.30, $-$0.03, +0.32   & $-$2.23, $-$1.85, $-$1.31 & $-$1.91, $-$1.72, $-$1.52          & $-$0.59, $-$0.11, +0.46    \\ 
3        & 0.718956       & $15.61 \pm 0.02$       & $-$0.92, $-$0.76, $-$0.59 & $-$2.64, $-$2.46, $-$2.31 & $-$1.92, $-$1.74, $-$1.53          & +0.34, +0.50, +0.64        \\ 
4        & 0.492632       & $15.62 \pm 0.05$       & +0.04, +0.14, +0.26       & $-$3.66, $-$3.39, $-$3.13 & $-$1.93, $-$1.77, $-$1.55          & $-$0.73, $-$0.40, $-$0.03  \\ 
5        & 0.615662       & $15.66 \pm 0.01$       & +0.04, +0.14, +0.26       & $-$2.62, $-$2.48, $-$2.36 & $-$1.91, $-$1.73, $-$1.54          & $-$0.58, $-$0.47, $-$0.37  \\ 
6        & 0.452768       & $15.68 \pm 0.01$       & $-$0.38, $-$0.19, $-$0.01 & $-$3.90, $-$3.76, $-$3.62 & $-$1.92, $-$1.74, $-$1.52          & \nodata                    \\ 
7        & 0.807754       & $15.71 \pm 0.01$       & $-$0.77, $-$0.63, $-$0.49 & $-$2.47, $-$2.30, $-$2.15 & $<$$-$1.95, $<$$-$1.71, $<$$-$1.46 & $-$0.39, $-$0.20, +0.05    \\ 
8        & 0.536490       & $15.76 \pm 0.02$       & $-$1.43, $-$1.07, $-$0.61 & $-$2.46, $-$2.18, $-$1.83 & $<$$-$1.94, $<$$-$1.71, $<$$-$1.46 & $-$0.59, $-$0.34, $-$0.11  \\ 
9        & 0.557529       & $15.79 \pm 0.02$       & $-$0.21, $-$0.06, +0.12   & $-$2.82, $-$2.65, $-$2.53 & $-$1.93, $-$1.76, $-$1.53          & $-$0.35, $-$0.18, $-$0.03  \\ 
10       & 0.876403       & $15.91 \pm 0.02$       & $-$0.62, $-$0.51, $-$0.40 & $-$2.94, $-$2.81, $-$2.67 & $<$$-$1.92, $<$$-$1.65, $<$$-$1.44 & $-$0.02, +0.11, +0.24      \\ 
11       & 0.516578       & $15.99 \pm 0.01$       & $-$1.09, $-$0.96, $-$0.82 & $-$2.50, $-$2.31, $-$2.15 & $<$$-$1.94, $<$$-$1.72, $<$$-$1.47 & \nodata                    \\ 
12       & 0.829173       & $16.02 \pm 0.01$       & $-$1.41, $-$1.26, $-$1.10 & $-$3.33, $-$3.18, $-$3.04 & $-$1.93, $-$1.77, $-$1.55          & $-$0.07, +0.03, +0.12      \\ 
13       & 0.481677       & $16.02 \pm 0.01$       & $-$0.84, $-$0.75, $-$0.65 & $-$3.14, $-$3.01, $-$2.89 & $<$$-$1.94, $<$$-$1.70, $<$$-$1.46 & $-$0.68, $-$0.58, $-$0.47  \\ 
14       & 0.621428       & $16.09 \pm 0.01$       & +0.07, +0.09, +0.13       & $-$3.00, $-$2.97, $-$2.94 & $-$1.99, $-$1.95, $-$1.88          & $-$0.20, $-$0.17, $-$0.15  \\ 
15       & 0.619130       & $16.10 \pm 0.01$       & $-$0.00, +0.38, +0.84     & $-$1.64, $-$1.30, $-$1.01 & $-$1.92, $-$1.75, $-$1.53          & $-$0.88, $-$0.61, $-$0.24  \\ 
16       & 0.558200       & $16.13 \pm 0.02$       & $-$1.02, $-$0.75, $-$0.36 & $-$4.12, $-$3.73, $-$3.35 & $<$$-$1.94, $<$$-$1.69, $<$$-$1.46 & $-$0.60, $-$0.21, +0.16    \\ 
17       & 0.688948       & $16.16 \pm 0.03$       & $-$0.44, $-$0.33, $-$0.21 & $-$2.62, $-$2.47, $-$2.33 & $<$$-$1.93, $<$$-$1.68, $<$$-$1.46 & $-$0.70, $-$0.44, $-$0.13  \\ 
18       & 0.533507       & $16.17 \pm 0.04$       & $-$0.06, +0.03, +0.14     & $-$2.95, $-$2.81, $-$2.72 & $-$1.94, $-$1.80, $-$1.60          & $-$0.91, $-$0.72, $-$0.45  \\ 
19       & 0.518500       & $16.22 \pm 0.02$       & $-$0.30, $-$0.27, $-$0.25 & $-$3.85, $-$3.81, $-$3.77 & $-$1.45, $-$1.42, $-$1.40          & $-$0.60, $-$0.56, $-$0.51  \\ 
20       & 0.837390       & $16.36 \pm 0.02$       & $-$1.69, $-$1.55, $-$1.39 & $-$2.72, $-$2.52, $-$2.35 & $-$1.93, $-$1.75, $-$1.52          & $-$0.25, $-$0.05, +0.19    \\ 
21       & 0.682684       & $16.41 \pm 0.01$       & $-$1.45, $-$1.31, $-$1.13 & $-$2.36, $-$2.15, $-$2.00 & $-$1.95, $-$1.82, $-$1.59          & $-$0.71, $-$0.70, $-$0.68  \\ 
22       & 0.468486       & $16.48 \pm 0.07$       & $-$0.65, $-$0.54, $-$0.38 & $-$3.37, $-$2.98, $-$2.73 & $-$1.95, $-$1.83, $-$1.64          & $-$0.18, $-$0.04, +0.10    \\ 
23       & 0.712976       & $16.48 \pm 0.02$       & $-$1.57, $-$1.43, $-$1.29 & $-$2.88, $-$2.70, $-$2.54 & $<$$-$1.94, $<$$-$1.71, $<$$-$1.46 & \nodata                    \\ 
24       & 0.686086       & $16.50 \pm 0.02$       & $-$0.29, $-$0.25, $-$0.22 & $-$2.91, $-$2.87, $-$2.83 & $-$1.99, $-$1.96, $-$1.91          & +0.00, +0.05, +0.11        \\ 
25       & 0.836989       & $16.52 \pm 0.02$       & $-$1.39, $-$1.29, $-$1.17 & $-$3.03, $-$2.89, $-$2.77 & $-$1.92, $-$1.74, $-$1.53          & $-$0.32, $-$0.19, $-$0.01  \\ 
26       & 0.491977       & $16.53 \pm 0.02$       & $-$0.24, $-$0.21, $-$0.18 & $-$3.12, $-$3.06, $-$2.99 & $-$1.99, $-$1.96, $-$1.90          & $-$0.25, $-$0.19, $-$0.14  \\ 
27       & 0.556684       & $16.54 \pm 0.03$       & $-$3.01, $-$2.66, $-$2.12 & $-$2.91, $-$2.17, $-$1.45 & $<$$-$1.94, $<$$-$1.69, $<$$-$1.46 & \nodata                    \\ 
28       & 0.728885       & $16.63 \pm 0.05$       & $-$2.82, $-$2.61, $-$2.25 & $-$1.91, $-$1.48, $-$1.13 & $-$1.87, $-$1.75, $-$1.59          & $-$0.64, $-$0.46, $-$0.19  \\ 
29       & 0.470800       & $16.64 \pm 0.03$       & $-$1.00, $-$0.92, $-$0.85 & $-$3.52, $-$3.36, $-$3.20 & $<$$-$1.70, $<$$-$1.50, $<$$-$1.42 & \nodata                    \\ 
30       & 0.738890       & $16.67 \pm 0.01$       & $-$1.07, $-$0.96, $-$0.86 & $-$2.91, $-$2.79, $-$2.65 & $<$$-$1.93, $<$$-$1.65, $<$$-$1.43 & $-$0.44, $-$0.36, $-$0.29  \\ 
31       & 0.507220       & $16.70 \pm 0.03$       & $-$1.19, $-$1.07, $-$0.94 & $-$3.18, $-$3.04, $-$2.90 & $<$$-$1.94, $<$$-$1.69, $<$$-$1.46 & $-$0.88, $-$0.61, $-$0.28  \\ 
32       & 0.689370       & $16.91 \pm 0.06$       & $-$0.33, $-$0.26, $-$0.21 & $-$3.13, $-$3.11, $-$3.09 & $-$1.43, $-$1.41, $-$1.40          & $-$0.60, $-$0.46, $-$0.32  \\ 
33       & 0.575283       & $17.15 \pm 0.15$       & $-$0.17, +0.52, +1.26     & $-$3.63, $-$3.47, $-$3.29 & $<$$-$1.94, $<$$-$1.70, $<$$-$1.46 & \nodata                    \\ 
34       & 0.552615       & $17.25 \pm 0.15$       & $-$1.37, $-$1.15, $-$0.92 & $-$3.32, $-$3.15, $-$2.98 & $-$1.92, $-$1.73, $-$1.52          & \nodata                    \\ 
\enddata
\tablenotetext{a}{The lower and upper bounds represent the 68\% CI for detections and the 80\% CI for upper and lower limits. The `$<$' symbols indicate that a value is an upper limit. }
\end{deluxetable*}

\end{center}

\subsection{The Metallicity-\alphaeuvb\ Relationship} \label{sec:metslope}

\begin{figure*}[tbp]
\epsscale{1.1}
\plotone{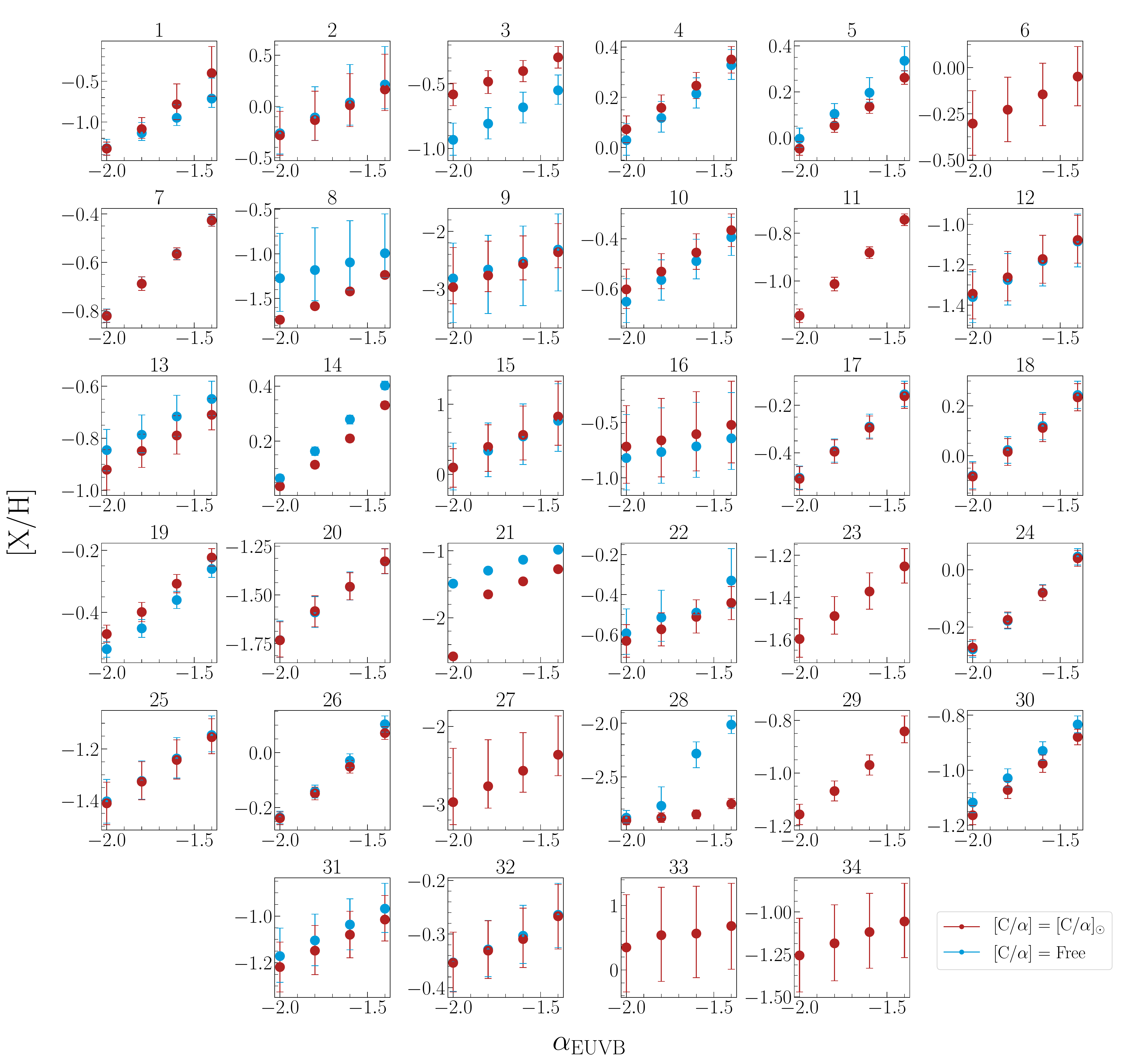}
\caption{The relationship between \xh\ and \alphaeuvb\ for the 34 absorbers in our sample. This illustrates the results of our single EUVB slope runs where we calculate the metallicity for \alphaeuvb\ values of $-2.0$, $-1.8$, $-1.6$, and $-1.4$. We show results for models with \ca\ fixed to solar (red) and models where we placed a flat prior on \ca\ (blue). Not all absorbers show flat prior results due to a lack of detected carbon and/or $\alpha$-elements. There is a clear trend of metallicity increasing as \alphaeuvb\ increases, and the two sets of models are statistically consistent in the majority of systems. For absorbers 21 and 28, the MCMC modeling is pushing the inferred \NHI\ inference away from the observed value for low and high values of \alphaeuvb\ respectively, leading to unreliable results.
\label{fig:metslope}
}
\end{figure*}

As stated above, we calculated four sets of models with fixed EUVB slopes spanning the full range of KS19 slope values ($\alphaeuvb = [-2.0, -1.8, -1.6, -1.4]$) to assess how the metallicity estimate depends on the slope of the EUVB. These results are summarized in Figure~\ref{fig:metslope}, where we show \xh\ as a function of \alphaeuvb\ for the 34 absorbers in our sample. The error bars span the 68\% confidence interval (16th--84th percentiles) of the metallicity posterior distribution functions (PDFs). The numbers above each panel correspond to the ID numbers shown in Tables~\ref{tab:sample_props} and \ref{tab:results_tab}. Red symbols refer to runs where \ca\ was fixed to solar, and blue symbols refer to runs where \ca\ was included as a free parameter with a flat prior. 

For the vast majority of the absorbers in our sample, the metallicity estimate between the two different \ca\ implementations are consistent with each other to within $1\sigma$, with some noticeable exceptions. Absorbers 3 and 14 are not consistent within $1\sigma$, and absorbers 21 and 28 have some large discrepancies, especially at low and high values of \alphaeuvb, respectively. In the case of absorber 3, the discrepancy is not that large (consistent to within $1.5$--$2\sigma$) and improves as \alphaeuvb\ increases. The errors on the metallicities of absorber 14 are very small, and the difference between the two metallicities is never larger than roughly 0.05 dex. In the case of absorbers 21 and 28, the MCMC modeling pushes the inferred \NHI\ value away from the observed value for certain values of \alphaeuvb. For absorber 21, this occurs for low values of \alphaeuvb\ and only for the model with \ca\ fixed to solar. It is in this low \alphaeuvb\ regime where the two models are most discrepant. Absorber 28 disagrees with the observed \NHI\ value at the lowest \alphaeuvb\ for the model with a flat prior on \ca\ and at high values of \alphaeuvb\ for the model with \ca\ fixed to solar. For both absorbers, this seems to be a result of the fixed background and does not occur when the EUVB slope is included as a free parameter in the modeling. In the discussion that follows we do not include the results for these two absorbers.

\begin{figure}[tbp]
\epsscale{1.15}
\plotone{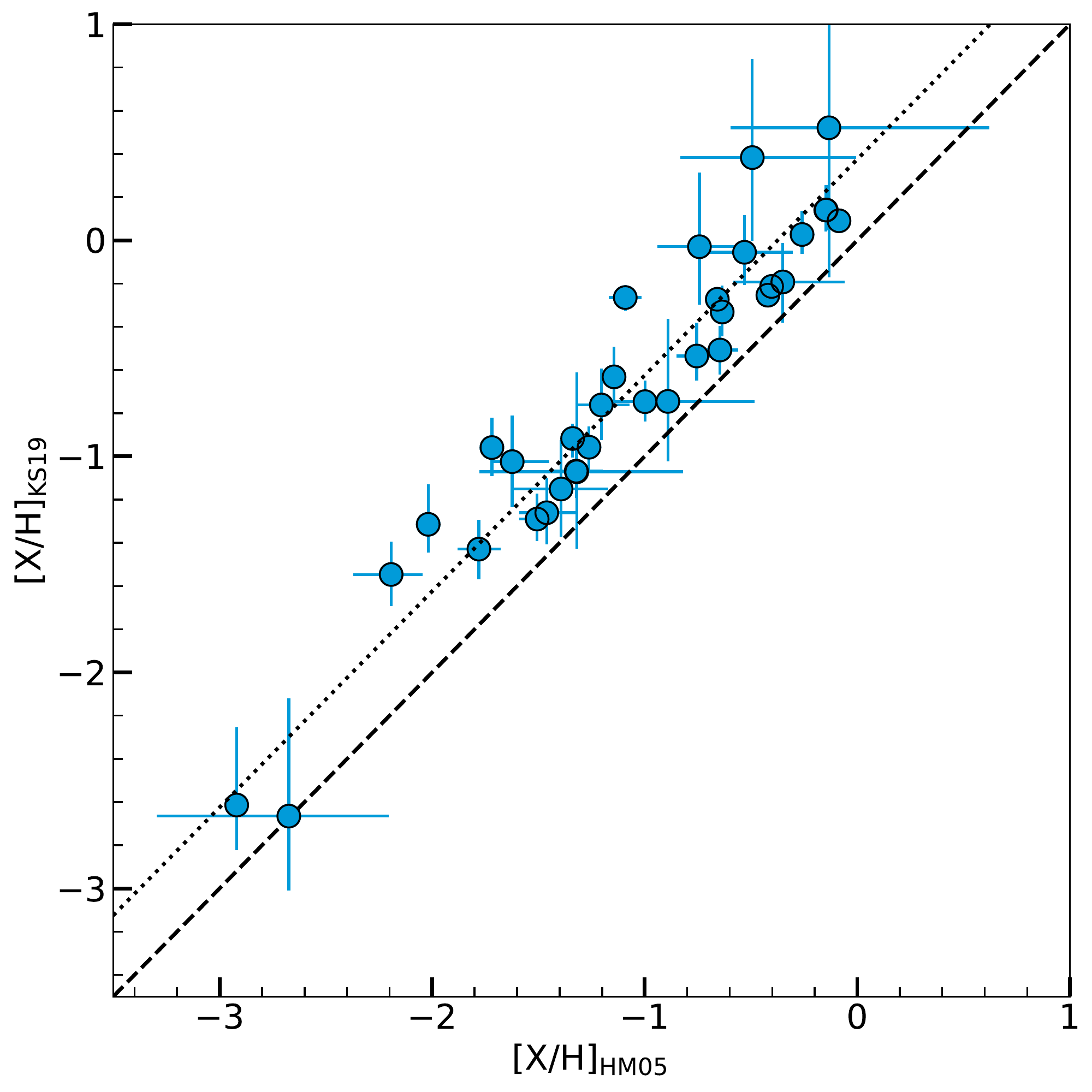}
\caption{Comparison of metallicities from our adopted model with the metallicities derived in \citetalias{wotta2019} and \citetalias{lehner2019}, which used the same model with the exception of the choice of EUVB. The dashed black line shows the one-to-one correspondence indicating that the metallicities derived in this work are larger than those derived with the \citetalias{haardt1996} background. The dotted line shows the median of the differences between the two metallicity estimates. The metallicities derived here are larger by about 0.3 dex on average compared to those derived with the HM05 background.
\label{fig:ksvshm05}
}
\end{figure}

The overall trend from Figure~\ref{fig:metslope} is that there is an obvious positive relationship between \xh\ and \alphaeuvb: the inferred metallicity increases as \alphaeuvb\ increases (i.e., as the radiation background becomes harder) regardless of the modeling of \ca. The magnitude of this increase varies from absorber to absorber but generally varies by about 0.2 dex to 0.4 dex. This trend has been observed before when comparing softer and harder and ionization backgrounds (see below, and e.g., \citealt{wotta16,chen17}; \citetalias{wotta2019}). In   Figure~\ref{fig:ksvshm05}, we compare $\xh_{\rm KS19}$ against $\xh_{\rm HM05}$, i.e., the metallicities derived here for our fiducial model versus those derived in \citetalias{wotta2019} and \citetalias{lehner2019}. The metallicities derived here are larger than those of \citetalias{wotta2019} and \citetalias{lehner2019} for all absorbers in our sample by $\approx 0.3$ dex on average. This is consistent with Figure~9 in \citetalias{wotta2019} that shows that by using a harder background, such as \citetalias{haardt2012}, the inferred metallicity increases (see also \citealt{zahedy19}).

To further quantify the increase of the metallicities with harder EUVBs, here we report the range in metallicities that are inferred over the range in \alphaeuvb\ explored here. On average, the metallicities span a range of 0.35 dex or a factor of $\approx 2.2$. The span in metallicity predictions are very similar regardless of the \ca\ implementation. The minimum range in metallicity prediction is the same for both choices of \ca\ modeling at 0.09 dex (a factor of 1.2). However, the maximum range is different at 0.66 dex for the \ca\ flat prior and 0.93 dex for the model with fixed \ca\ (factors of 4.6 and 8.5, respectively). We discuss these trends more in Section~\ref{sec:error_disc}.  

\begin{figure}[tbp]
\epsscale{1.20}
\plotone{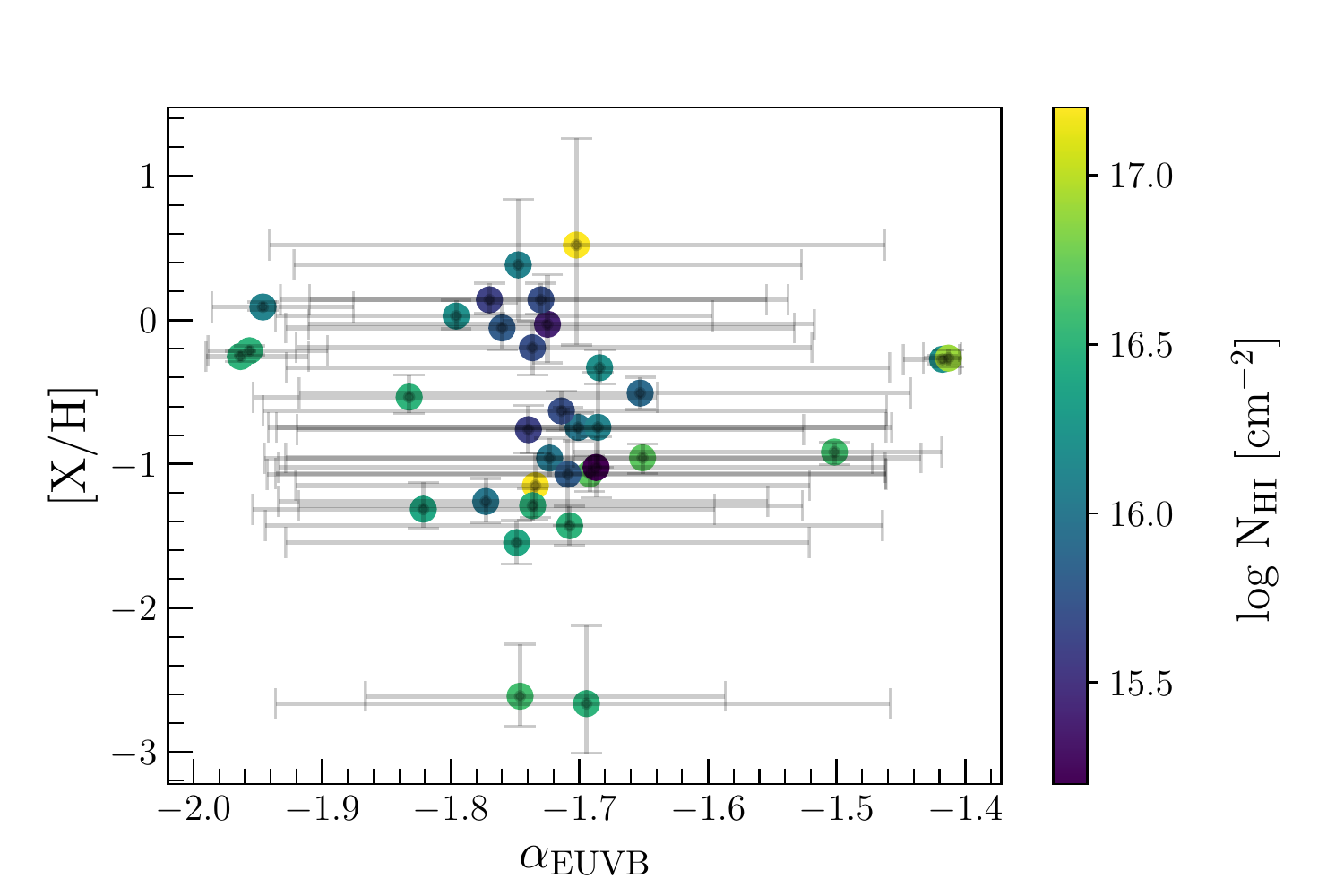}
\caption{The relationship between \xh\ and \alphaeuvb\ for the 34 absorption systems in our sample when \alphaeuvb\ is included in our model described with a flat prior. The error bars represent the 16th and 84th percentiles of their respective posterior probability distribution functions. The absorbers are colored by observed \NHI\ to assess any trends with \HI\ column density. Although there is a clear clustering for values of \alphaeuvb\ around $-1.7$ to $1.8$, the uncertainties here are large and is likely an artifact of being in the center of the \alphaeuvb\ distribution. The six absorbers at the edge of the \alphaeuvb\ range have probability distributions strongly skewed toward one bound and will be discussed in more detail in Section~\ref{sec:disc}.
\label{fig:metalphaadopted}
}
\end{figure}

\begin{figure}[tbp]
\epsscale{1.15}
\plotone{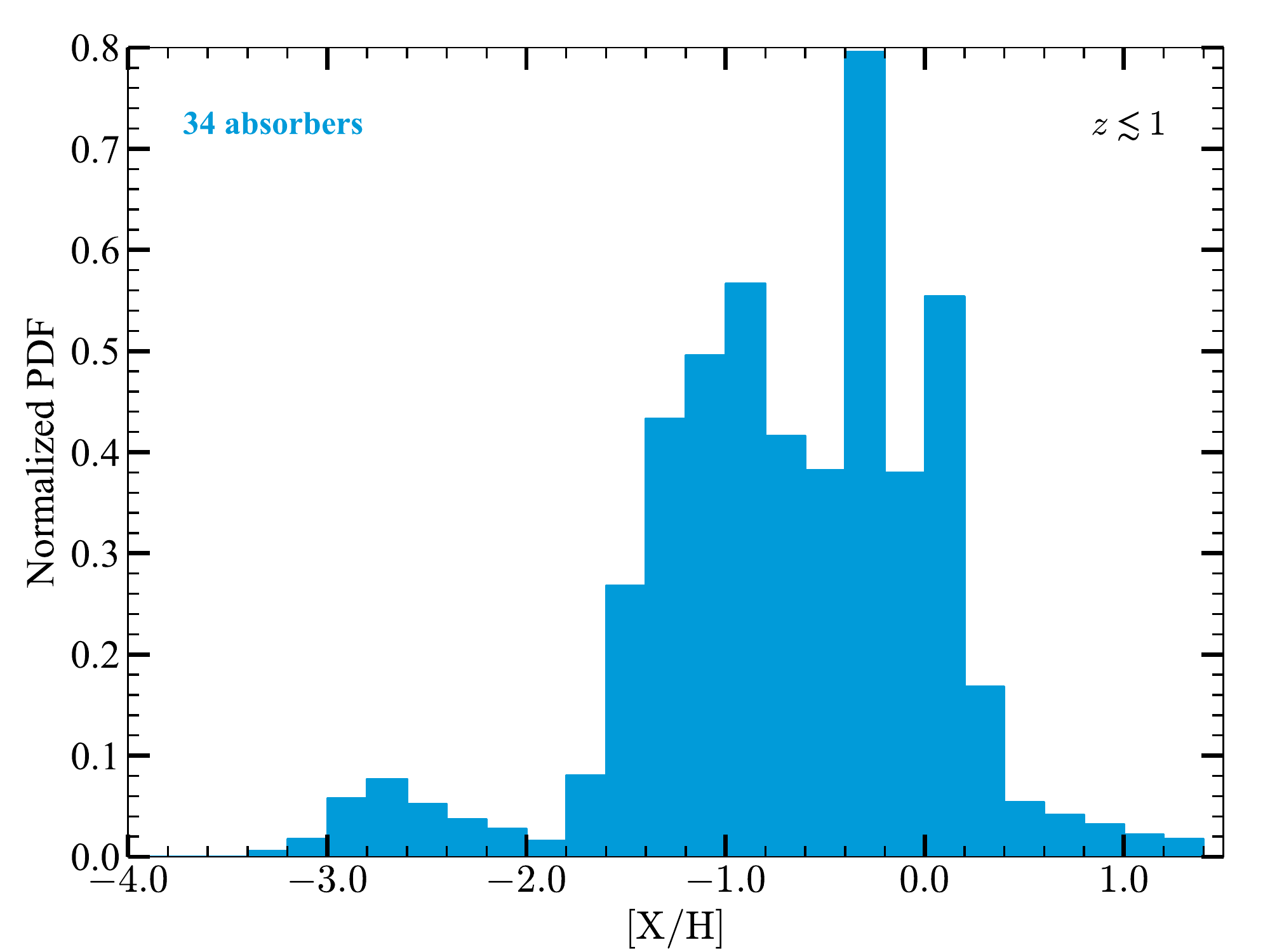}
\caption{Normalized metallicity distribution for the 34 absorbers in our sample showing that our sample of absorbers tend to have high metallicities.
\label{fig:metpdf}
}
\end{figure}

In Figure~\ref{fig:metalphaadopted}, we show \xh\ versus \alphaeuvb\ for our adopted model where the EUVB slope is included as a free parameter modeled with a flat prior. From this figure it can be seen that the constraints on \alphaeuvb\ from the ions accessible for these absorbers are crude, giving large errors, especially when compared to the errors on \xh\ (the error bars show the 68\% confidence interval). Despite the generally large uncertainties on \alphaeuvb, there is a clear clustering around $-1.8$ (though this may be due to this value being in the middle of our explored range of \alphaeuvb\ values), with a range of metallicities for each \alphaeuvb. The coloring indicates the \NHI\ value of the absorber, and there is no clear relationship between \NHI, \alphaeuvb, or \xh. However, it should be noted that our sample is not necessarily representative of the low-redshift CGM given that our absorbers tend to have higher metallcities and, given the EUV ions we targeted, higher redshifts (see Figure~\ref{fig:metpdf}). Interestingly, the relationship between \xh\ and \alphaeuvb\ is only seen on an absorber-by-absorber basis and does not exist within the entire sample when \alphaeuvb\ is allowed to vary. This shows that the derived metallicities (the median of the posterior PDF) are not strongly affected by including \alphaeuvb\ as a free parameter and instead are driven more by other factors such as the observed ionic column densities and ionic ratios. However, the large uncertainties on \alphaeuvb\ leave open the possibility of a relationship between \alphaeuvb\ and \xh\ to be uncovered in the future if \alphaeuvb\ values can be better constrained.

\subsection{Impact of Marginalizing over \alphaeuvb}\label{sec:met_results_robust}
\begin{figure}[tbp]
\epsscale{1.15}
\plotone{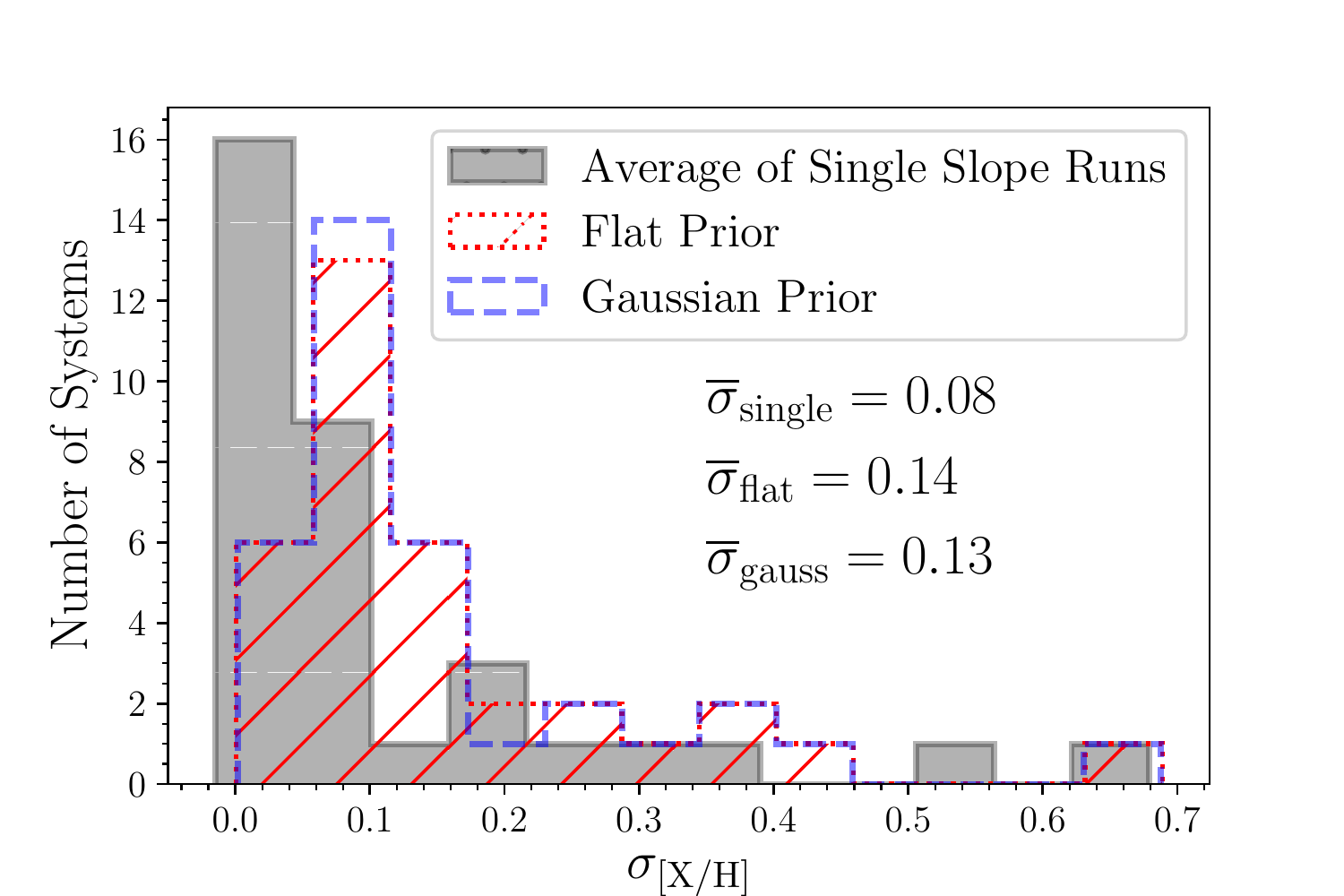}
\caption{Frequency of the metallicity uncertainties for our three different models expressed as number of systems per 1-$\sigma$ uncertainty given by the 84th percentile minus the 16th percentile divided by two. To summarize the eight different metallicity estimates from the runs with fixed \alphaeuvb\ we averaged the uncertainties of each of these runs (gray histogram). The flat prior on \alphaeuvb\ (red histogram) and Gaussian prior on \alphaeuvb\ (blue histogram) give almost identical results. The average uncertainty for each model run are shown below the figure legend. Uncertainties are smaller on average when using a single EUVB slope value, but marginalizing over uncertainties in \alphaeuvb\ generally does not increase the metallicity uncertainties by a large amount.
\label{fig:errorcomp}
}
\end{figure}

An important remaining question is whether and to what extent including \alphaeuvb\ as a free parameter affects the robustness of metallicity estimates. To this end, in Figure~\ref{fig:errorcomp}, we show a histogram of the 1-$\sigma$ uncertainties, in derived metallicities, approximated for our absorbers as the 84th percentile of the distribution minus the 16th percentile of the distribution divided by two. Given that there are eight different metallicity estimates for the single slope runs (four \alphaeuvb\ values and two \ca\ implementations), we estimate the uncertainties for these runs by taking the average of the absorber uncertainties for each run. It is immediately clear that the uncertainty distribution for the Gaussian prior on \alphaeuvb\ and the flat prior on \alphaeuvb\ are virtually identical. Thus, using a broad Gaussian prior on \alphaeuvb\ does not lead to better constrained metallicities than those derived by using a flat prior on \alphaeuvb. We will discuss the role of this prior choice in more detail in the next section.

Compared to the Gaussian and flat prior on \alphaeuvb, the runs that assume a single EUVB slope have smaller uncertainties on average with 22 of the 34 absorbers having uncertainties less than 0.1 dex, whereas the Gaussian and flat runs have ten and eight absorbers with uncertainties less than 0.1 dex, respectively. We also show the standard deviations of the metallicity uncertainties for each class of models, which are 0.08, 0.14, and 0.13 dex for the single slope runs, flat prior run, and Gaussian prior run, respectively. This implies that the uncertainties in the underlying EUVB actually dominate the error budget. To go from an uncertainty of 0.08 to 0.14 dex for normally distributed variables would imply a metallicity uncertainty from the EUVB slope alone of 0.11 dex, which is greater than the uncertainty of 0.08 dex arising from our data and modeling. So, while there is not a critical difference between the two types of models (single EUVB slope and allowing the EUVB slope to vary) in terms of the uncertainties they impart on metallicity estimates, the uncertainty in the EUVB slope is an important, if not dominant, inclusion.

We note that another potential source of uncertainty in the shape of the radiation field is the impact of leaking radiation from local (i.e., within the host galaxy) ionizing radiation sources. Several studies have found that, within the Milky Way, local sources of radiation can actually dominate the ionization state of absorbers over radiation from the EUVB if they lie within $\approx 50$--100 kpc of the Galaxy \citep{giroux97, bland-hawthorn99,fox05}, or about $\la 1/4 R_{\rm vir}$ (where $R_{\rm vir}$ is the virial radius, about 300 kpc for the Milky Way). Other works have noted that local sources of ionizing radiation may be important for observed absorption systems at higher redshifts, $z \approx 1$--3 \citep{dodorico01, dessauges03, meiring09},  but these are typically much stronger \HI\ absorbers and closer to their host galaxies (see \citealt{lehner2013}, M.~Berg et al.~2021, in prep.). Our sample is predominantly comprised of low density SLFSs and pLLSs, which are typically located farther than 50--100 kpc from their host galaxies or beyond $\ga 1/4 R_{\rm vir}$ (M.~Berg et al.~2021, in prep.), and therefore unlikely to be strongly affected by the radiation coming from the host galaxy. Furthermore, insights as to the importance of local radiation sources at higher redshifts are not as conclusive as studies of the Galaxy have been. \citet{fumagalli16} studied a sample of 157 absorption systems between redshifts of 1.8 and 4.4. In their work, they tested whether local contributions from a galaxy or QSO led to tighter constraints on the physical properties of their absorber sample. The found that adding local sources of ionizing radiation did not lead to a better agreement between their model column densities and observational column densities and that a large contribution from local galaxies or QSOs had a low probability. Therefore, local ionizing sources  are unlikely to greatly affect the overall EUVB slope for the absorbers considered here and hence the metallicities.

\subsection{Constraints on the EUVB Slope} \label{sec:slope_constraints}

\subsubsection{Constraints for Individual Absorption Systems} \label{sec:ind_systems}

\begin{figure*}[tbp]
\epsscale{1.1}
\plotone{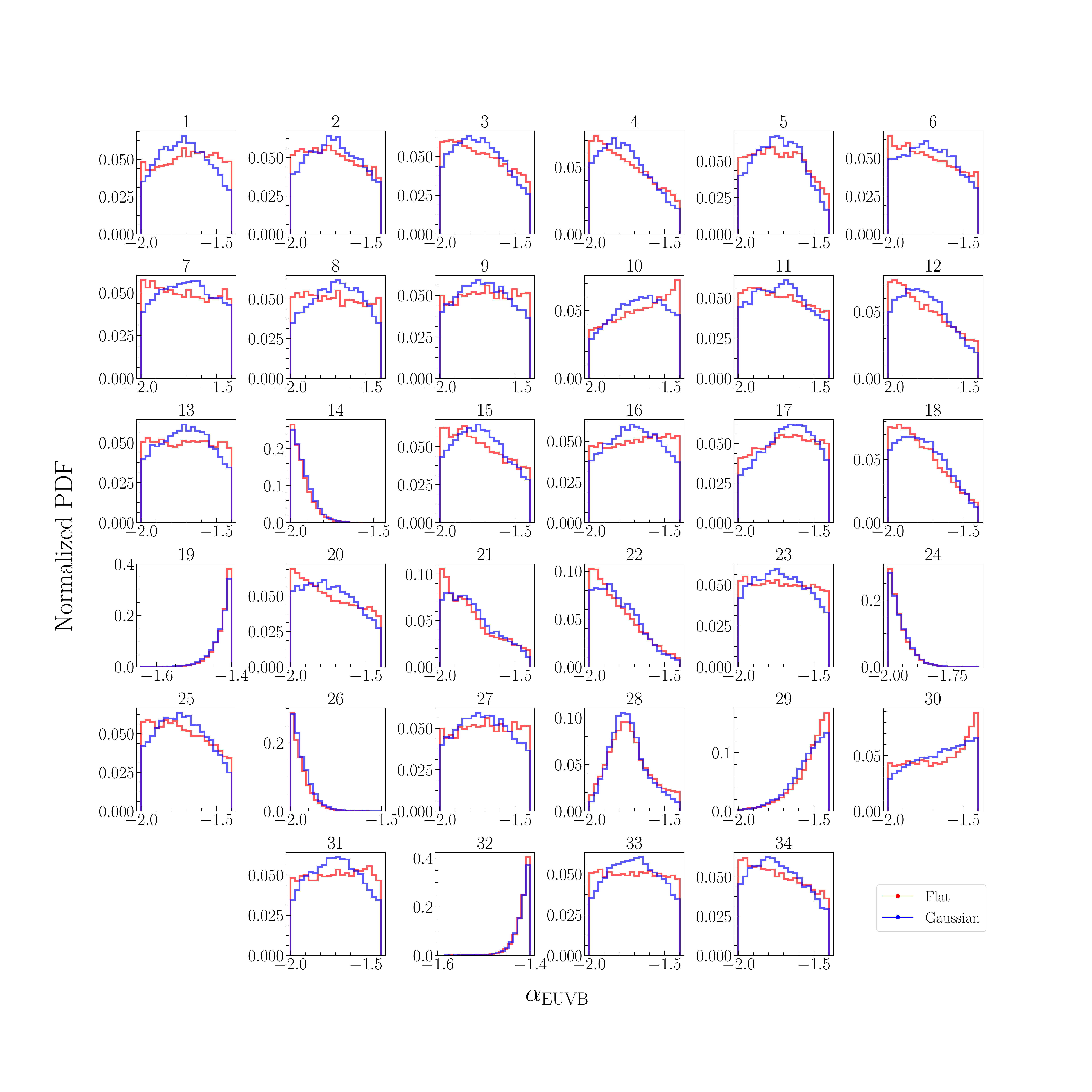}
\caption{Individual \alphaeuvb\ posterior probability distributions for the 34 absorption systems in our sample. Red histograms show the results when assuming a flat prior on \alphaeuvb, and the blue histograms show the results for a Gaussian prior. There are three noticeable `types' of \alphaeuvb\ posteriors ranging from broad/flat, broad/skewed, and skewed toward either bound of the \alphaeuvb\ range. Switching to a Gaussian prior generally lowers the probability at the edges of flat distributions but does little for the posteriors that were strongly skewed with a flat prior. Note that out of all of the absorbers, absorber 28 is the only one with a quasi-Gaussian shape fairly well centered on $\alphaeuvb = -1.8$.
\label{fig:slope_posterior}
}
\end{figure*}

Another goal of this work is to arrive at constraints on the value of \alphaeuvb\ or to, at least, assess how well \alphaeuvb\ can be constrained given the large uncertainties known to exist in regard to the EUVB. We selected a sample of absorbers most likely to have some constraining power over the EUVB slope, at least over the redshift and column density range explored here. We have already seen in Figure~\ref{fig:metalphaadopted} that the derived uncertainties on \alphaeuvb\ as expressed by 84th and 16th percentiles of the \alphaeuvb\ posterior probability distribution are large on average. In order to more clearly show the constraints we derive on \alphaeuvb, we show the individual \alphaeuvb\ posterior probability distribution functions for the 34 absorbers in our sample in Figure~\ref{fig:slope_posterior}. The Gaussian prior on \alphaeuvb\ results are shown in blue, and the flat prior results are shown in red.

Looking first at the results derived with a flat prior, we can see that for the vast majority of absorbers the EUVB slope is not well constrained as evidenced by the broad and flat appearance of the distributions. Not all of the broad posteriors are perfectly flat with some of the distributions, such as those for absorbers 4, 10, 18, and 20, having excess probability at either end of the \alphaeuvb\ range, while still spanning the full range. There are a handful of systems that have distributions strongly skewed toward one of the \alphaeuvb\ bounds and little to no probability density at the other bound. Absorbers 14, 24, and 26 are skewed toward the lower bound ($\alphaeuvb = -2.0$) while absorbers 19, 29, and 32 are skewed toward the upper bound at ($\alphaeuvb = -1.4$), which could indicate an insufficient range in \alphaeuvb. These six outliers will be discussed in more detail in Section~\ref{sec:disc}. 

Turning to the effects that switching to a Gaussian prior has on the posterior distributions for \alphaeuvb, we see the effects that using a slightly more informative prior have on \alphaeuvb\ determinations and whether it leads to more robust inferences on \alphaeuvb. Looking first at the absorbers that have broad and flat \alphaeuvb\ posteriors (absorbers 1, 9, 27, 31, etc.),~we see that the Gaussian prior does tend to produce more convex distributions as compared to the flat prior distributions. In other words, the probability density at the bounds of the distribution decreases, while increasing slightly toward the middle of the \alphaeuvb\ range at values between $-1.7$ to $-1.8$. For the absorbers that are still fairly broad, but not necessarily flat, the Gaussian prior tends to lower the probability density at the end of the flat distribution with excess probability. Thus, for these absorbers the distributions are still fairly broad, but the maximum probability density is no longer as strongly skewed toward one bound. For the six absorbers with strongly skewed probability distributions when using a flat prior, the effect of switching to a Gaussian prior is negligible. Given the almost identical result for these six absorbers between the two priors, we explore the meaning of this strong and persistent skew in Section~\ref{sec:slope_disc}.

There are three main types of \alphaeuvb\ posterior probability distributions: flat and broad, broad with a skew, and strongly skewed. Absorber 28, however, is markedly different with a fairly well-constrained \alphaeuvb\ as shown by the Gaussian shape of its PDF. For the flat prior on \alphaeuvb, the median value of the slope is $-1.75$ with 16\% and 84\% confidence limits of $-1.87$ and $-1.59$. The shape of this absorber’s posterior is largely independent of the choice of \alphaeuvb\ prior, as both the flat and the Gaussian prior lead to a tightly constrained \alphaeuvb\ value. The exact reason why this absorber is so well constrained is not immediately clear given that the metal ions detected for this absorber are not unique. It is only one of two absorbers with low metallicities ($\xh < -2$), but the other low metallicity system does not have as many detected metal ions, providing fewer constraints to the ionization model. We explore what makes absorber 28 unique in Section~\ref{sec:slope_disc}.  

\subsubsection{Combined Posterior on the EUVB Slope} \label{sec:tot_post}

In the previous section, we explored the \alphaeuvb\ posterior distributions of our 34 absorption systems noting the general shapes exhibited by the PDFs of those systems. Generally, we saw that the majority of the absorbers exhibited \alphaeuvb\ posteriors that are both broad and flat indicating that, for those systems, the slope of the ionizing background inferred is not well constrained between the \alphaeuvb\ range explored here. Other systems exhibited broad \alphaeuvb\ posteriors that had slight excess probability on either one of the bounds of \alphaeuvb\ values showing that the inferred EUVB slope in these absorbers was still poorly constrained but favored harder or softer backgrounds. A smaller number showed a strong skew toward either the hard or soft edge of the \alphaeuvb\ range and one absorber had a quasi-Gaussian posterior with a mean of $\alphaeuvb = -1.75$. Given this diversity in the \alphaeuvb\ posteriors of individual absorbers, what, if anything, can be said about the total posterior on \alphaeuvb\ inferred from the 34 individual \alphaeuvb\ posteriors in our sample?

In an ideal scenario, there would be more absorption systems with Gaussian-like posteriors and it would be clearer if a certain value of \alphaeuvb\ was favored, such as $\alphaeuvb = -1.8$, the fiducial value of \citetalias{KS19}. It is interesting to note that the majority of our systems have inferred median values around $\approx -1.8$ (as does absorber 28), but this is likely a function of being near the middle of the \alphaeuvb\ range we explore. Regardless, we can combine the individual inferences on \alphaeuvb\ to arrive at a net PDF on \alphaeuvb\ taking into account the information from the individual absorbers. The information that we have are the individual PDFs for \alphaeuvb, based on a total of 100,000 walkers, each giving a value of \alphaeuvb. These absorbers are a small sub-sample of the thousands of known absorption systems in the  $z< 1$ circumgalactic medium meaning that the inferred \alphaeuvb\ values from our sample may not be representative of the intrinsic \alphaeuvb\ distribution for all CGM absorbers (as discussed in Section~\ref{sec:data}). Simply taking the mean \alphaeuvb\ values from each of the 100,000 values in the posteriors and treating that as the intrinsic \alphaeuvb\ distribution for CGM absorbers does not actually inform us about the true value of \alphaeuvb. Instead, to come to conclusions about the intrinsic distribution of \alphaeuvb\ values, we need to make some assumptions about the parent distribution they were sampled from in conjunction with the PDFs from our individual absorbers.  

Hierarchical Bayesian modeling (HBM) does exactly that by using Bayesian inference and MCMC techniques to sample from the posterior of the parameters of the intrinsic distribution \citep{betancourt2015}. In other words, in HBM, the intrinsic/parent distribution of \alphaeuvb\ values is assumed to be characterized by some probability distribution, such as a Gaussian, with unknown properties, and this assumption is combined with our sample of individual \alphaeuvb\ PDFs to infer the properties of the parent distribution, which, in the case of a Gaussian parent distribution, would be the mean and standard deviation. By assigning priors to the mean and standard deviation of the parent distribution the problem can be solved using Bayesian inference. The likelihood for each absorber can be written by taking the product of the assumed parent distribution and the \alphaeuvb\ posterior from that absorber. This general procedure is described in detail in the appendix of \citet{Baronchelli2020}. Different distributions can be assumed to represent the intrinsic distribution of \alphaeuvb. For our purposes, we use the Python module $\tt{PosteriorStacker}$\footnote{\url{https://github.com/JohannesBuchner/PosteriorStacker}} to perform the MCMC sampling of the parent \alphaeuvb\ distribution using the individual \alphaeuvb\ PDFs. $\tt{PosteriorStacker}$ takes in a sample of posterior distributions (e.g., the 100,000 \alphaeuvb\ positions per absorber) and uses $\tt{UltraNest}$ \citep{buchner2021} to perform the HBM of the intrinsic parameter distribution. $\tt{PosteriorStacker}$ performs two separate HBM iterations with one assuming a Gaussian prior on the parent distribution and the other assuming a flat Dirichlet prior on the parent distribution. The Dirichlet prior is non-parametric with a user-defined number of bins (we choose 20 bins) over the range of \alphaeuvb. In both cases, $\tt{PosteriorStacker}$ uses an importance sampling assumption to express the likelihoods as:

\begin{equation} \label{eqn:prob_eqn}
\mathcal{L} \approx \Pi_i\  \Sigma_j\  N(\alpha_{\rm EUVB}^{i,j} \vert \mu, \sigma)
\end{equation}
(see A2 of \citealt{Baronchelli2020}). Here $\{i,j\}$ correspond to the absorbers and to the individual \alphaeuvb\ walker positions for each absorber, respectively. The symbol `$N$' indicates that the distribution is Gaussian. The $\mu$ and $\sigma$ are the unknown properties of the assumed intrinsic Gaussian distribution. When this likelihood is combined with priors on $\mu$ and $\sigma$, the posterior on $\mu$ and $\sigma$ can be sampled. In the case of the Dirichlet model, Equation~\ref{eqn:prob_eqn} would use the Dirichlet distribution and the sampling would determine the probability density in each bin.

\begin{figure}[tbp]
\epsscale{1.15}
\plotone{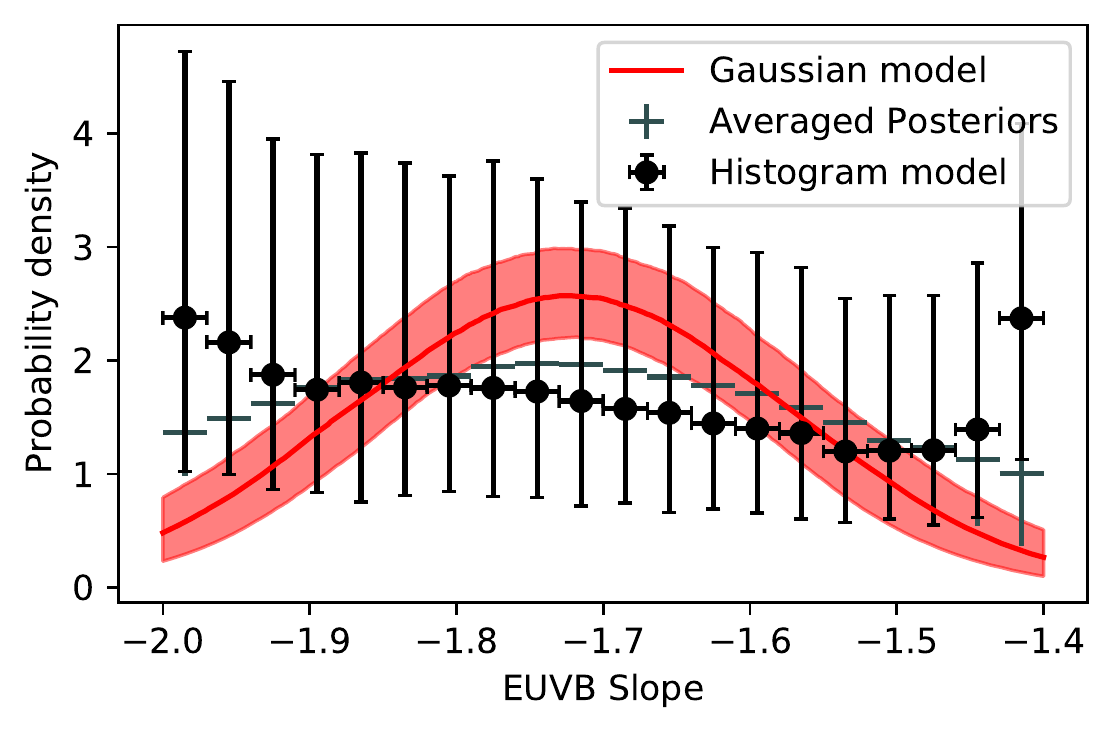}
\caption{The results of running $\tt{PosteriorStacker}$ (described in the text) on the \alphaeuvb\ posteriors from our run where we assumed a Gaussian prior on \alphaeuvb. $\tt{PosteriorStacker}$ perform two MCMC runs using UltraNest \citep{buchner2021} to infer the properties of the parent \alphaeuvb\ distribution: one where the parent distribution is assumed to be a Gaussian (red) and one with a less informative Dirichlet assumption (black symbols). The gray symbols show the distribution of mean \alphaeuvb\ values for our sample. 
\label{fig:posterior_comp}
}
\end{figure}

We present the results of our $\tt{PosteriorStacker}$ run in Figure~\ref{fig:posterior_comp}. The only inputs we provide are the individual \alphaeuvb\ posteriors for each absorber (for the model where \alphaeuvb\ is modeled as a Gaussian), the number of bins for the Dirichlet prior, and the \alphaeuvb\ range covered. In Figure~\ref{fig:posterior_comp}, the red line shows the best inferred parent Gaussian distribution with the red shading indicating the 1-$\sigma$ uncertainties on this inference. The black symbols indicate the results when assuming a Dirichlet prior on the parent \alphaeuvb\ distribution along with the 16th and 84th percentiles and the gray symbols show the distribution of mean \alphaeuvb\ values from the individual absorbers in 20 bins. As mentioned before, the gray symbols are not representative of the parent \alphaeuvb\ distribution but do reflect the distribution of \alphaeuvb\ means from our sample. The resulting mean and variance on the parent \alphaeuvb\ distribution inferred by $\tt{PosteriorStacker}$ are $\mu = -1.72 \pm 0.04$ and $\sigma = 0.15 \pm 0.02$, which we can compare with the Gaussian prior we placed on \alphaeuvb\ where we used $\mu = -1.7$ and $\sigma = 0.3$. The main effect of combining individual \alphaeuvb\ posteriors with the assumption that the parent \alphaeuvb\ distribution is Gaussian is to narrow the distribution by a factor of $\approx 2$.

The black symbols, indicating the Dirichlet model, tell a different story. This model fits for the probability density in each of the 20 bins spanning our range in \alphaeuvb\ values. It is immediately clear that the uncertainties with this model are much larger, reflecting the stronger influence from the individual \alphaeuvb\ posteriors (with large uncertainties) and the lack of influence from the assumed Dirichlet prior, which is not as informative as the Gaussian prior. The probability density is also highest on the edges of the \alphaeuvb\ distribution reflecting some of the trends apparent in the individual \alphaeuvb\ posteriors. Referring back to Figure~\ref{fig:slope_posterior}, there are many absorbers whose probabilities are strongly skewed toward the edges and a larger number that are still fairly broad but with a skew toward either edge. The majority of our systems show this with a slightly smaller number showing both broad and flat posteriors. The individual Gaussian prior posteriors show more convex appearances as compared to the flat prior posteriors as discussed in Section~\ref{sec:ind_systems}. This change leads to a higher probability density toward the middle of the \alphaeuvb\ distributions, but not by a large amount. The Dirichlet model maintains larger probability densities at the edges with roughly constant probability densities in-between. 

Taken together, the Dirichlet model and the Gaussian model give very different inferences about the parent distribution of \alphaeuvb\ values inferred from CGM absorption systems. Assuming a Gaussian distribution and inferring the intrinsic mean and variance of that distribution naturally leads to the net Gaussian posterior shown which is only weakly informed by our individual \alphaeuvb\ posteriors, which are not well-constrained. On the other hand, the weakly-informative Dirichlet prior leads to a clearly non-Gaussian total posterior reflecting the makeup of the individual \alphaeuvb\ posteriors. With more constrained individual \alphaeuvb\ posteriors (such as that of absorber 28), the Dirichlet model results should begin to approach the Gaussian model results, or if the intrinsic \alphaeuvb\ distribution deviates from a pure Gaussian (i.e., with any skew or asymmetries), the Dirichlet model should show that as well. Ultimately, the set of absorbers used here do not strongly constrain the shape of the EUVB with the notable exception of absorber 28. Future work will explore different and larger samples of absorption systems to understand what absorber properties are best suited to constrain \alphaeuvb.

\section{Discussion} \label{sec:disc}

We have explored the effects that varying the slope of the EUVB component of the EGB---which is largely responsible for ionizing circumgalactic gas---has on metallicity determinations for SLFs, pLLs, and one LLS at $z< 1$. Additionally, we explored to what extent observations of ionic column densities in our sample of absorbers can constrain the slope of the EUVB. As shown in Figure~\ref{fig:slope_posterior}, \alphaeuvb\ is generally not constrained by our sample of absorbers, but there are some exceptions such as absorber 28. On the other hand, including constraints on \alphaeuvb\ into our MCMC photionization models of these absorbers does not adversely affect our ability to constrain metallicities (along with other properties such as the hydrogen number density, $\nH$). This is seen in Figures~\ref{fig:ksvshm05} and \ref{fig:errorcomp}, where we show that the metallicities presented here are consistent with previous determinations, and that the errors on \xh\ are increased on average from 0.08 to 0.14 dex when including \alphaeuvb\ as a free parameter. Here, we elaborate on these two results by discussing to what extent the EUVB slope can be constrained (Section~\ref{sec:slope_disc}) and the impact of including uncertainties in \alphaeuvb\ on metallicity determinations (Section~\ref{sec:error_disc}). 

\subsection{How can the EUVB slope be constrained?} \label{sec:slope_disc}

A goal of this work is to provide constraints on the EUVB slope. Tentatively, we can see from Figure~\ref{fig:posterior_comp} that when including the information from all of the absorption systems and assuming a Gaussian parent distribution on \alphaeuvb, values of $\alphaeuvb \approx -1.7$ are preferred. However, when modeling the \alphaeuvb\ parent distribution with a flat Dirichlet prior, the total \alphaeuvb\ posterior remains largely unconstrained. Both of these results ultimately show that our sample of absorption systems is currently insufficient to constrain the slope of the EUVB. The Gaussian model result (red line of Figure~\ref{fig:posterior_comp}) is largely a result of the assumed Gaussian prior on the parent distribution of \alphaeuvb. If the Dirichlet prior model (black symbols of Figure~\ref{fig:posterior_comp}) more closely resembled the Gaussian prior model, this would lend more support for \alphaeuvb\ values of $\approx -1.7$. This is not the case, as the Dirichlet prior model posterior distribution on \alphaeuvb\ is essentially flat across the whole range of \alphaeuvb\ values, reflecting the results of the individual \alphaeuvb\ PDFs. The question we wish to address here is: what would it take to better constrain the EUVB slope? More specifically, what properties should CGM absorption systems have to enable constraints on \alphaeuvb?

To address this question, we explored the differences in metallicity and detected ions within our sample of absorption systems, paying particular attention to the properties of absorbers with strongly skewed \alphaeuvb\ posteriors (numbers 14, 19, 24, 26, 29, and 32) and to absorber 28 with a Gaussian posterior on \alphaeuvb. Recall, from the discussion in Section~\ref{sec:ind_systems} that absorbers 14, 24, and 26 have \alphaeuvb\ posteriors skewed toward EUVB slope values of $\approx -2.0$ (softer radiation backgrounds) and absorbers 19, 29, and 32 have \alphaeuvb\ posteriors skewed toward $\alphaeuvb \approx -1.4$ (harder radiation backgrounds). The extreme skew seen in these absorbers does not seem to be entirely the result of using a flat prior with bounds at $-1.4$ and $-2.0$, as the majority of the absorbers do not show this trend and switching to a Gaussian prior on \alphaeuvb\ has little effect on the \alphaeuvb\ posterior of these absorbers.

In short, we find no clear trends between the properties of our absorption systems and their resulting \alphaeuvb\ posterior. There is no clear difference in the ions detected, metallicity, or \HI\ column density for absorber 28 and the other absorbers in our sample whose EUVB slope is poorly constrained. Looking at the six absorbers with skewed \alphaeuvb\ posteriors, there are no clear differences in their properties either. In fact, the detected metal ions are very consistent between absorbers skewing toward $\alphaeuvb = -2.0$ and $\alphaeuvb = -1.4$. 

As discussed in Section~\ref{sec:data}, we selected the absorbers in our sample so that they have coverage of several species in multiple ionization levels (e.g., \OI, \OII, \OIII, \OIV). A priori, this should have facilitated more direct inferences on \alphaeuvb. A posteriori, this is not the case since  \alphaeuvb\ is not constrained for the majority of the absorbers. We note that several absorbers have lower or upper limits, which are typically less constraining than unsaturated detections. However, considering absorber 28, the main ions constraining the ionization models are \CIII, \OIII, and \OIV. Absorber 21 has detections of these ions plus \SIII\ and constraining upper limits on \CII\ and \OII, and yet \alphaeuvb\ remains poorly constrained in that case. 

A larger sample will be required to further explore the constraining power that absorber ionic properties may have over the shape of the EUVB, but it appears that a larger number of ionic column density \emph{ratios} with less saturated data may provide the diagnostic power for the value of \alphaeuvb\ and the shapes of the EUVB slope posteriors. As shown in Fig.~\ref{fig:metpdf}, the absorbers in our sample are biased toward metal-enriched systems compared to the metallicity distribution of the same absorbers selected solely based on their \HI\ content (see \citetalias{lehner2019}). This results in absorbers with absorption in key ions (e.g., \OII, \OIII) often being saturated. A sample with low metallicity absorbers may result with a somewhat less broad suite of ions, but may provide stronger constraints on \alphaeuvb\ because saturated lines (i.e., lower limits on $N$) will be absent or less present for those absorbers. For example, \citet{agafonova2005} and \citet{agafonova2013} reconstructed the incident EUVBs at high redshift for a set of absorption systems by adjusting certain features of the EUVB in order to reproduce various measured ionic column density ratios. We will explore this further in future work, with a larger set of absorption systems spanning a more diverse range in CGM properties, in order to arrive at a better understanding of the types of absorption systems and column density ratios with the most constraining power over the shape of the EUVB.  

\subsection{Nature of the Systematic Errors on the EUVB Slope} \label{sec:error_disc}

A key result in this work is that even if the slope of the EUVB is included as a free parameter in the modeling, robust metallicities can be derived and are consistent with those previously accounting for the systematic errors associated using a softer/harder EUVB. In fact, using the suite of backgrounds from \citetalias{KS19} essentially amounts to the difference between switching from the \citetalias{haardt1996} background to the \citetalias{haardt2012} background as discussed above (see also \citetalias{wotta2019}). As shown in Figure~\ref{fig:errorcomp}, the uncertainty on the metallicity determination is not much larger when switching from an assumed EUVB slope to a variable \alphaeuvb\ so as to render the derived metallicities unusable. The same is true for other properties such as \nH\ and \ca. Furthermore, allowing the shape of the EUVB to vary does not have any negative impacts on the more directly observable properties such as redshift and \NHI, which is not always the case for runs with a single \alphaeuvb\ value as discussed in Section~\ref{sec:metslope}, where sometimes the MCMC solution pushed the inferred \NHI\ value away from the measured \NHI\ value by up to 3--4$\sigma$. 

To put this result into a broader context, here we compare the uncertainties of our metallicity estimates with those of \citet{acharya2021} who explored the effects of varying ionization backgrounds on metallicity (and density) inferences in the CGM using toy models of metal-enriched absorption systems. They adopt the same seven backgrounds as those presented here with the addition of the background from \citet{fauchergiguere20} and \citet{puchwein2019}. They consider both photoionized absorbers and collisionally ionized absorbers with densities spanning $-5 < \log \nH < -3$ and metallicities spanning $-2 < \xh < 0$. They report uncertainties for both types of absorption systems when considering all nine of their backgrounds and when just considering the seven \citetalias{KS19} backgrounds as in this work (see Sections 2 and 3 of \citealt{acharya2021} for more detail). 

Similar to the results shown in Figure~\ref{fig:metslope}, \citeauthor{acharya2021} also see a trend of increasing \xh\ inference as the hardness (slope) of the EUVB is increased. They also observe a trend of increasing \nH\ with increasing EUVB slope that we find as well. An explanation for harder fields leading to higher metallicities is that for a harder field to have the same intensity at an EUVB energy to produce a metal ion ratio (e.g., $\OIII$ ionized to $\OIV$ at 55 eV), the harder field has a lower intensity at 13.6 eV to ionize $\HI$ leading to a higher neutral fraction, less total hydrogen overall, and therefore a higher metallicity.   In terms of metallicity uncertainties, \citet{acharya2021} report their uncertainties as the difference between the metallicity predicted by the $\alphaeuvb = -1.4$ background and the $\alphaeuvb = -2.0$ background, so to compare our results with theirs we compare the range in metallicities inferred between this work and theirs. The difference between the metallicities they infer with the $\alphaeuvb = -1.4$ background and the metallicities they infer with the $\alphaeuvb = -2.0$ background range from 0.24 dex to 0.36 dex. In comparison, our differences in inferred metallicity range from 0.09 dex to 0.66 dex (0.93 dex when \ca\ is fixed) with an average difference of 0.34 dex. The uncertainties in metallicity are clearly similar when looking at our averages, but our range in metallicity values is larger as can be seen in Figure~\ref{fig:metslope}. The main reason for this is that we explore different ranges in \nH\ and \xh\ with the majority of our absorbers having inferred \nH\ around $\nH = 10^{-2}~{\rm cm^{-2}}$ which is larger by a factor of ten than the densest absorbers they consider. Our metallicities also span a slightly larger range varying from approximately $-2.6$ on the low end to +0.6 on the high end.

The main takeaway is that while the inclusion of the EUVB slope as a free parameter leads to larger uncertainties on CGM properties, such as the metallicity explored here, explicitly including this variable still allows metallicities to be constrained with reasonable precision. The uncertainties for individual absorption systems can be large when including \alphaeuvb\ as a free parameter, but on average including this parameter only increases the average uncertainty from 0.08 to 0.14 dex (Figure~\ref{fig:errorcomp}). Furthermore, as seen in Figure~\ref{fig:ksvshm05}, the uncertainties on metallicities derived with a single background are not noticeable different from the metallicities derived when the EUVB slope is free to vary (within the range explored in this work).

\section{Conclusions} \label{sec:concl}

We perform a Bayesian analysis using MCMC sampling to infer the properties of a sample of $z < 1$ CGM absorption systems probing the \HI\ column density range $15.25 \le \log \NHI \le 17.25$. We include variations in the slope of the EUVB component of the EGB, which suffers from large uncertainties, but is crucial for maintaining the ionization state in the CGM. Our goal is to assess the impacts of uncertainties in the radiation field on the inferences of physical properties in these systems and to provide constraints on \alphaeuvb. We summarize our main results below. 

\begin{enumerate}
    
\item We use a series of models with fixed \alphaeuvb\ to explore the effects that increasing the EUVB slope had on metallicity determinations, using radiation fields from \citetalias{KS19}. As shown in Figure~\ref{fig:metslope}, the inferred metallicity increases with the hardness of the EUVB. This relationship persists when \ca\ is also treated as a free parameter, and the inferred metallicities are the same in this case for most absorbers in our sample.

\item We include the EUVB power-law slope, \alphaeuvb, as a free parameter in photoionization models, using these models to explore the effects this parameter has on inferred metallicities and to provide inferences on \alphaeuvb. The slope \alphaeuvb\ is largely unconstrained by the current sample of absorbers, which gives median values around $\alphaeuvb = -1.7$ with large uncertainties. There is no clear relationship between \xh, \alphaeuvb, and \NHI\ in the sample as a whole (Figure~\ref{fig:metalphaadopted}). 

\item The median metallicities we derive increase by 0.35 dex, on average, compared to the metallicities previously determined in \citetalias{wotta2019} and \citetalias{lehner2019} for the same absorption systems (Figure~\ref{fig:ksvshm05}). This increase is similar to the metallicity increase reported in \citetalias{wotta2019} when switching from the softer \citetalias{haardt1996} background to the harder \citetalias{haardt2012} background. This reflects that the \citetalias{KS19} backgrounds are, for most of the EUVB slopes considered, harder than the \citetalias{haardt1996} background. The trend of increasing metallicity with increasing EUVB power-law slope has been noted before with various backgrounds (\citealt{wotta16,chen17}; \citetalias{wotta2019}). The metallicity increase is on average 0.2 to 0.4 dex depending on the \HI\ column density of the absorption system and is seen in both high and low metallicity systems.   

\item The average uncertainty of our inferred metallicities increased from 0.08 dex to 0.14 dex when switching from models with a fixed EUVB slope to those with \alphaeuvb\ as a free parameter varying over $-2.0$ to $-1.4$ (though it can be less than 0.1~dex; see Figure~\ref{fig:errorcomp}). This increase indicates that uncertainties in \alphaeuvb\ dominate the overall metallicity uncertainty, rather than uncertainties in column density measurements or modeling approach. Thus, the uncertainty in the shape of the EUVB can be included in photoionization models of CGM absorption systems and still yield useful and well-constrained metallicity estimates. 

\item The slope of the EUVB is largely unconstrained by our sample of 34 absorption systems (Figure~\ref{fig:slope_posterior}). Combining the results from our sample as a whole hints at an underlying Gaussian distribution centered at $\alphaeuvb = -1.7$, but the large uncertainties in the individual absorbers make this difficult to confirm (Figure~\ref{fig:posterior_comp}).

\end{enumerate}
The relatively small uncertainties that variations in the EUVB impart onto the absorption systems studied here also have implications for the broader CCC sample of 224 absorption systems. Our absorbers were selected from that parent sample and modeled the same as previous work, with the exception of including \alphaeuvb\ as a fixed parameter. Thus, it would seem that previous results derived for the parent sample of absorbers are also robust to variations in the EUVB, with only slight increases to the overall uncertainty. Future work will explore the entire CCC sample of absorption systems to assess this result and seek to identify more absorption systems with well constrained EUVB slopes in order to more accurately characterize the underlying \alphaeuvb\ distribution. 

\acknowledgments
Support for this research was provided by NASA through grant HST-AR-15634 from the Space Telescope Science Institute (STScI), which is operated by the Association of Universities for Research in Astronomy, Incorporated, under NASA contract NAS5-26555. Data presented in this work were obtained from CCC, which was funded through NASA grant HST-AR-12854 from STScI. The ground-based data presented in this work were obtained from KODIAQ, which was funded through NASA ADAP Grant Nos. NNX10AE84G and NNX16AF52G. This research has made use of the KOA, which is operated by the W. M. Keck Observatory and the NASA Exoplanet Science Institute (NExScI), under contract with NASA. The authors wish to recognize and acknowledge the very significant cultural role and reverence that the summit of Maunakea has always had within the indigenous Hawaiian community. Based also on ground-based observations collected at the European Southern Observatory under ESO program 0100.A-0483(A,B) and archival programs 076.A-0860(A), 075.A-0841(A), and 293.A-5038(A). This research was supported by the Notre Dame Center for Research Computing through the Grid Engine software and, together with the Notre Dame Cooperative Computing Lab, through the HTCondor software. KLC acknowledges partial support from NSF AST-1615296.


\textit{Facilities:} HST(COS), Keck(HIRES), VLT(UVES)

\software{astropy \citep{astropy},  Cloudy \citep{cloudy17}, emcee \citep{foreman2013}, Matplotlib \citep{hunter07}, PyIGM \citep{prochaska17a}}

\bibliography{CCCIV_refs}{}
\bibliographystyle{aasjournal}

\appendix
\section{Full Set of Corner and Comparison Plots}
In this Appendix, we show the full set of corner and model-data comparison plots for each of the absorption systems in our sample. Corner plots show the 2D probability distributions for each parameter combination in our model as well as the marginalized PDFs for each parameter. Comparison plots juxtapose observationally determined column densities and model-predicted column densities for the ions of each absorber. The dashed lines in each marginalized PDF of the corner plot indicate the median value of that parameter as well as the 68\% CI for detections and 80\% CI for upper and lower limits. In the comparison plots, blue symbols are the model column densities and red symbols are our observations where a circle represents a detection, upwards triangles are 1-$\sigma$ lower limits, and downward triangles are 1-$\sigma$ upper limits. 

\begin{figure}[tbp]
\epsscale{0.5}
\plotone{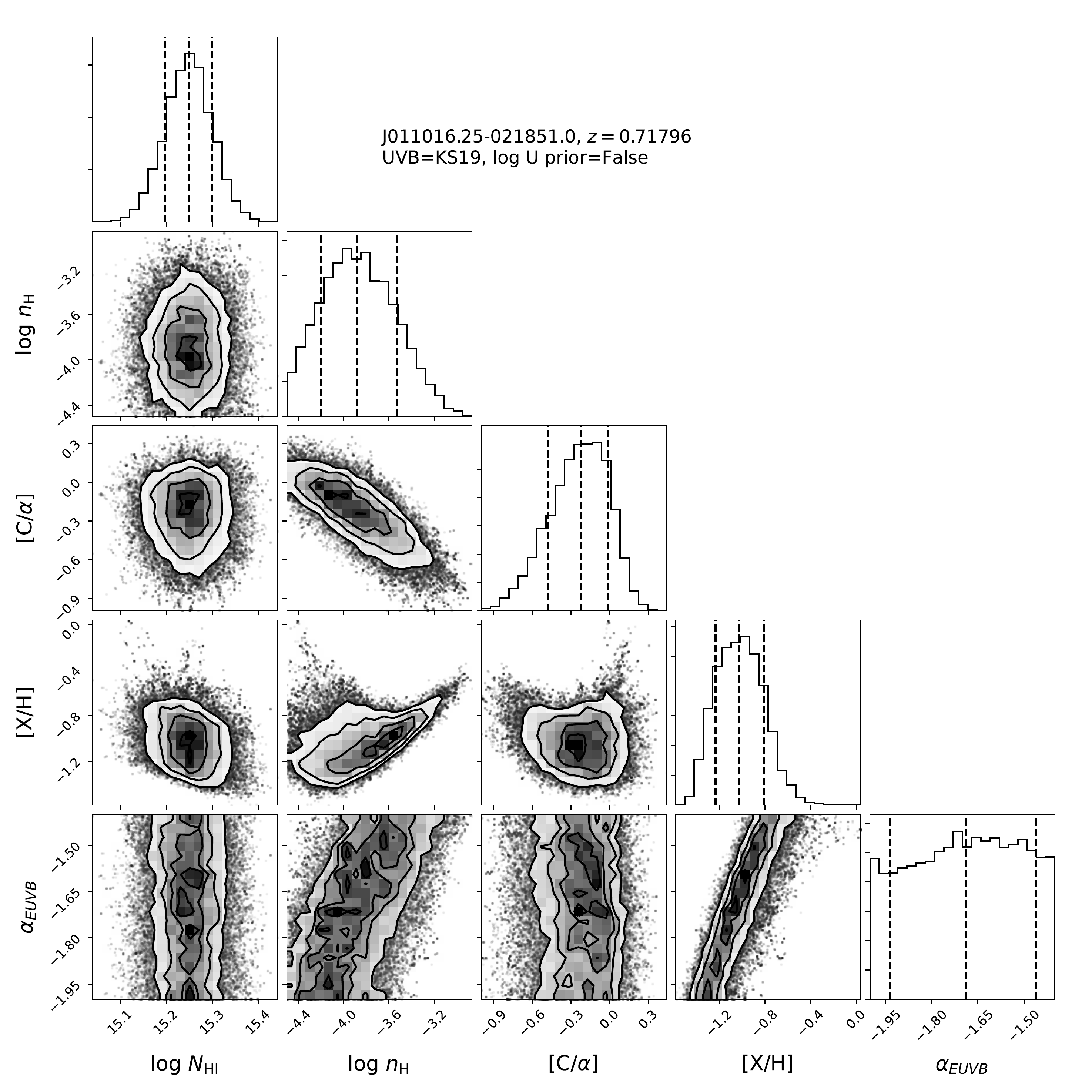}
\plotone{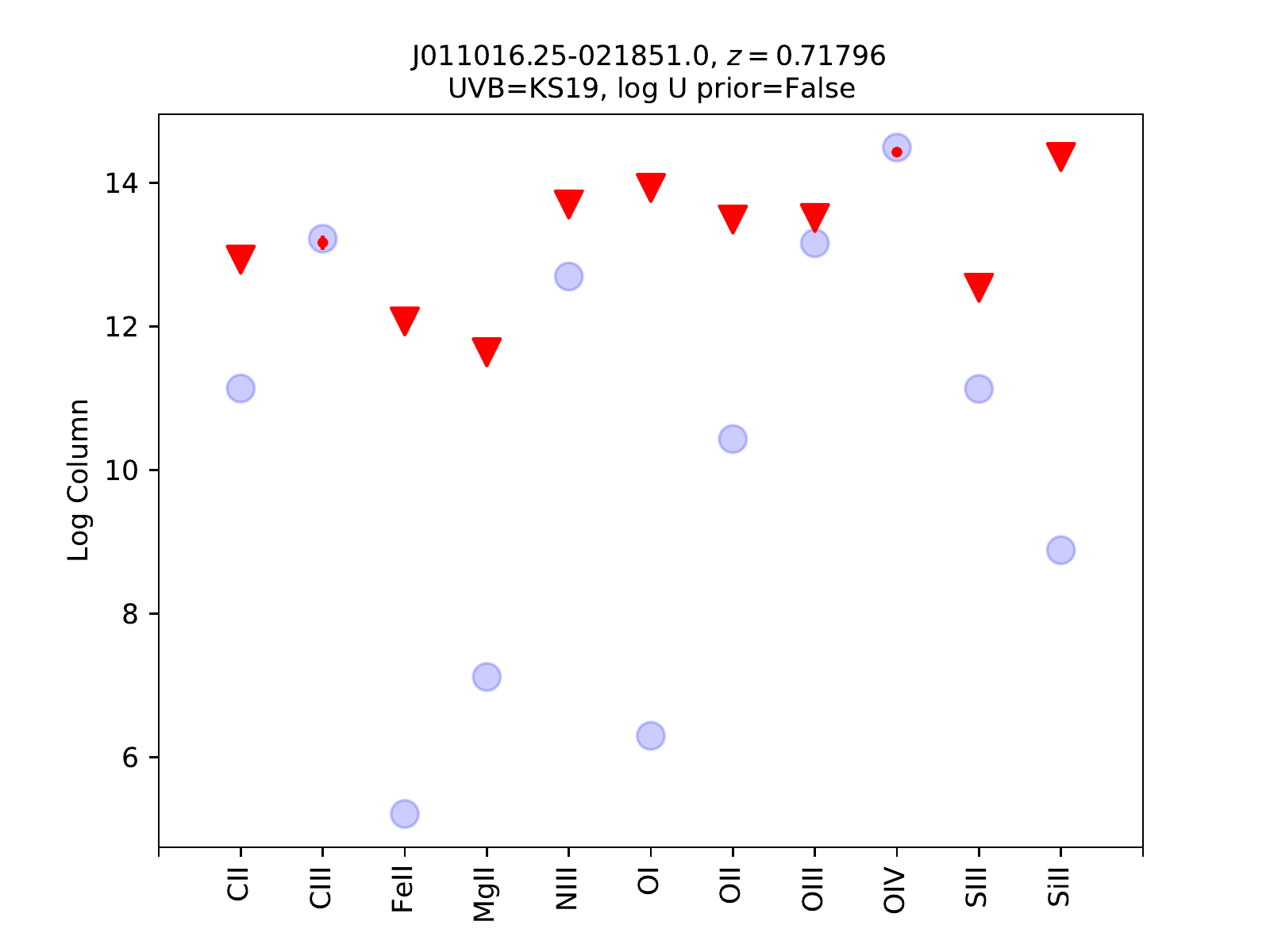}
\caption{Corner and comparison plot for absorber 1 in our sample. More information about specific values for observations and parameter values can be found in Table~\ref{tab:sample_props} and Table~\ref{tab:results_tab} .
\label{fig:A1}
}
\end{figure}

\begin{figure}[tbp]
\epsscale{0.5}
\plotone{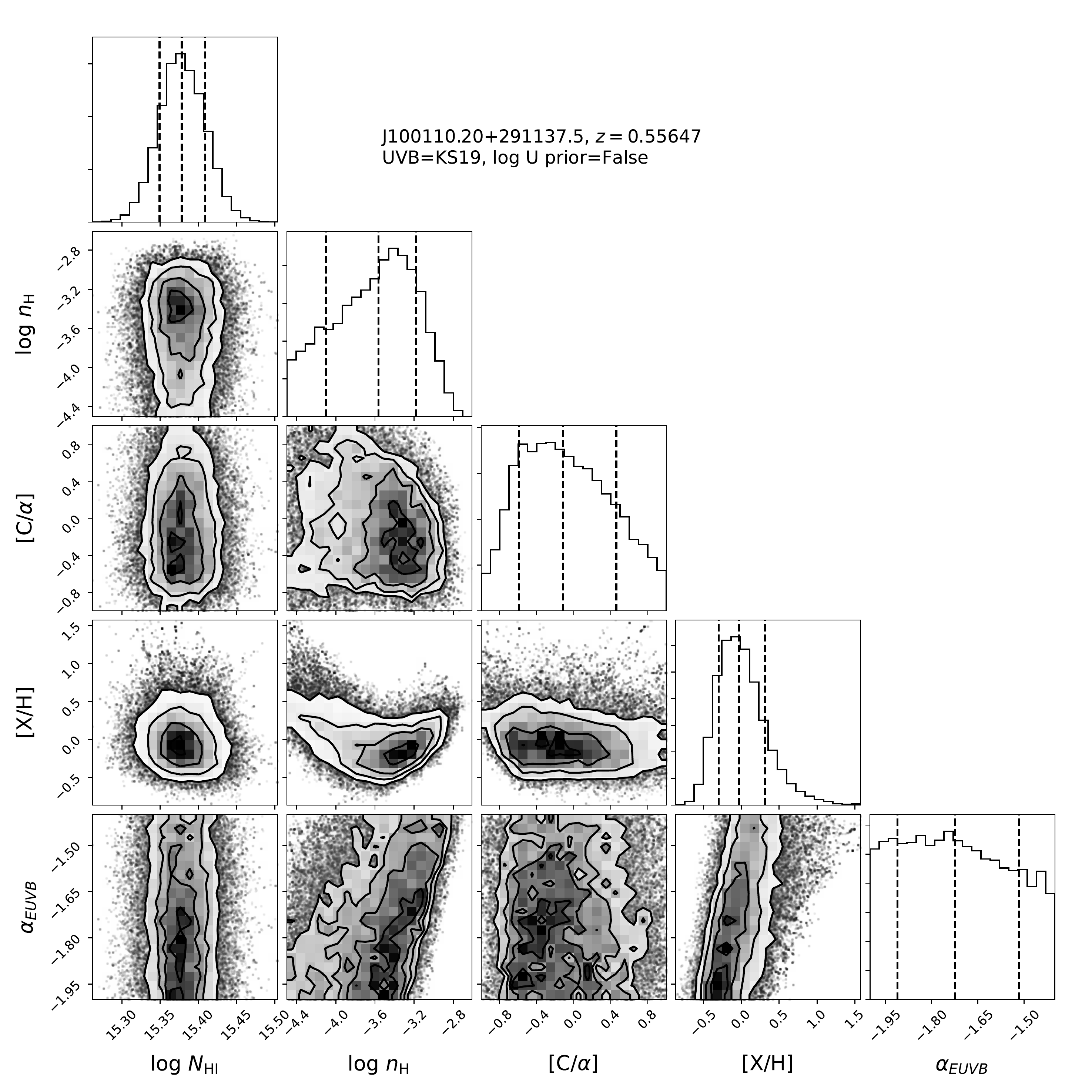}
\plotone{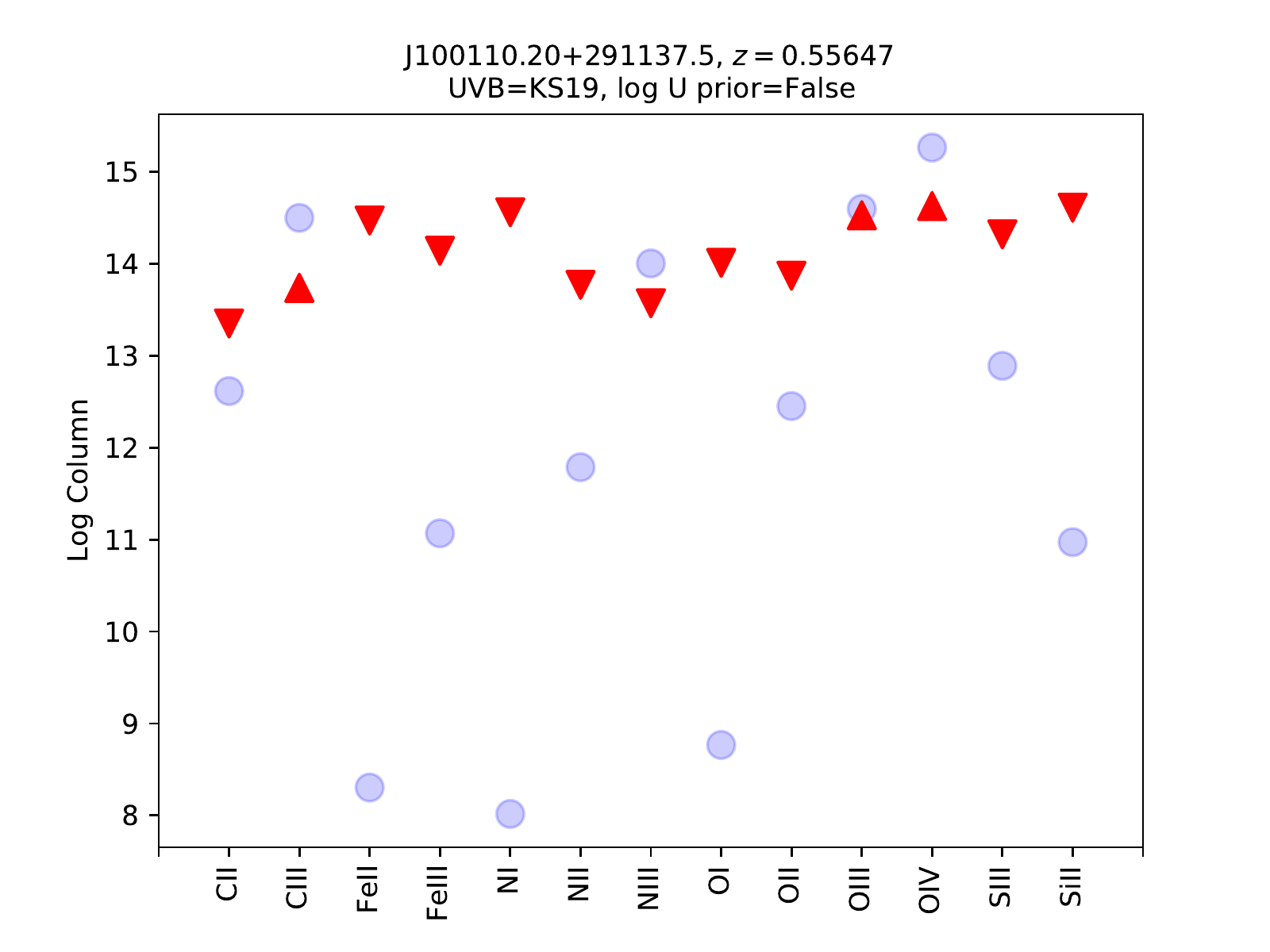}
\caption{Same as Figure~\ref{fig:A1}, but for absorber 2.
\label{fig:A2}
}
\end{figure}

\begin{figure}[tbp]
\epsscale{0.5}
\plotone{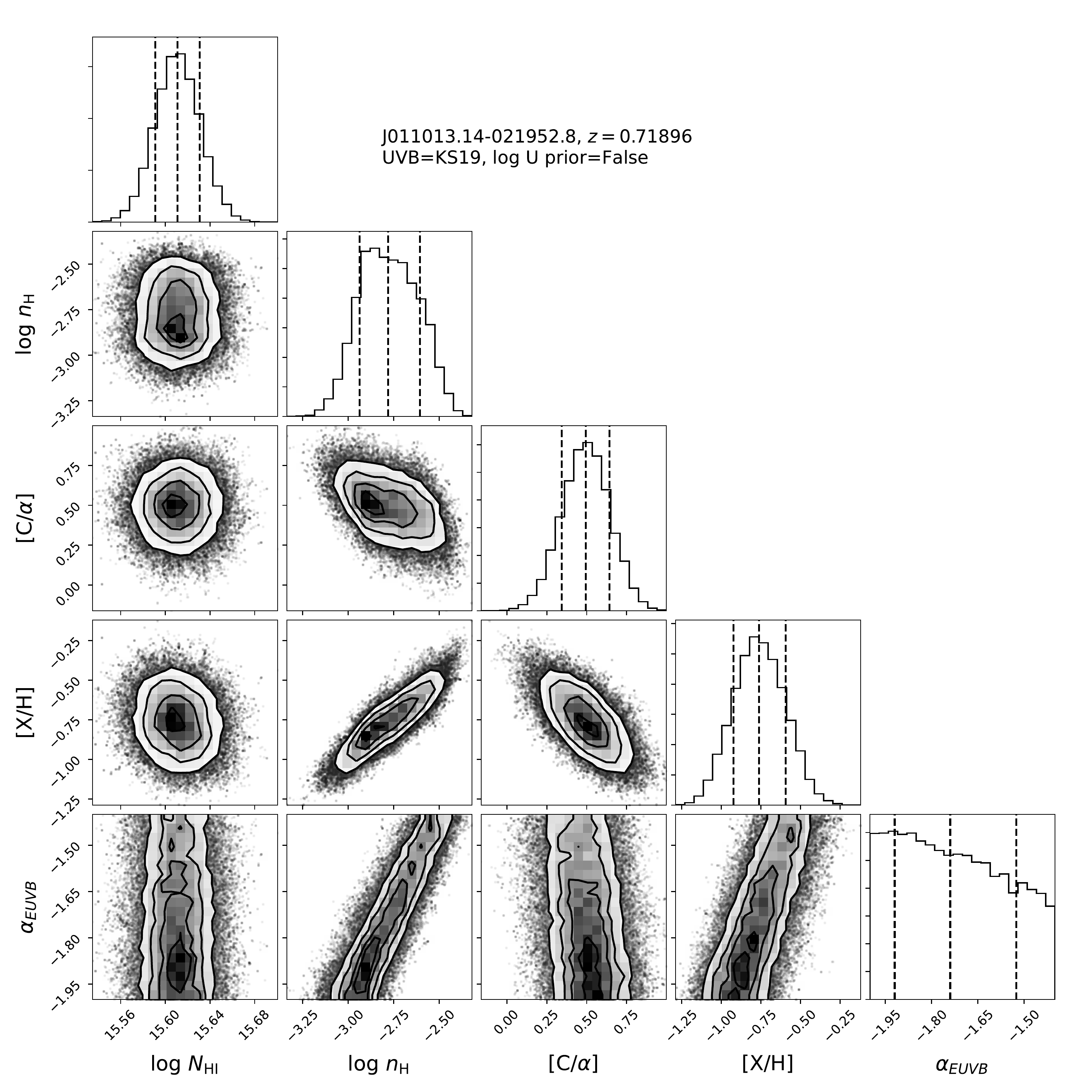}
\plotone{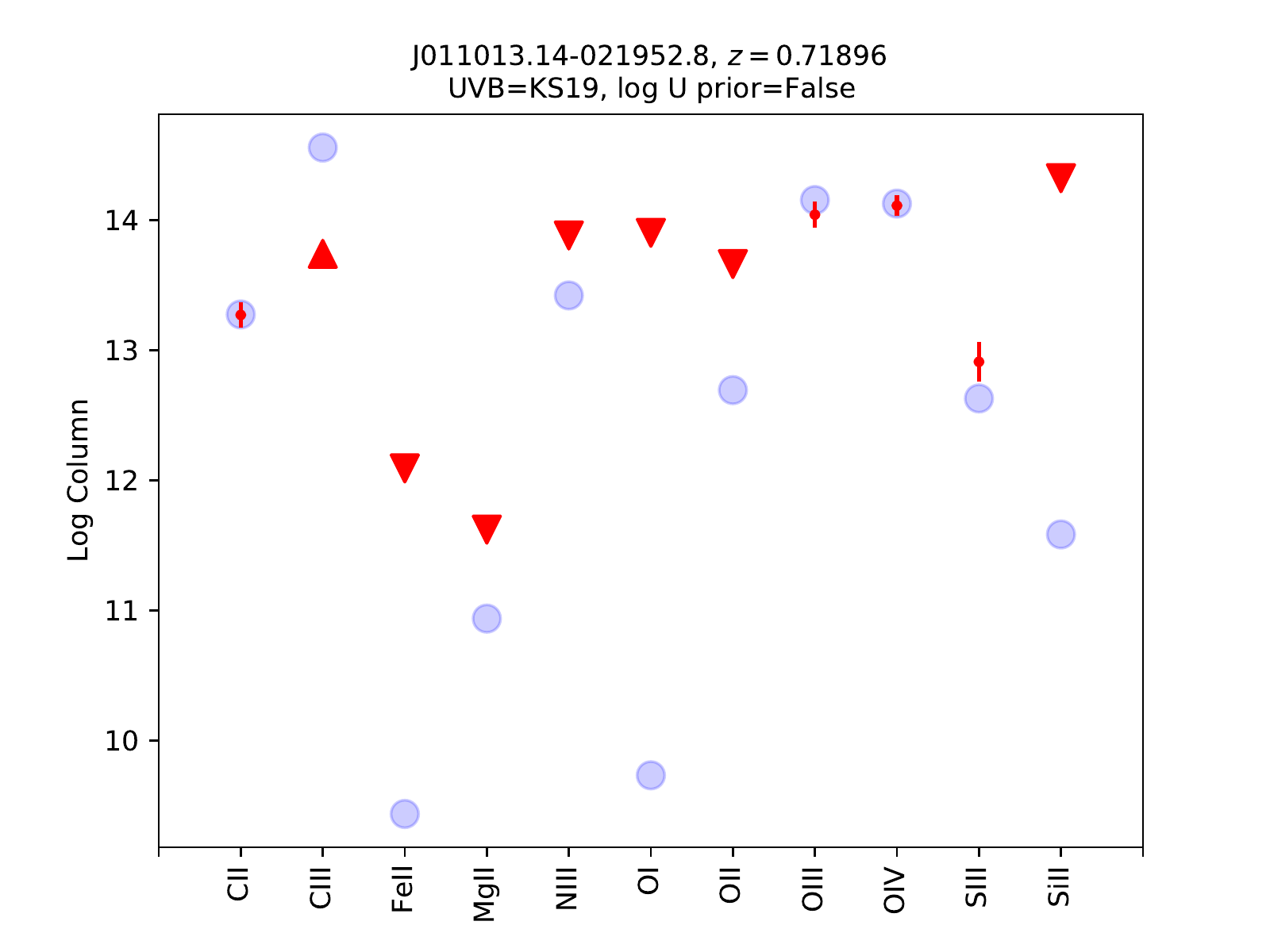}
\caption{Same as Figure~\ref{fig:A1}, but for absorber 3.
\label{fig:A3}
}
\end{figure}

\begin{figure}[tbp]
\epsscale{0.5}
\plotone{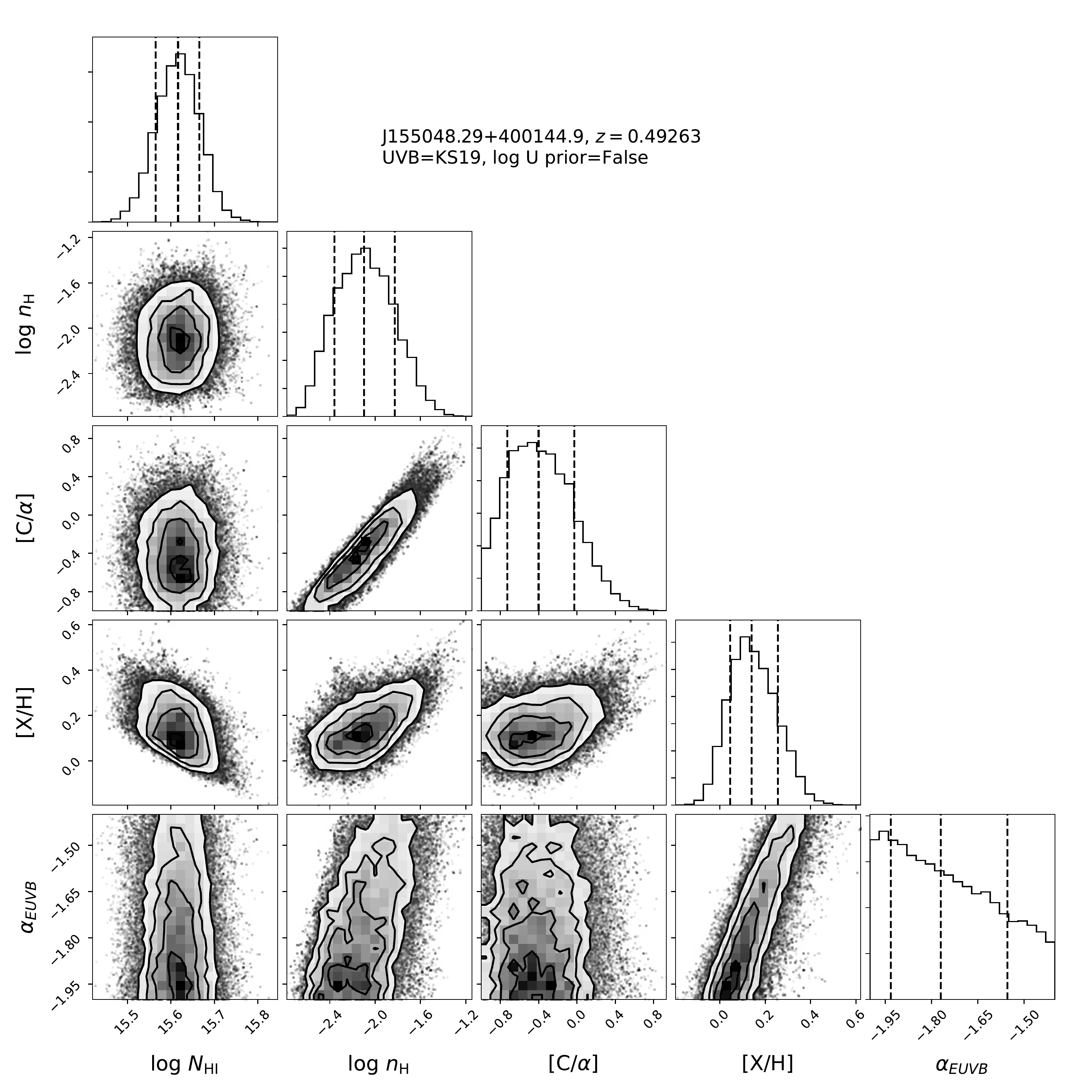}
\plotone{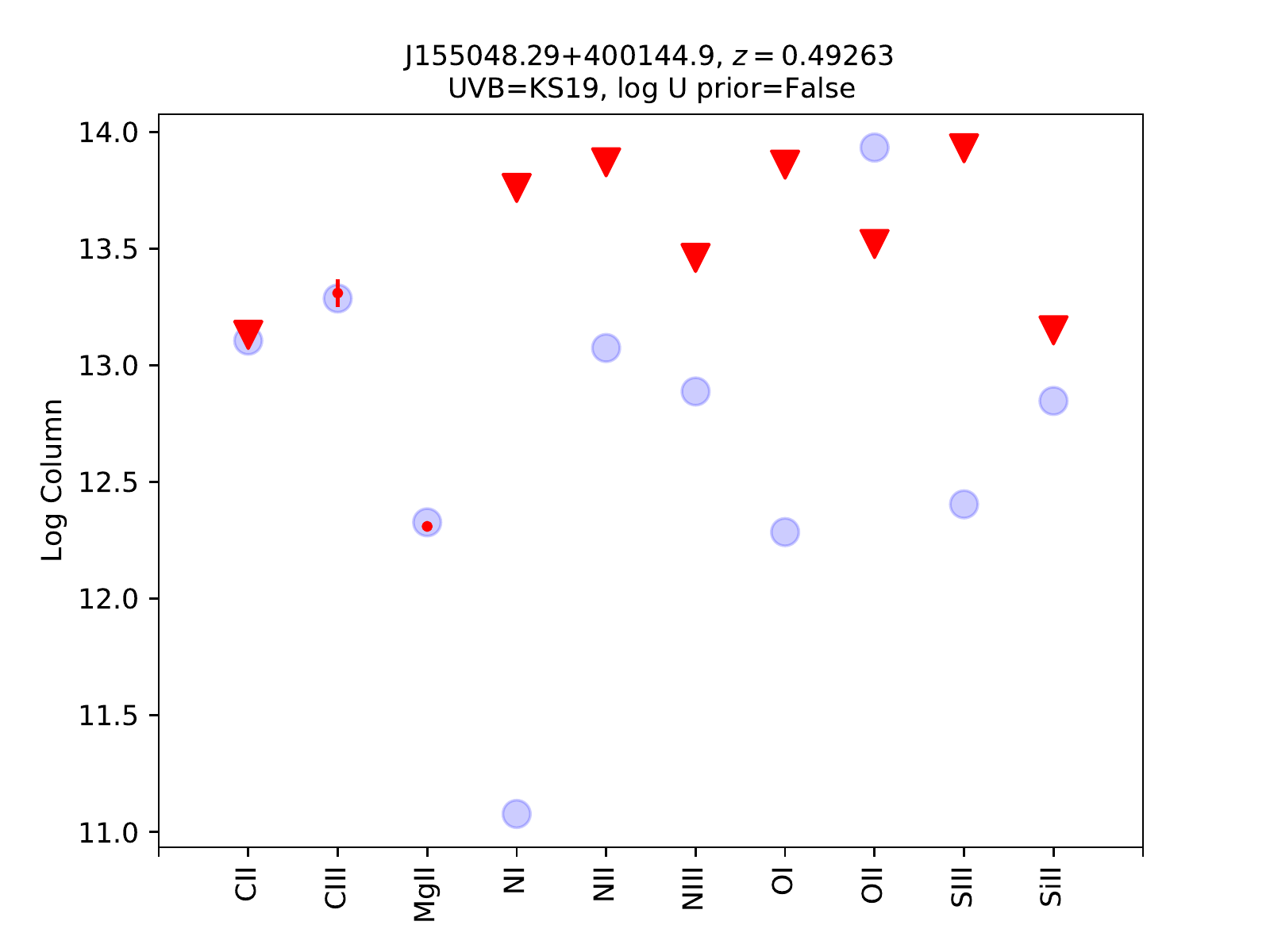}
\caption{Same as Figure~\ref{fig:A1}, but for absorber 4.
\label{fig:A4}
}
\end{figure}

\begin{figure}[tbp]
\epsscale{0.5}
\plotone{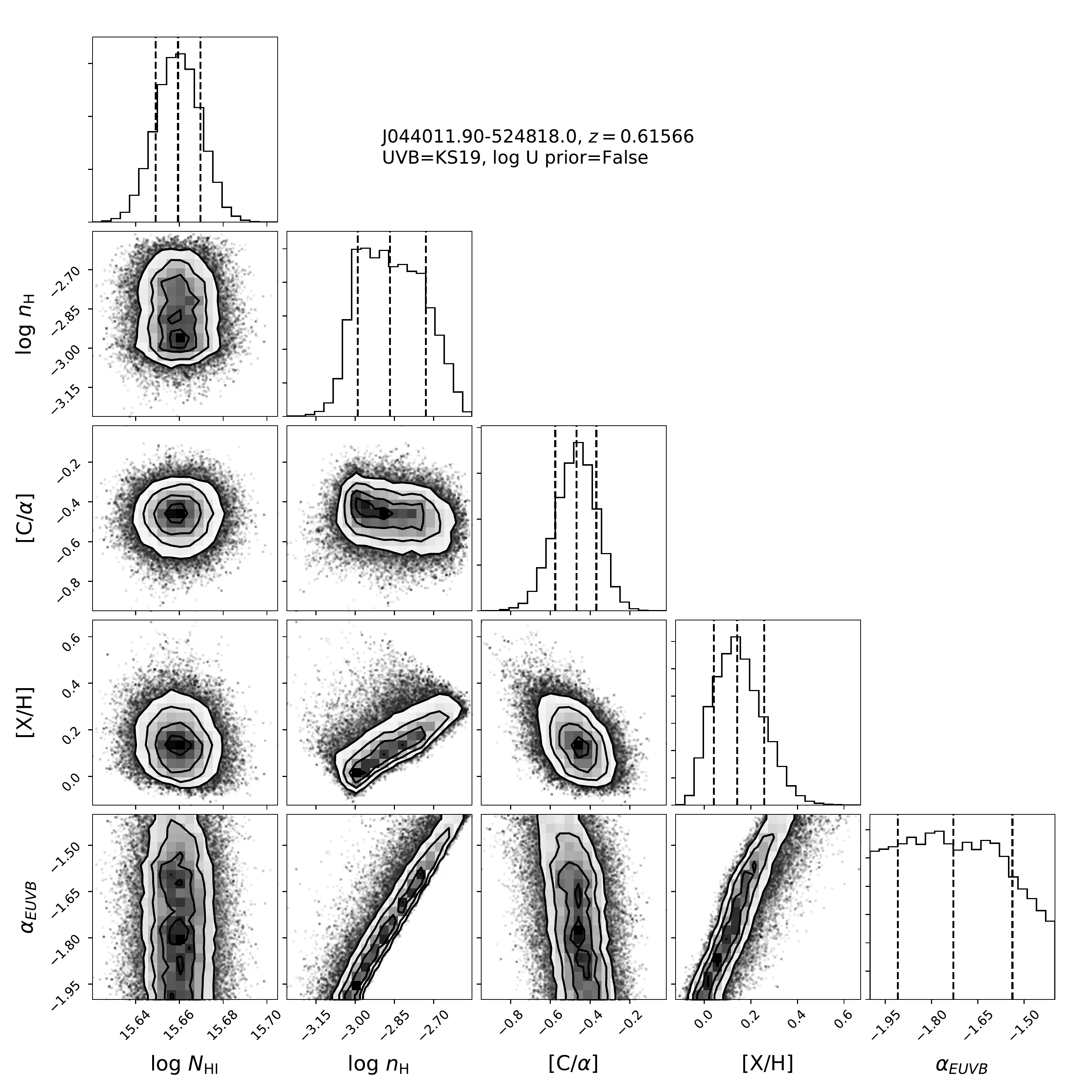}
\plotone{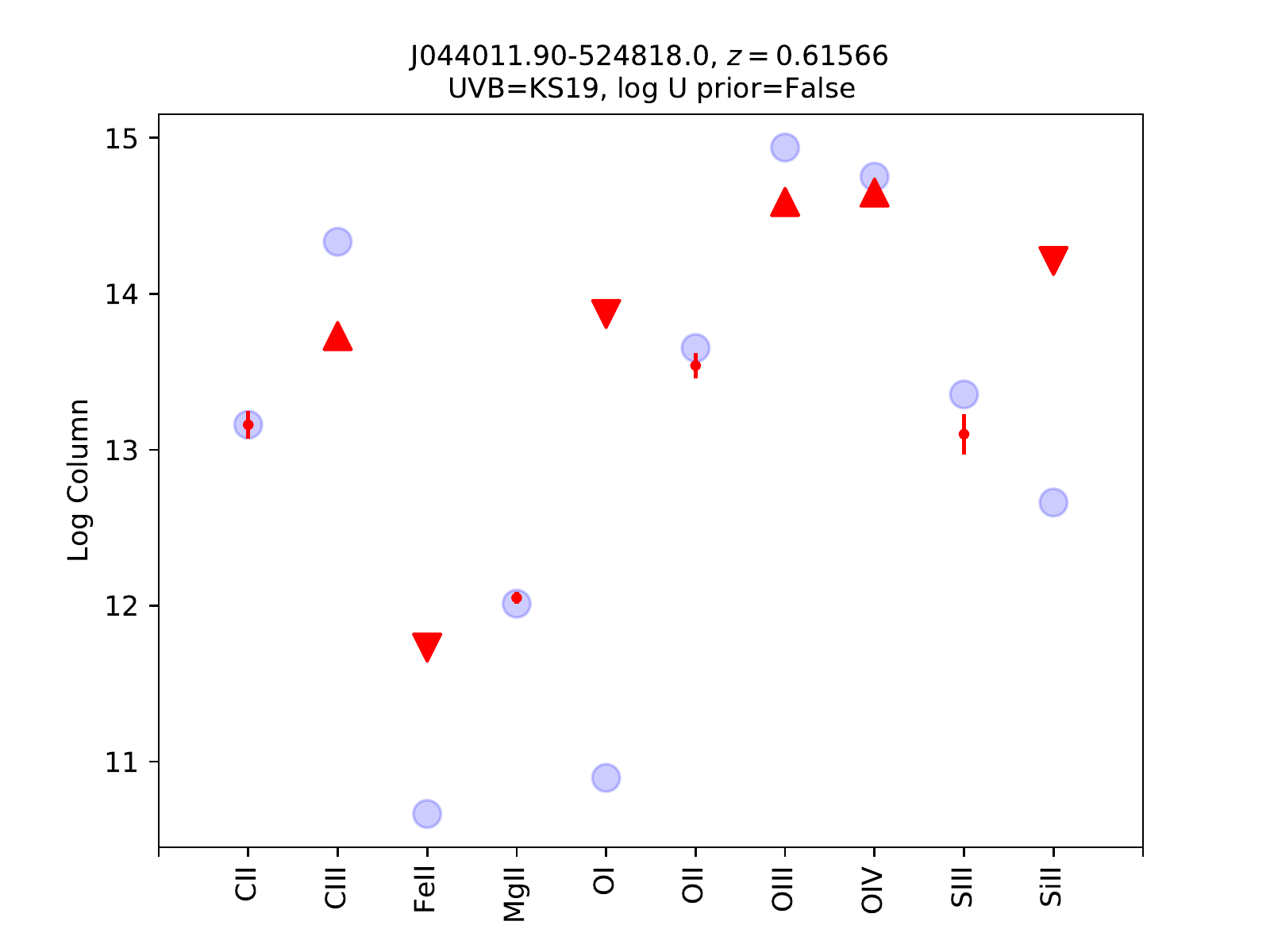}
\caption{Same as Figure~\ref{fig:A1}, but for absorber 5.
\label{fig:A5}
}
\end{figure}

\begin{figure}[tbp]
\epsscale{0.5}
\plotone{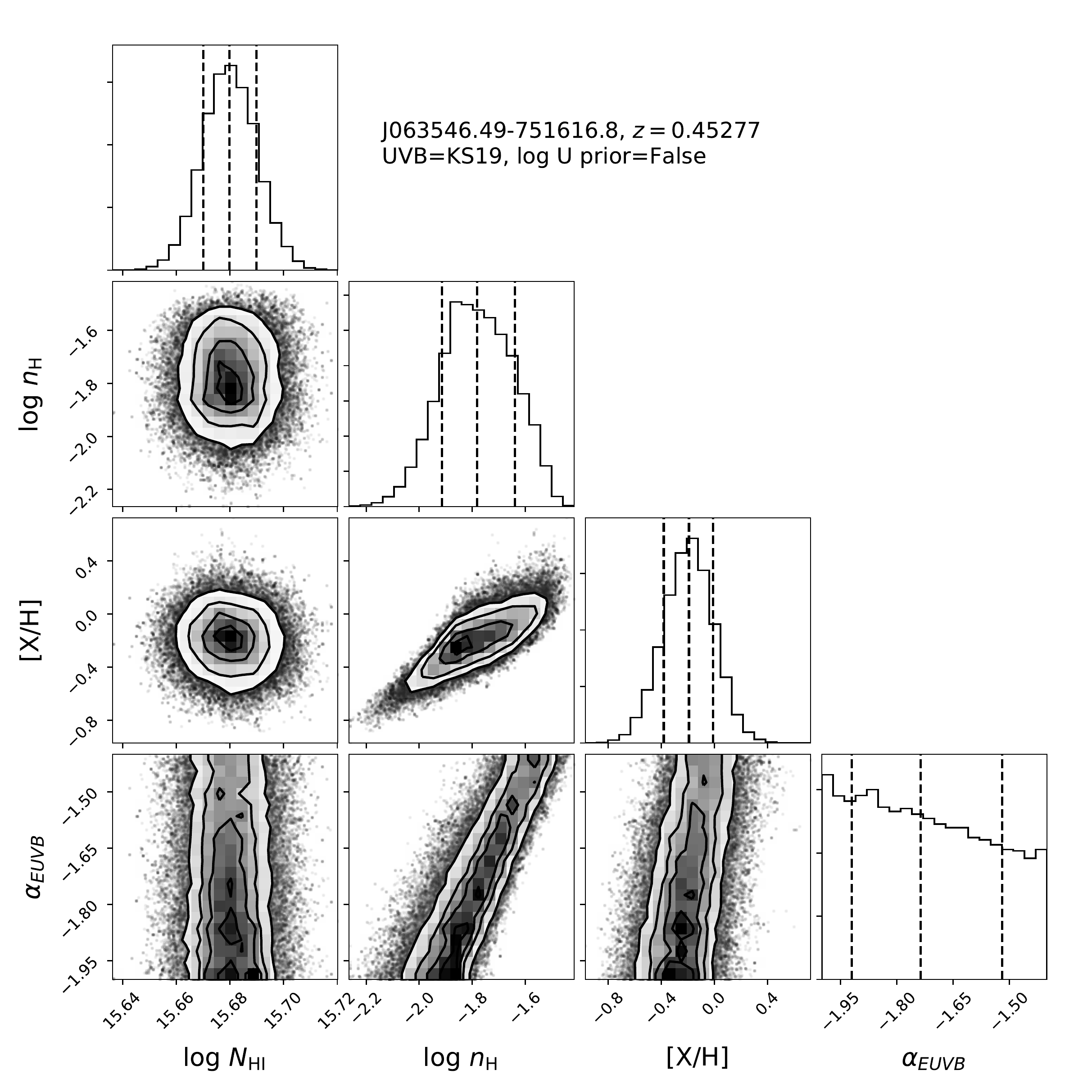}
\plotone{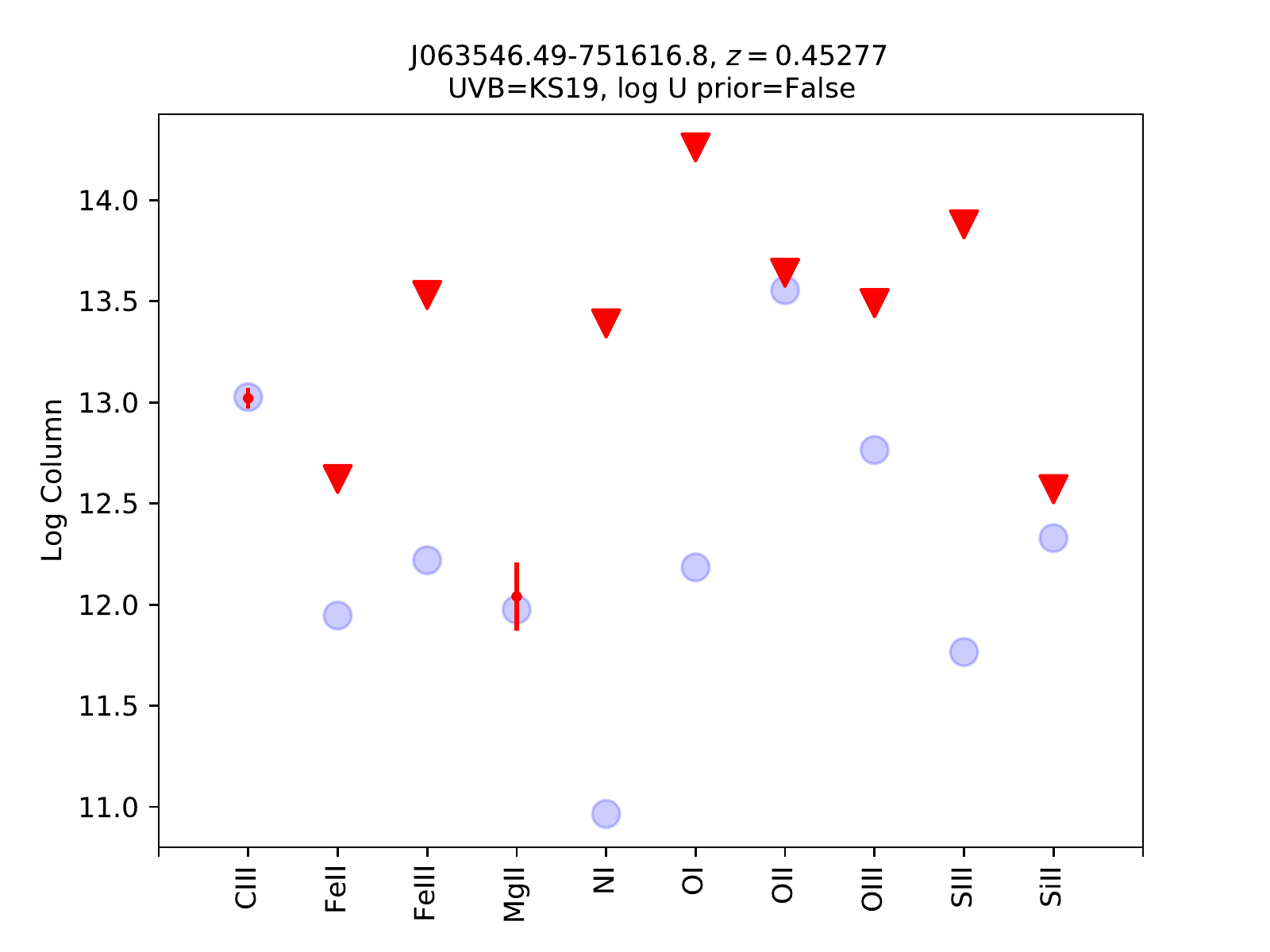}
\caption{Same as Figure~\ref{fig:A1}, but for absorber 6.
\label{fig:A6}
}
\end{figure}

\clearpage

\begin{figure}[tbp]
\epsscale{0.5}
\plotone{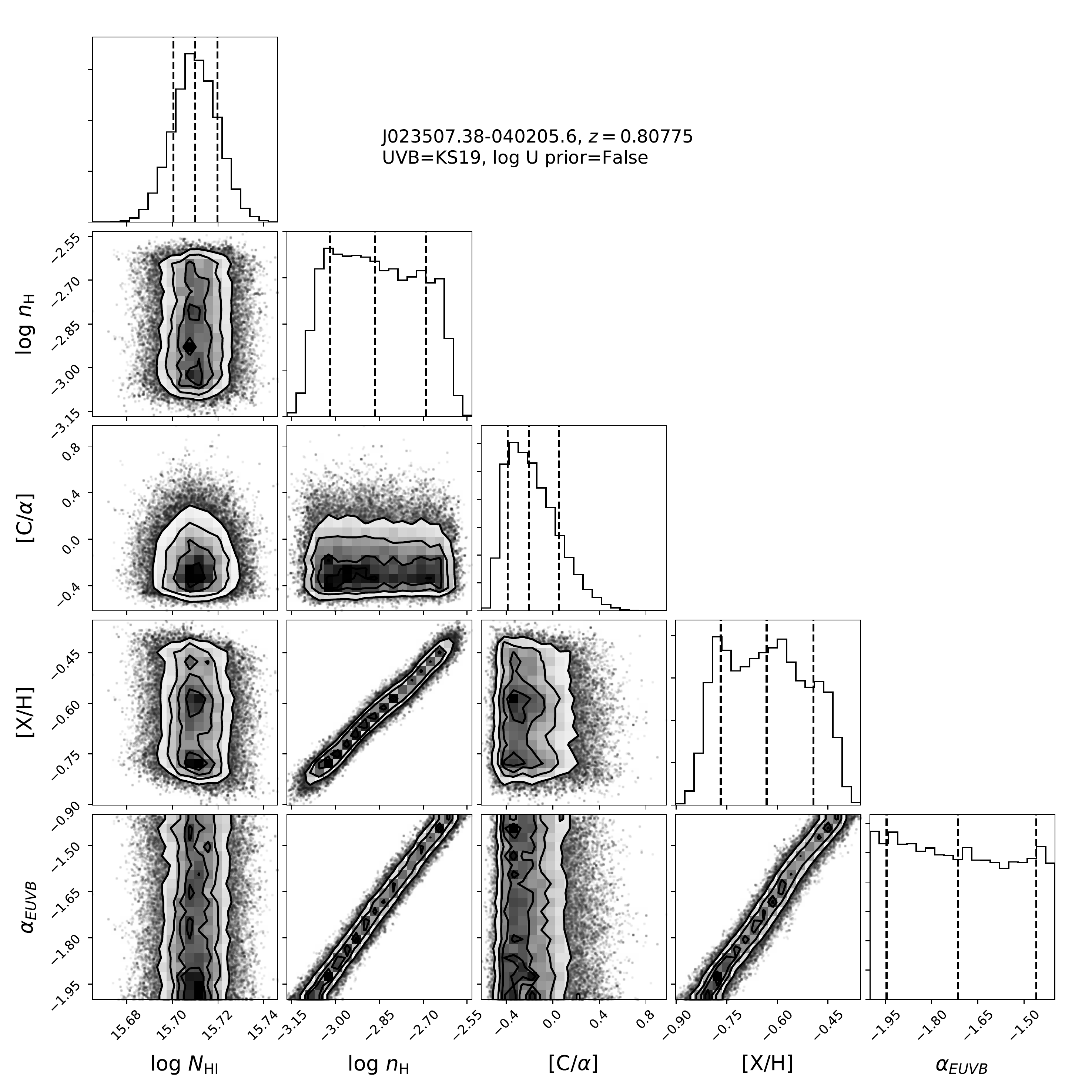}
\plotone{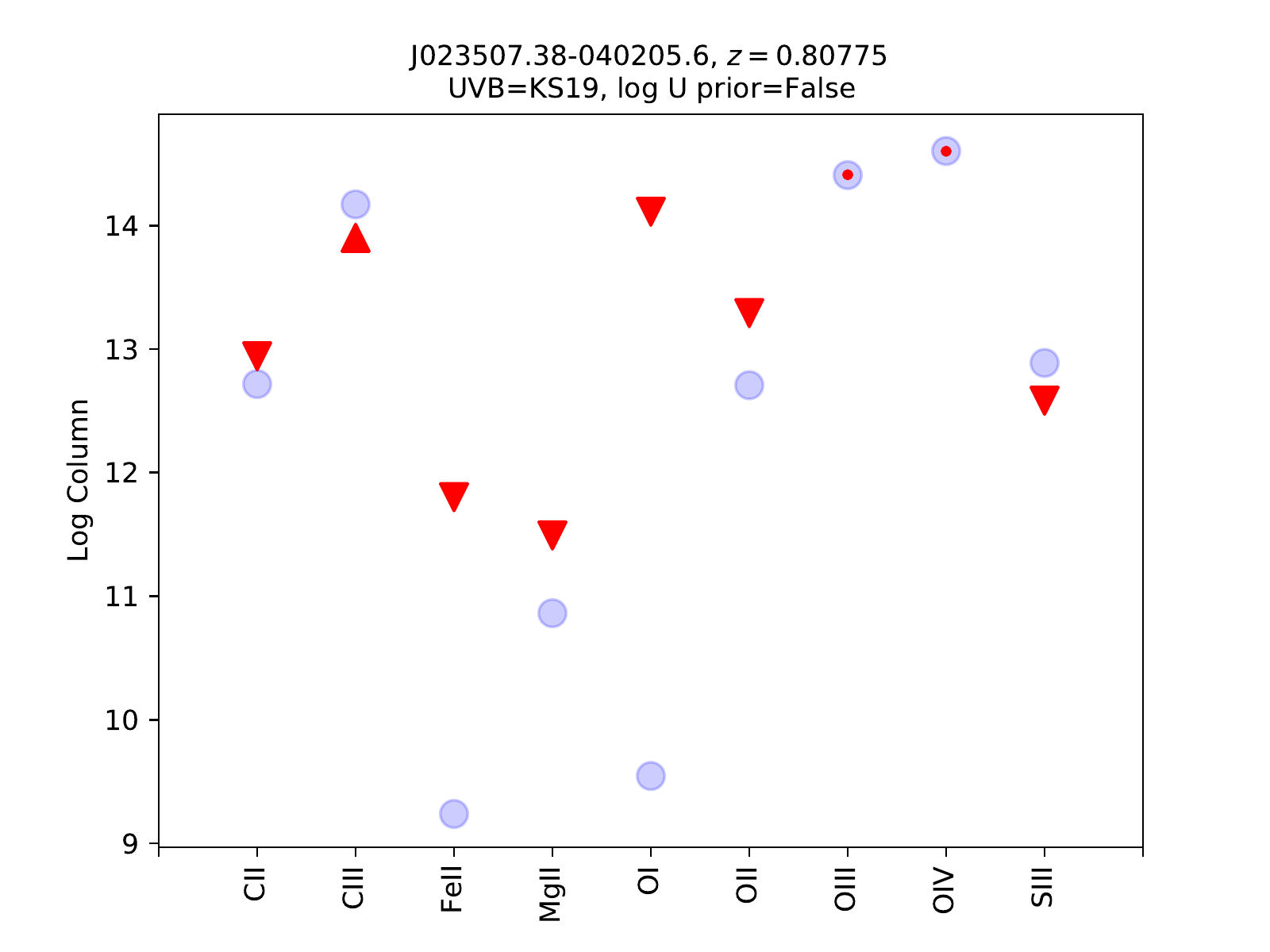}
\caption{Same as Figure~\ref{fig:A1}, but for absorber 7.
\label{fig:A7}
}
\end{figure}

\begin{figure}[tbp]
\epsscale{0.5}
\plotone{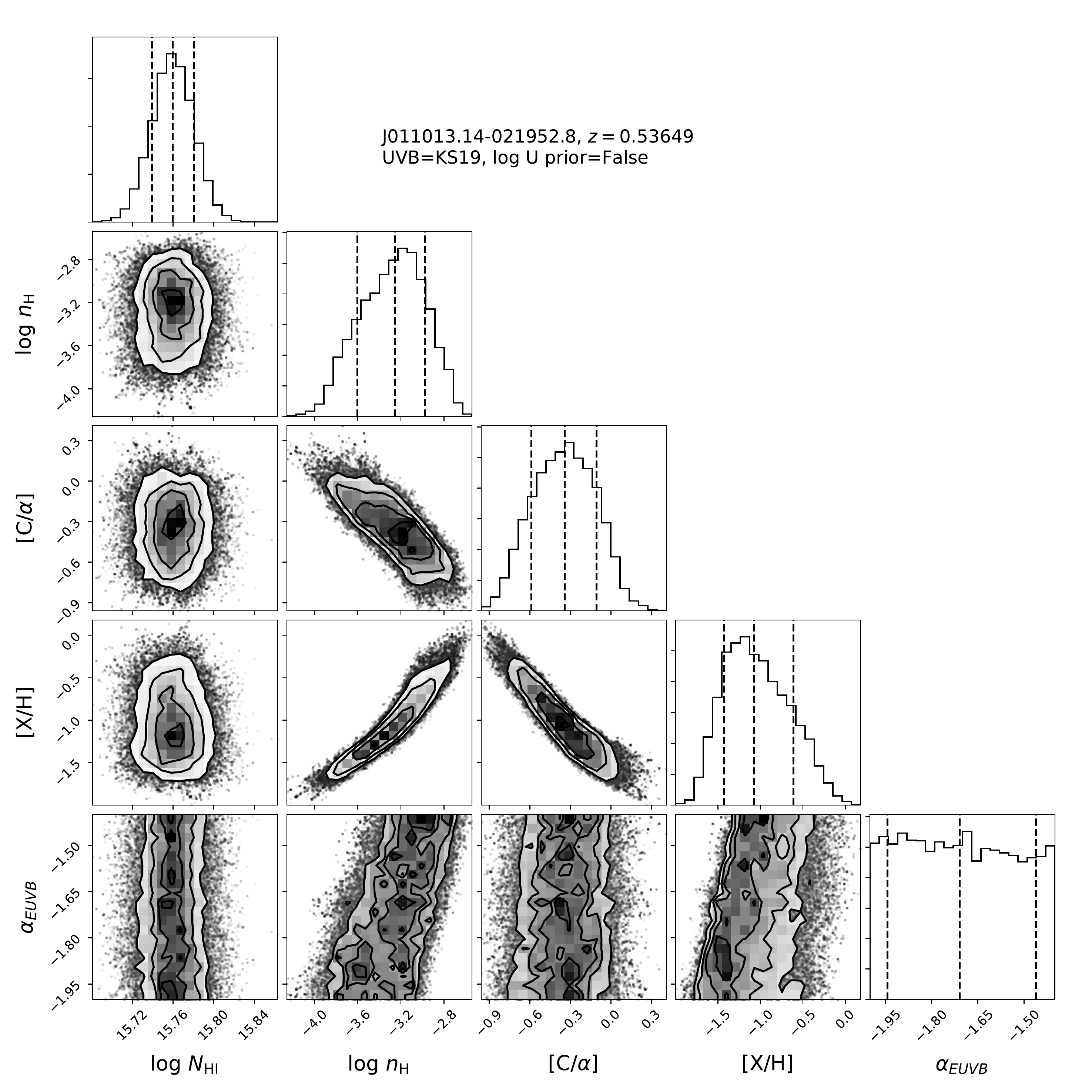}
\plotone{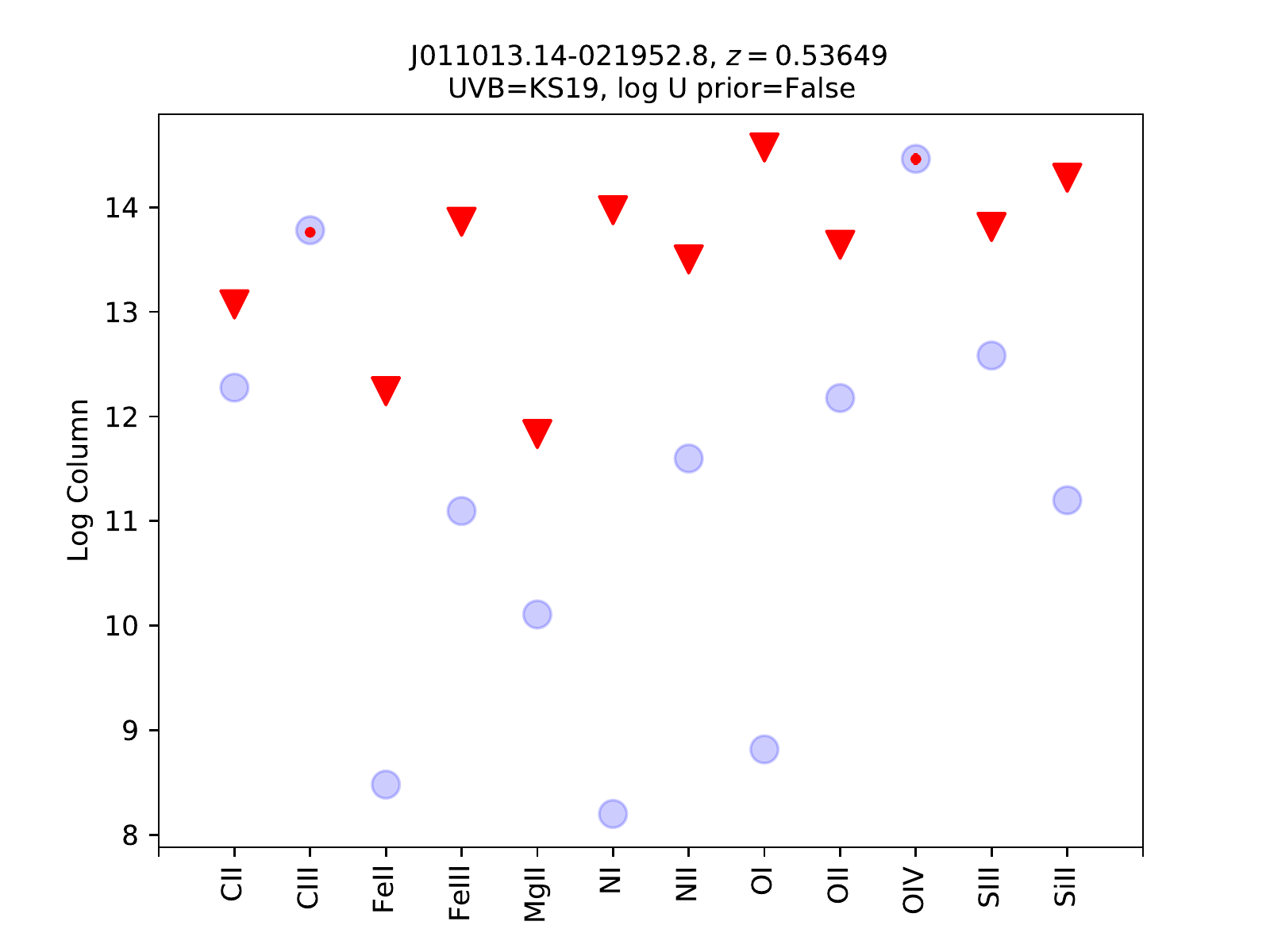}
\caption{Same as Figure~\ref{fig:A1}, but for absorber 8.
\label{fig:A8}
}
\end{figure}

\begin{figure}[tbp]
\epsscale{0.5}
\plotone{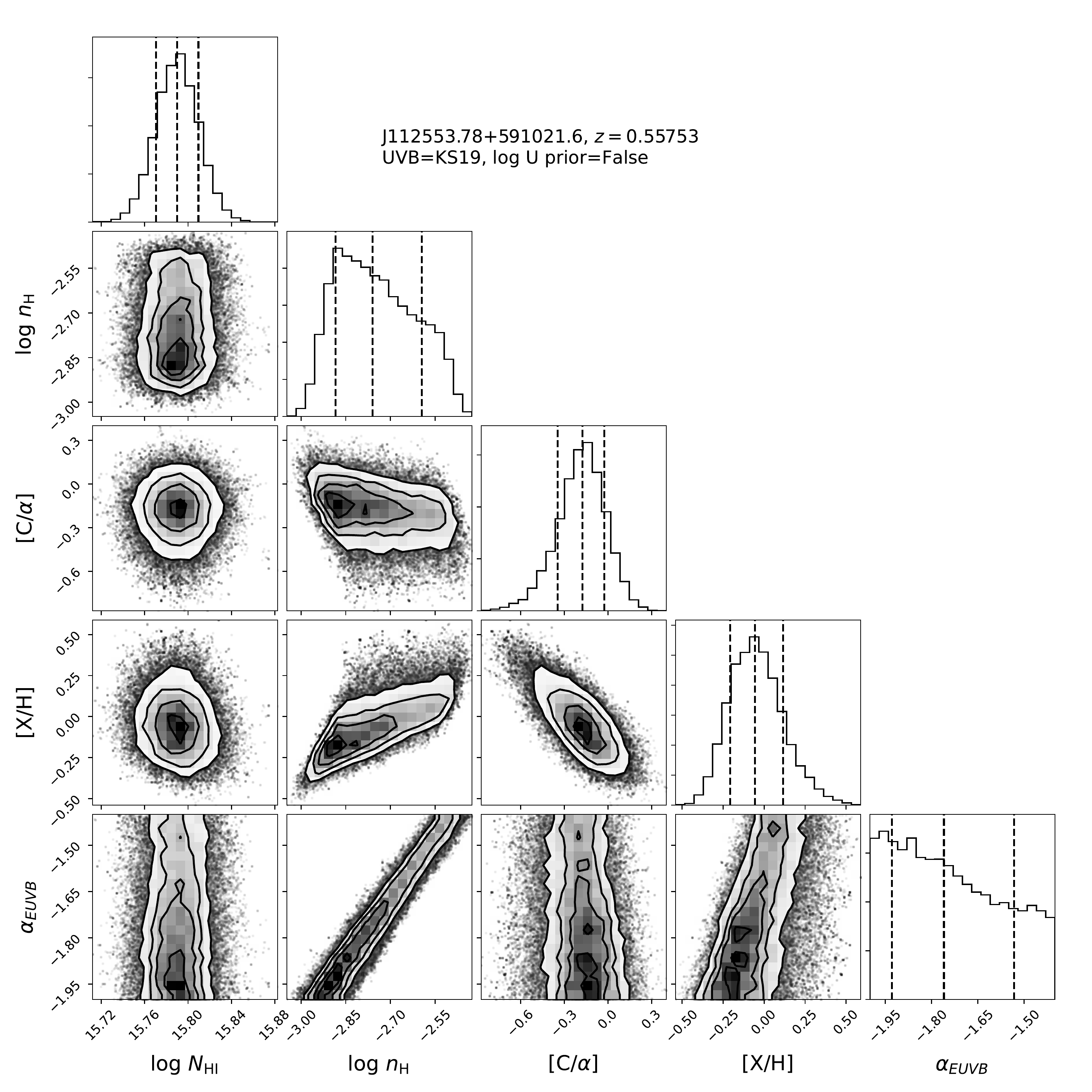}
\plotone{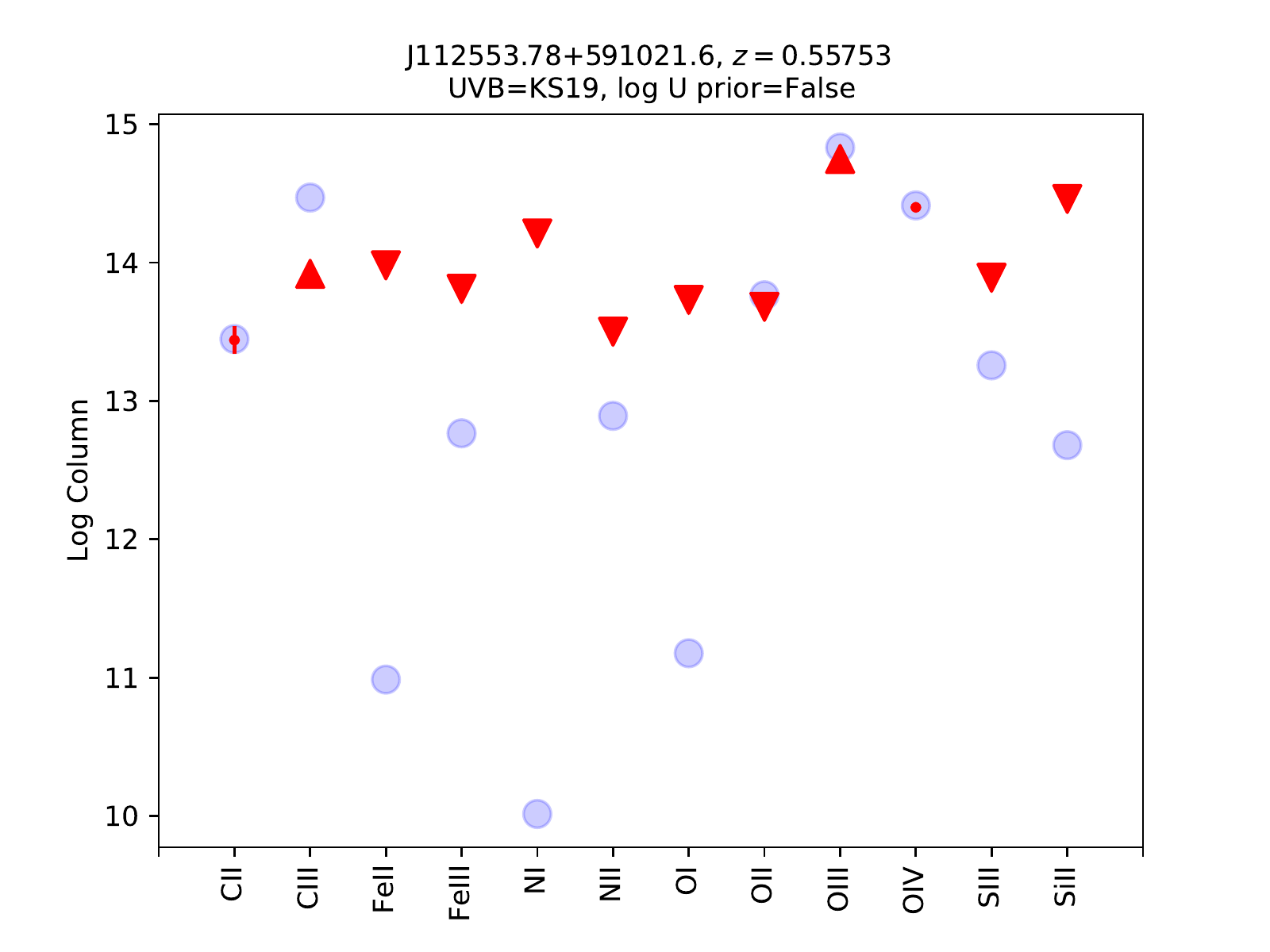}
\caption{Same as Figure~\ref{fig:A1}, but for absorber 9.
\label{fig:A9}
}
\end{figure}

\begin{figure}[tbp]
\epsscale{0.5}
\plotone{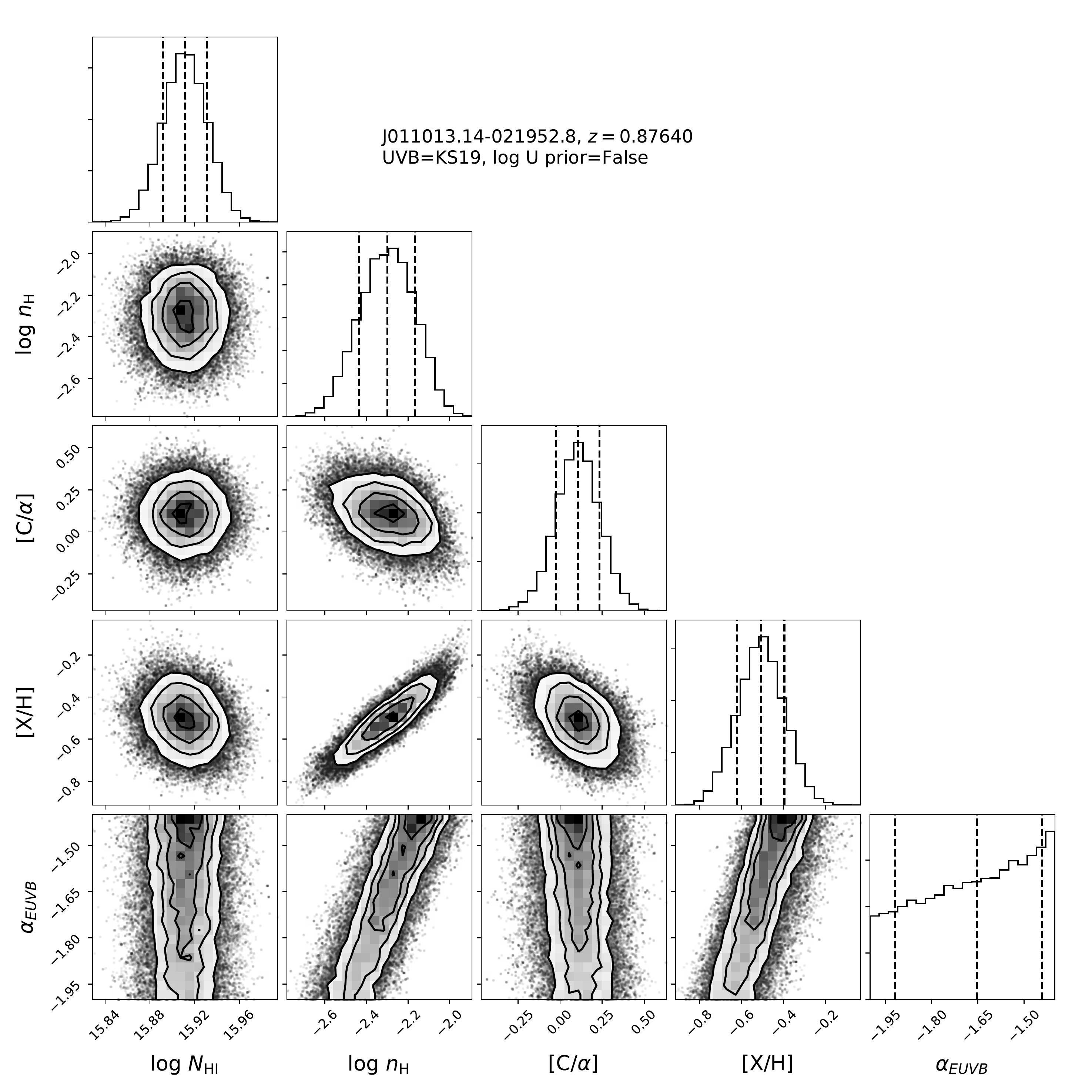}
\plotone{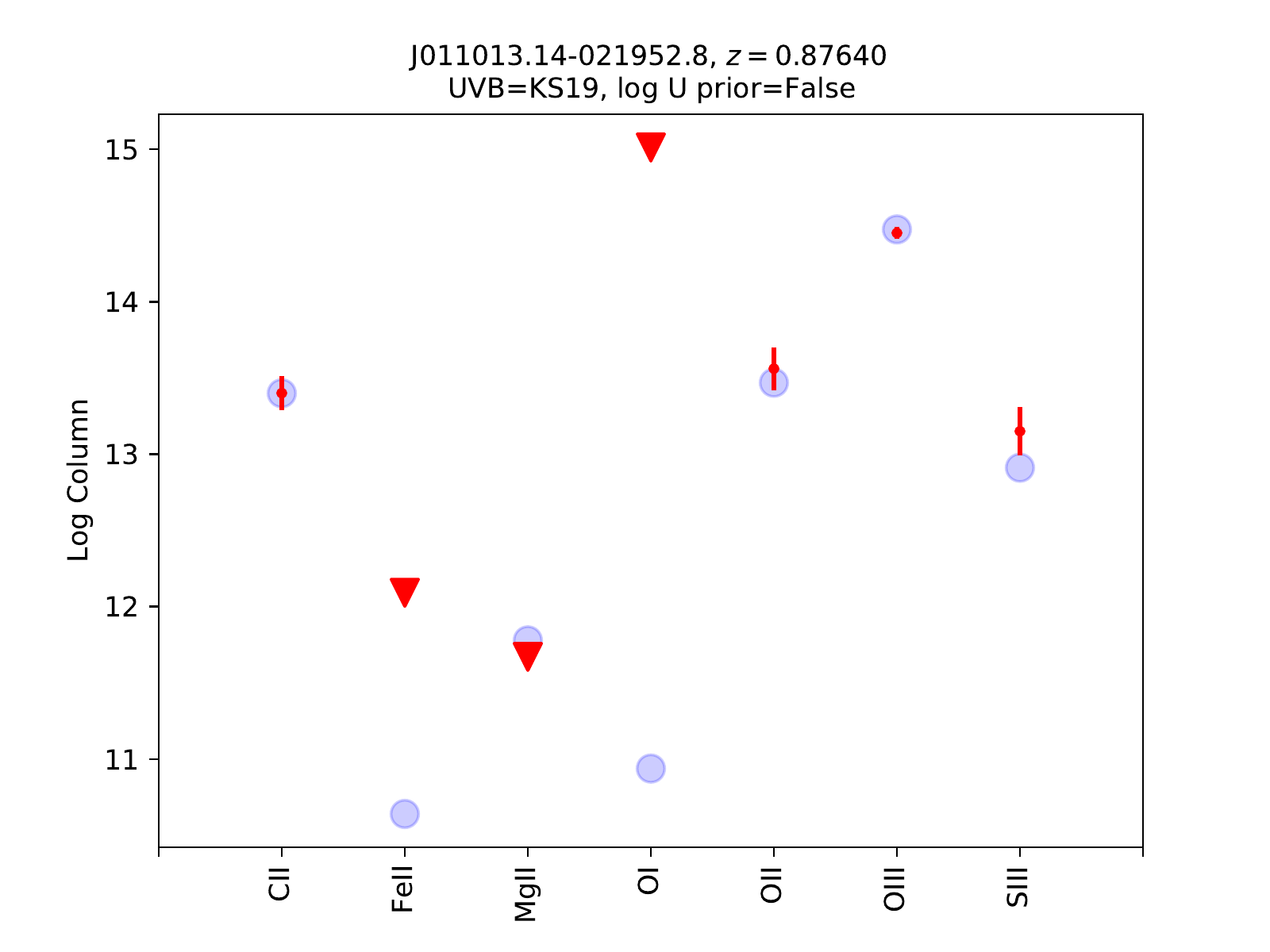}
\caption{Same as Figure~\ref{fig:A1}, but for absorber 10.
\label{fig:A10}
}
\end{figure}

\begin{figure}[tbp]
\epsscale{0.5}
\plotone{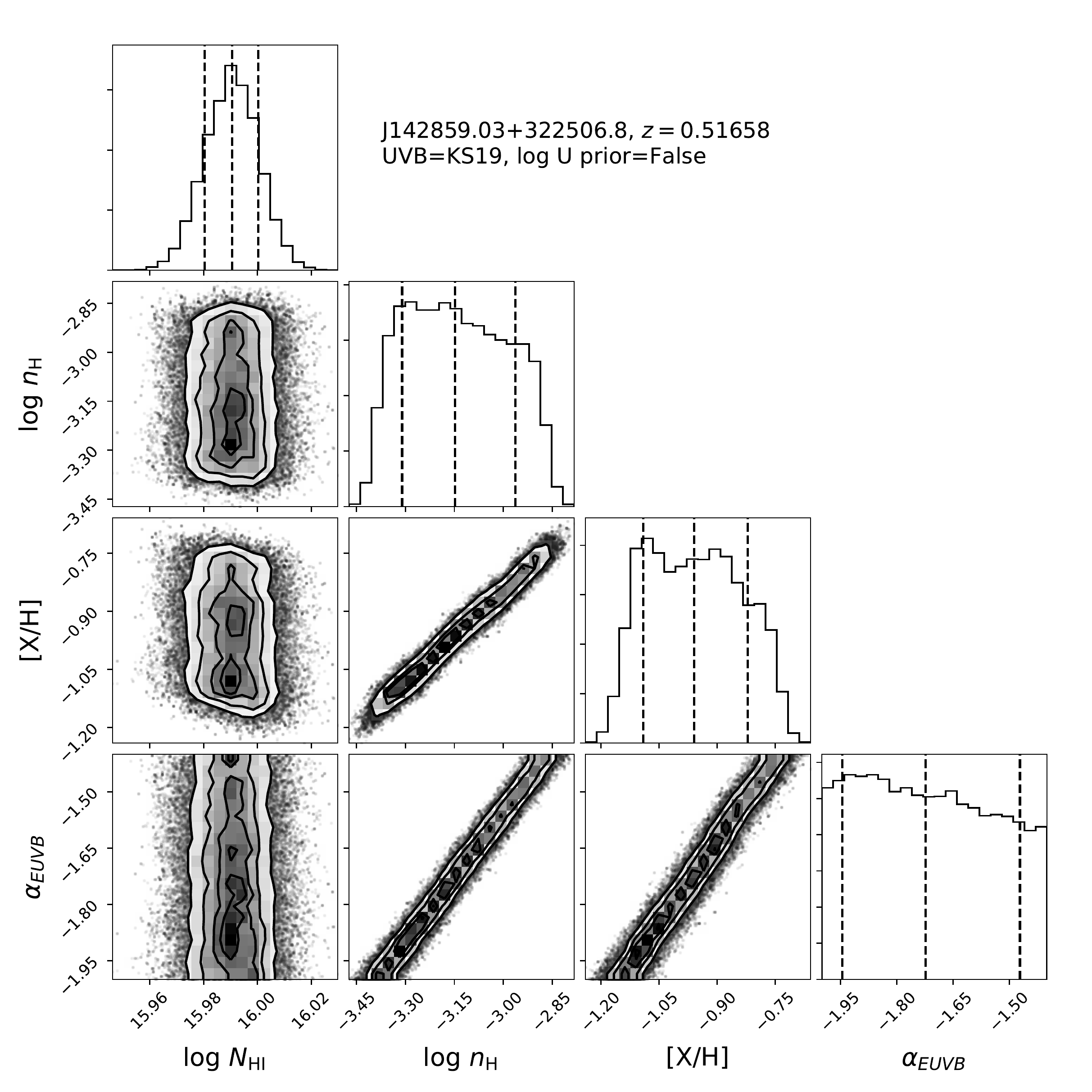}
\plotone{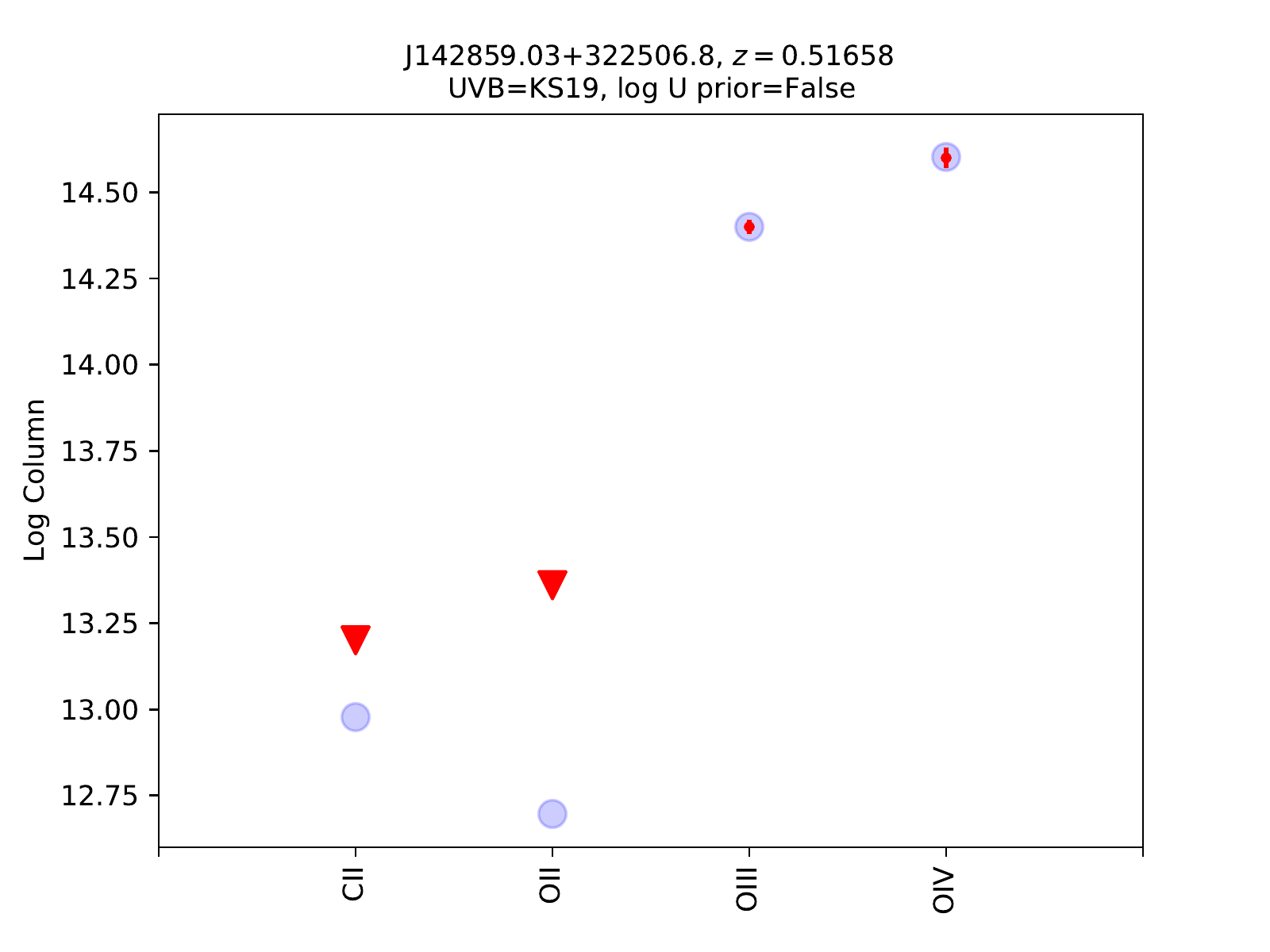}
\caption{Same as Figure~\ref{fig:A1}, but for absorber 11.
\label{fig:A11}
}
\end{figure}

\begin{figure}[tbp]
\epsscale{0.5}
\plotone{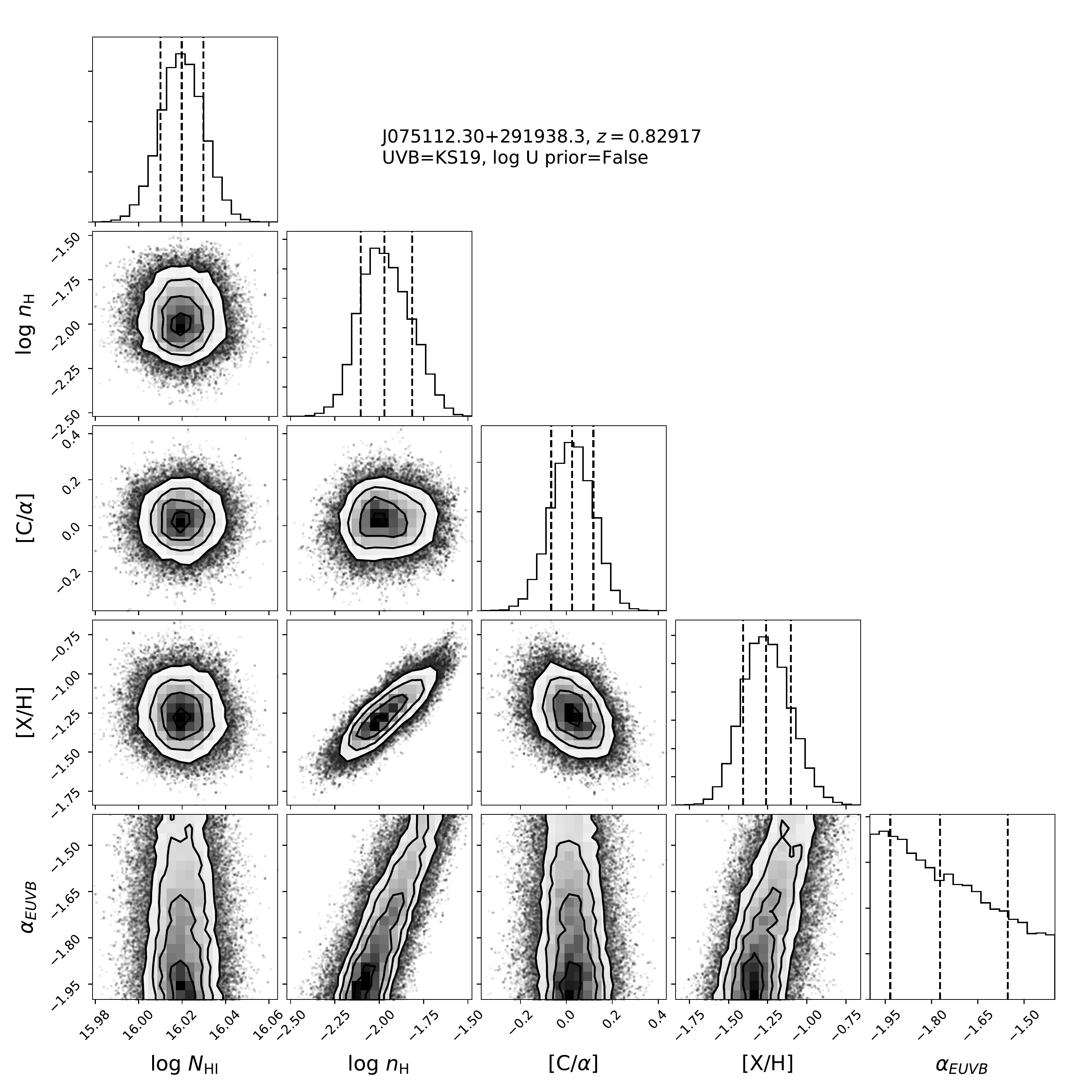}
\plotone{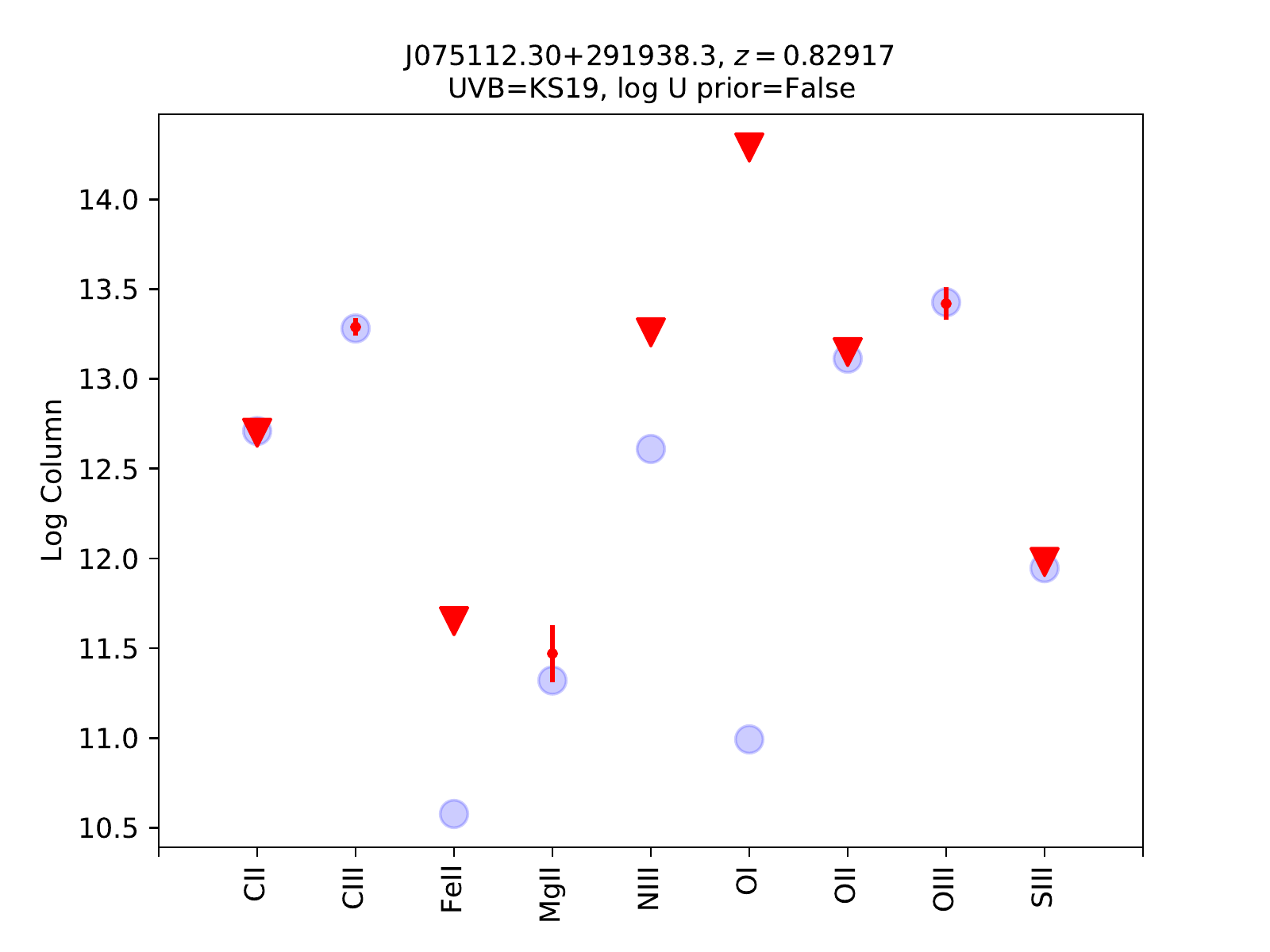}
\caption{Same as Figure~\ref{fig:A1}, but for absorber 12.
\label{fig:A12}
}
\end{figure}

\begin{figure}[tbp]
\epsscale{0.5}
\plotone{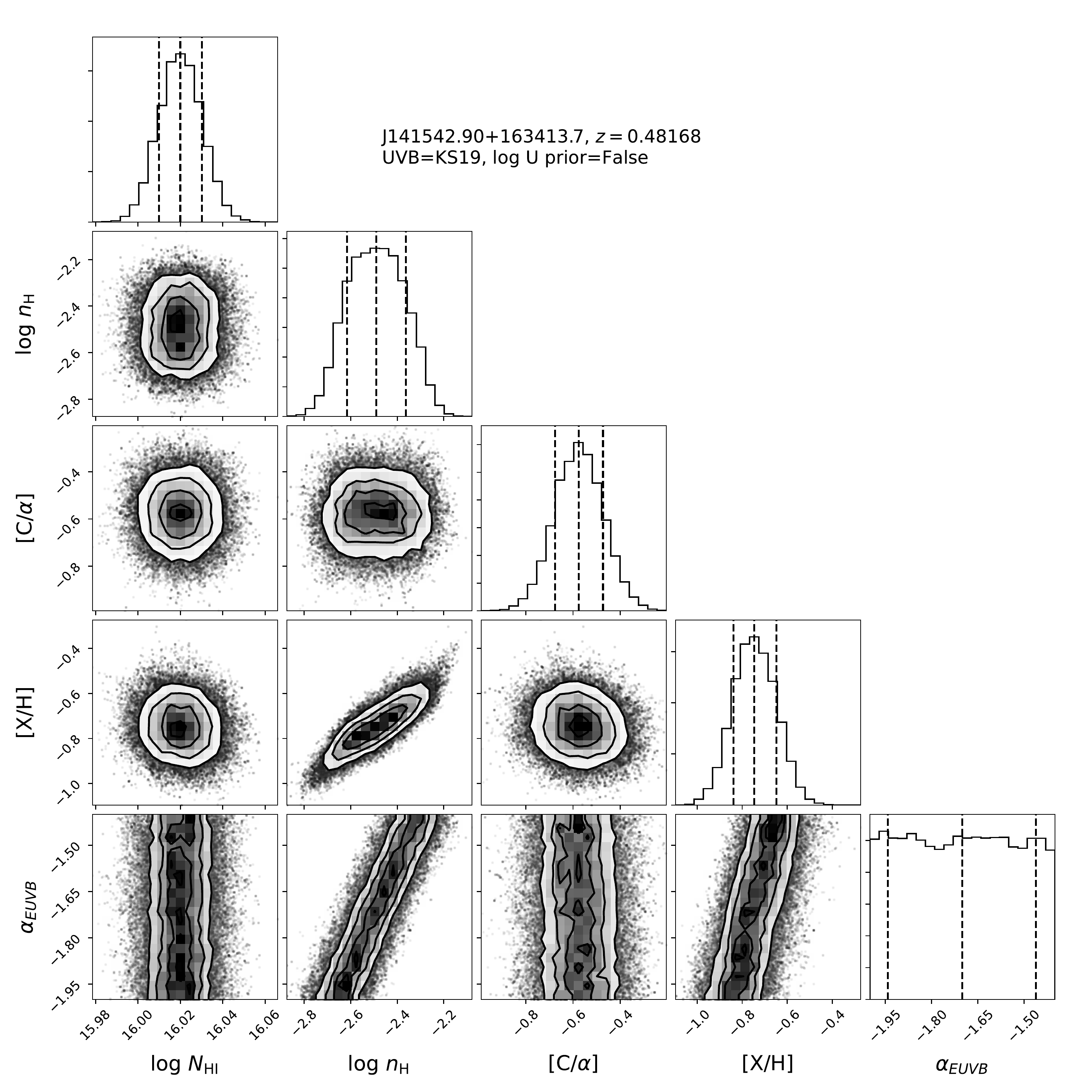}
\plotone{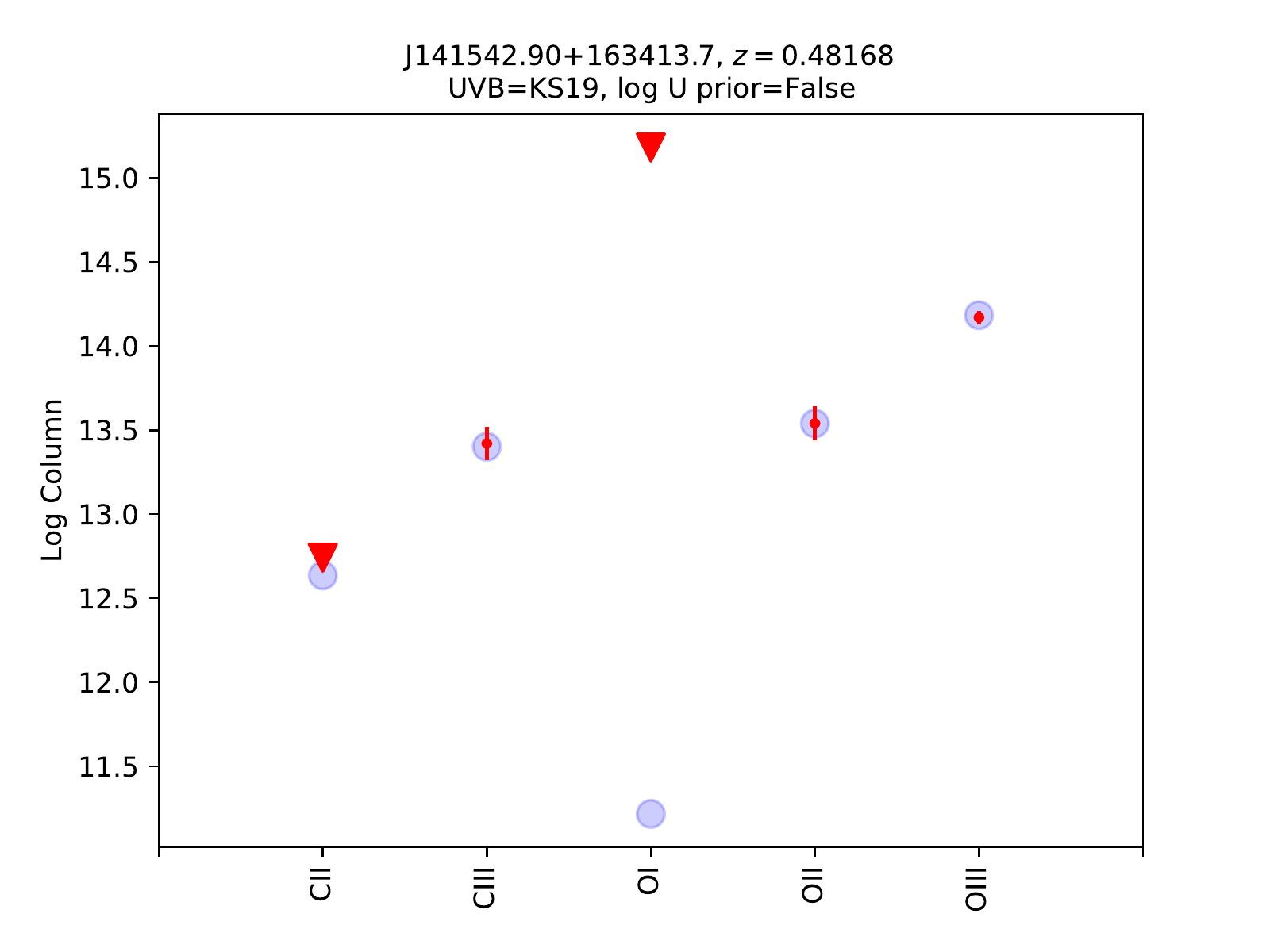}
\caption{Same as Figure~\ref{fig:A1}, but for absorber 13.
\label{fig:A13}
}
\end{figure}

\begin{figure}[tbp]
\epsscale{0.5}
\plotone{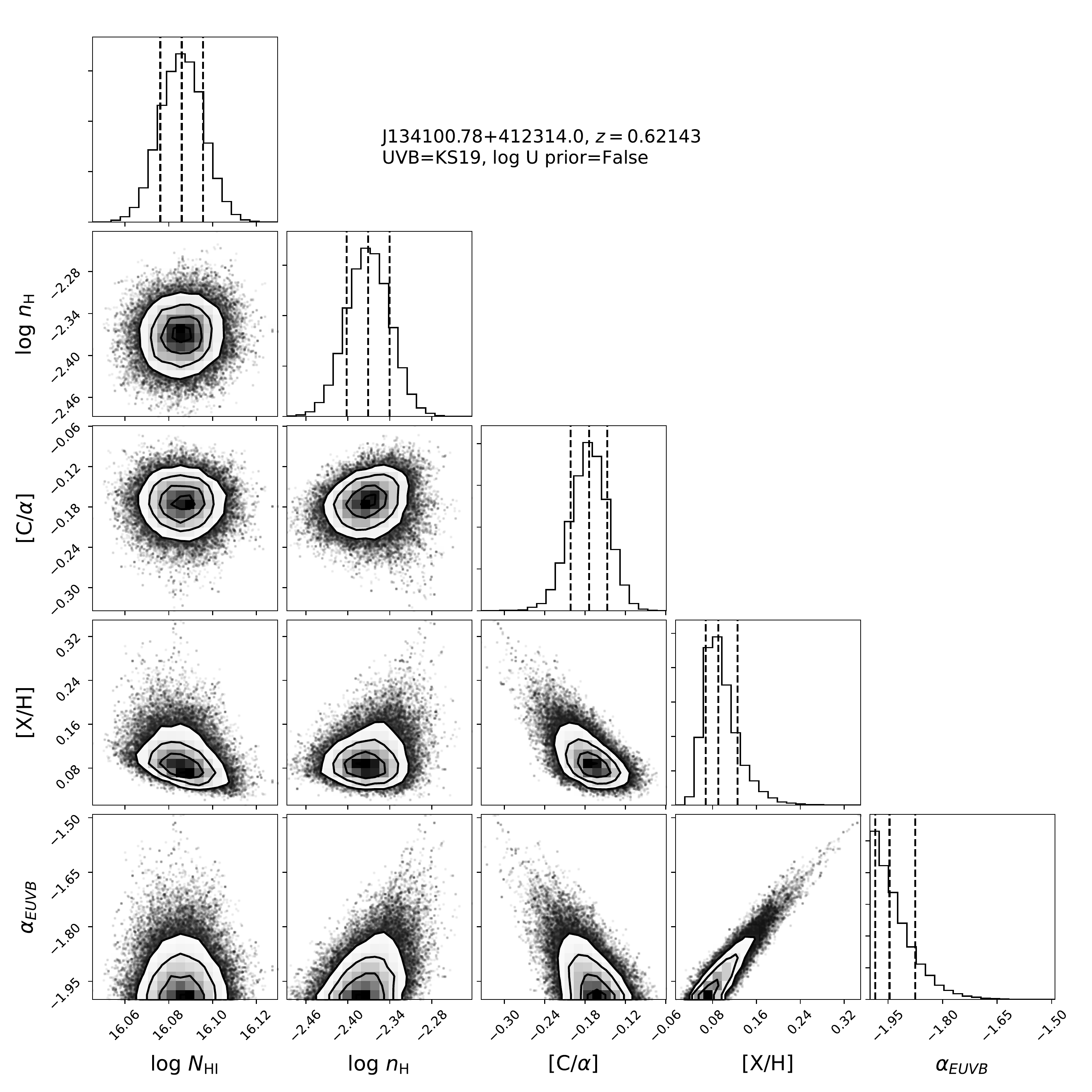}
\plotone{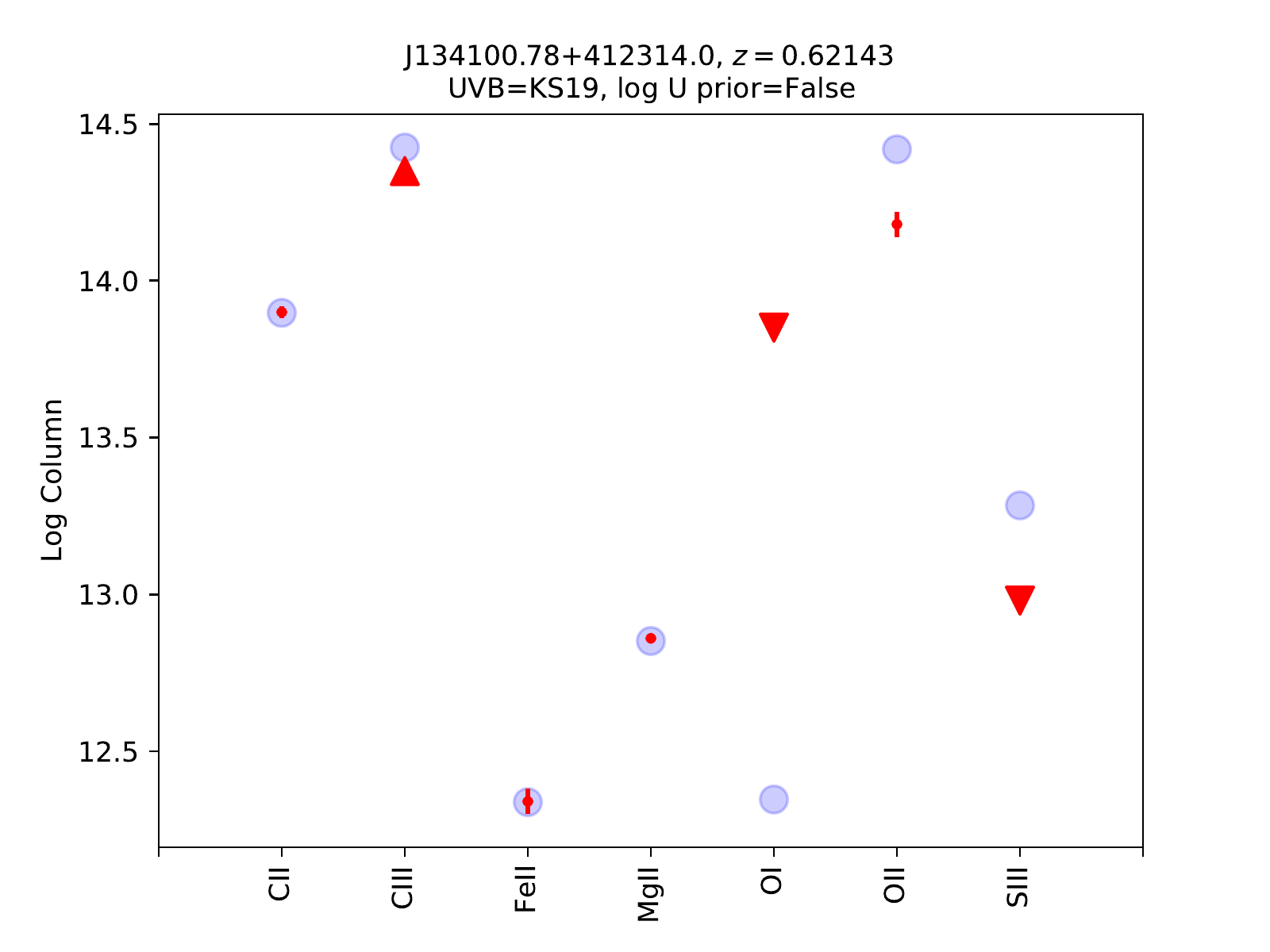}
\caption{Same as Figure~\ref{fig:A1}, but for absorber 14.
\label{fig:A14}
}
\end{figure}

\clearpage

\begin{figure}[tbp]
\epsscale{0.5}
\plotone{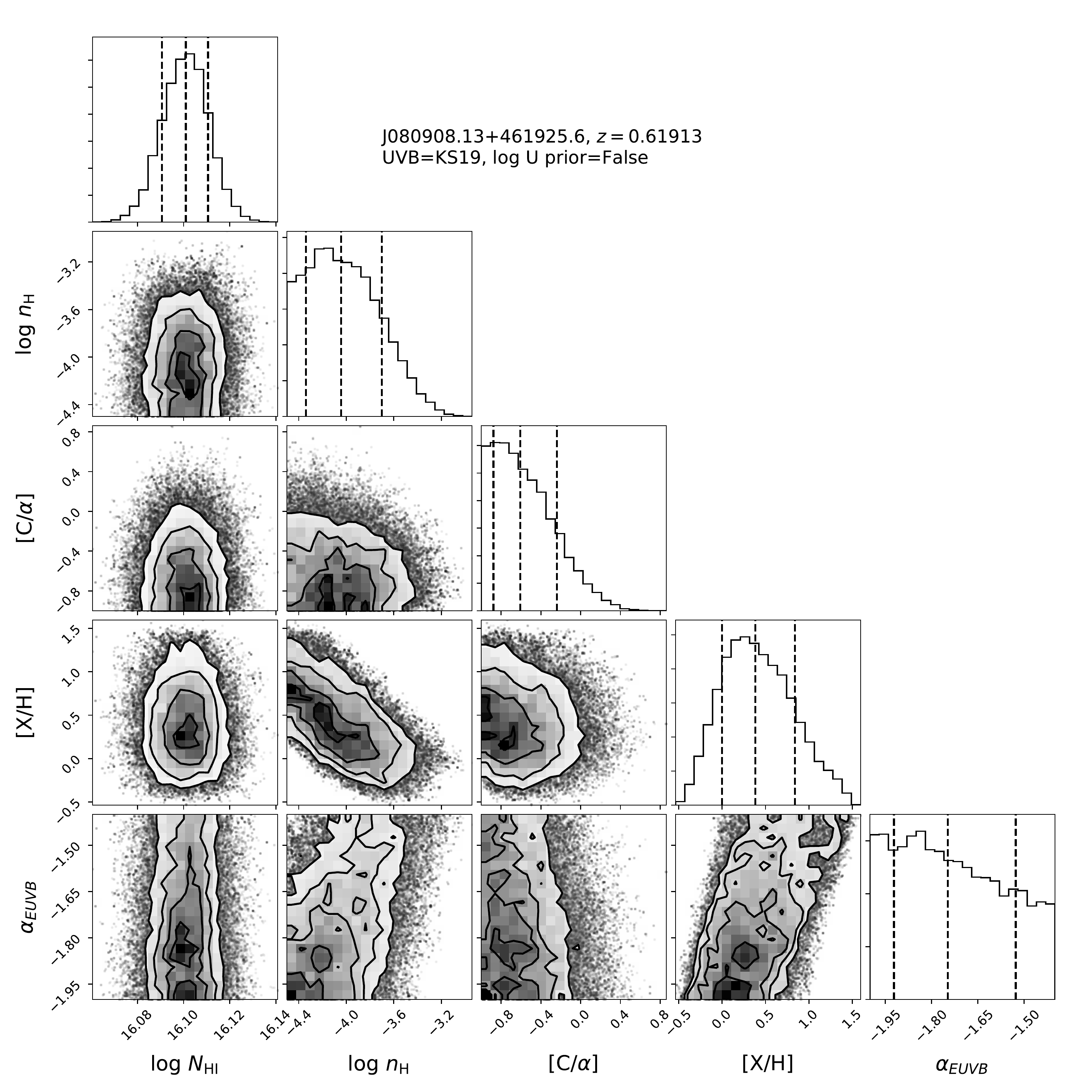}
\plotone{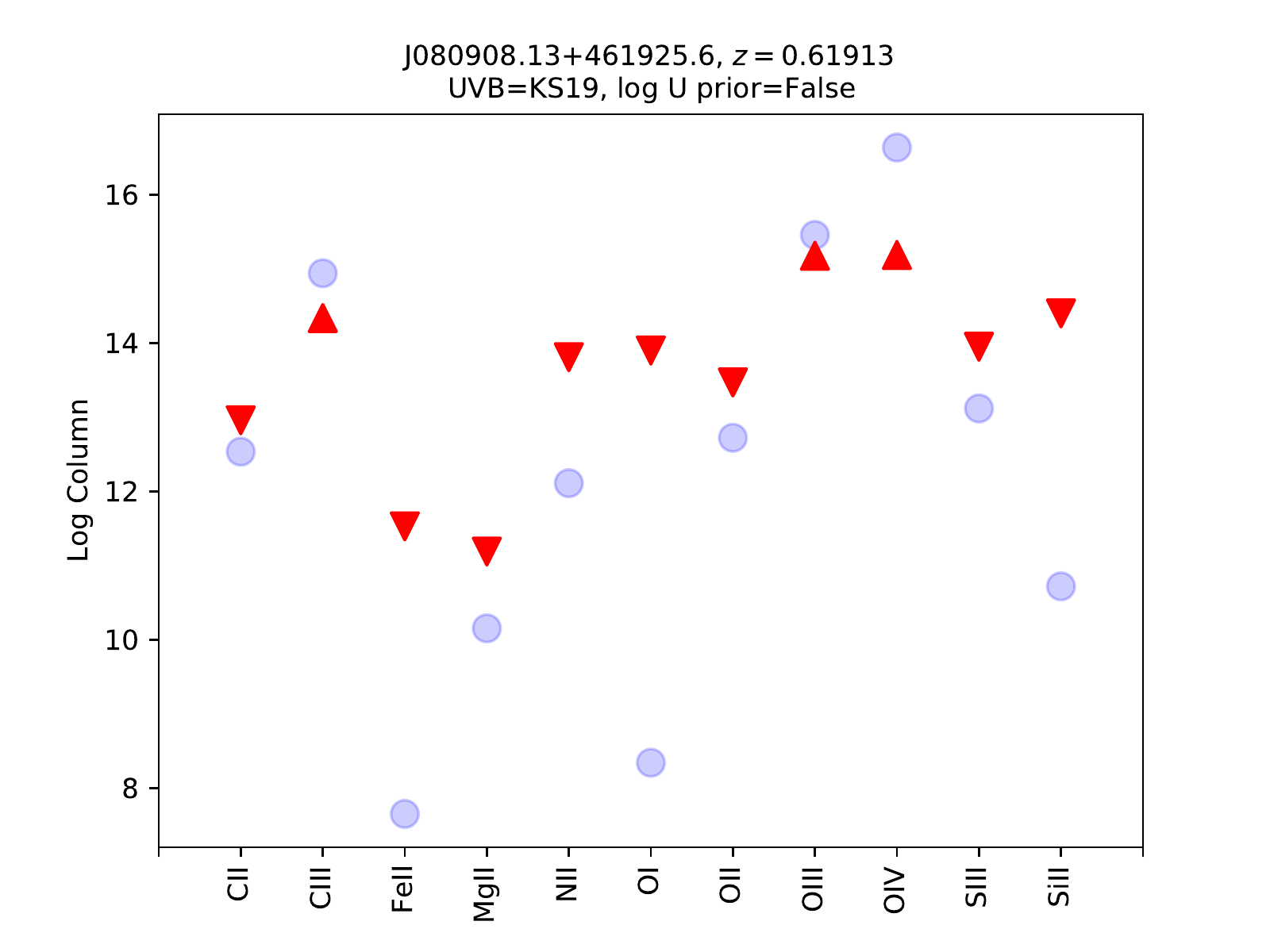}
\caption{Same as Figure~\ref{fig:A1}, but for absorber 15.
\label{fig:A15}
}
\end{figure}

\begin{figure}[tbp]
\epsscale{0.5}
\plotone{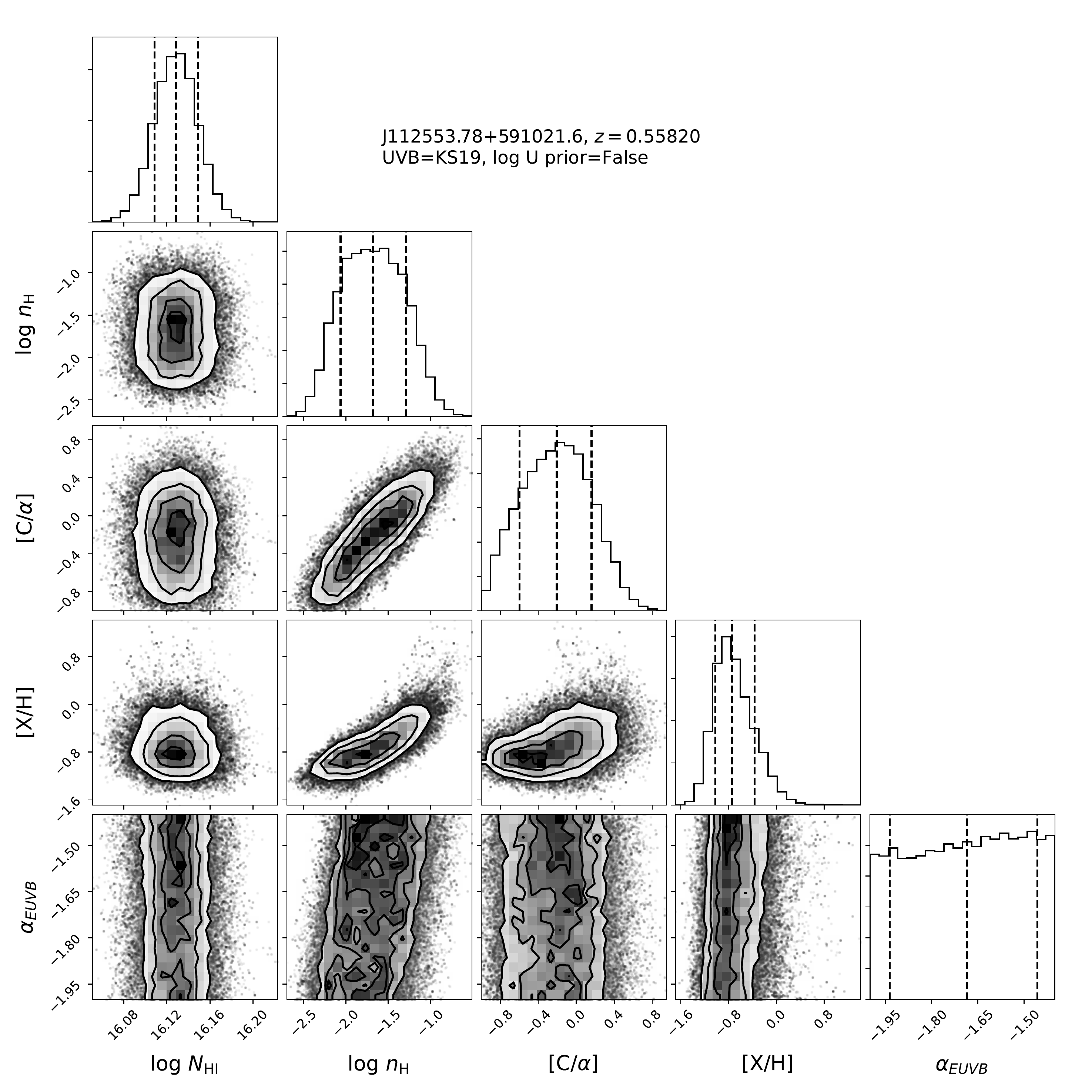}
\plotone{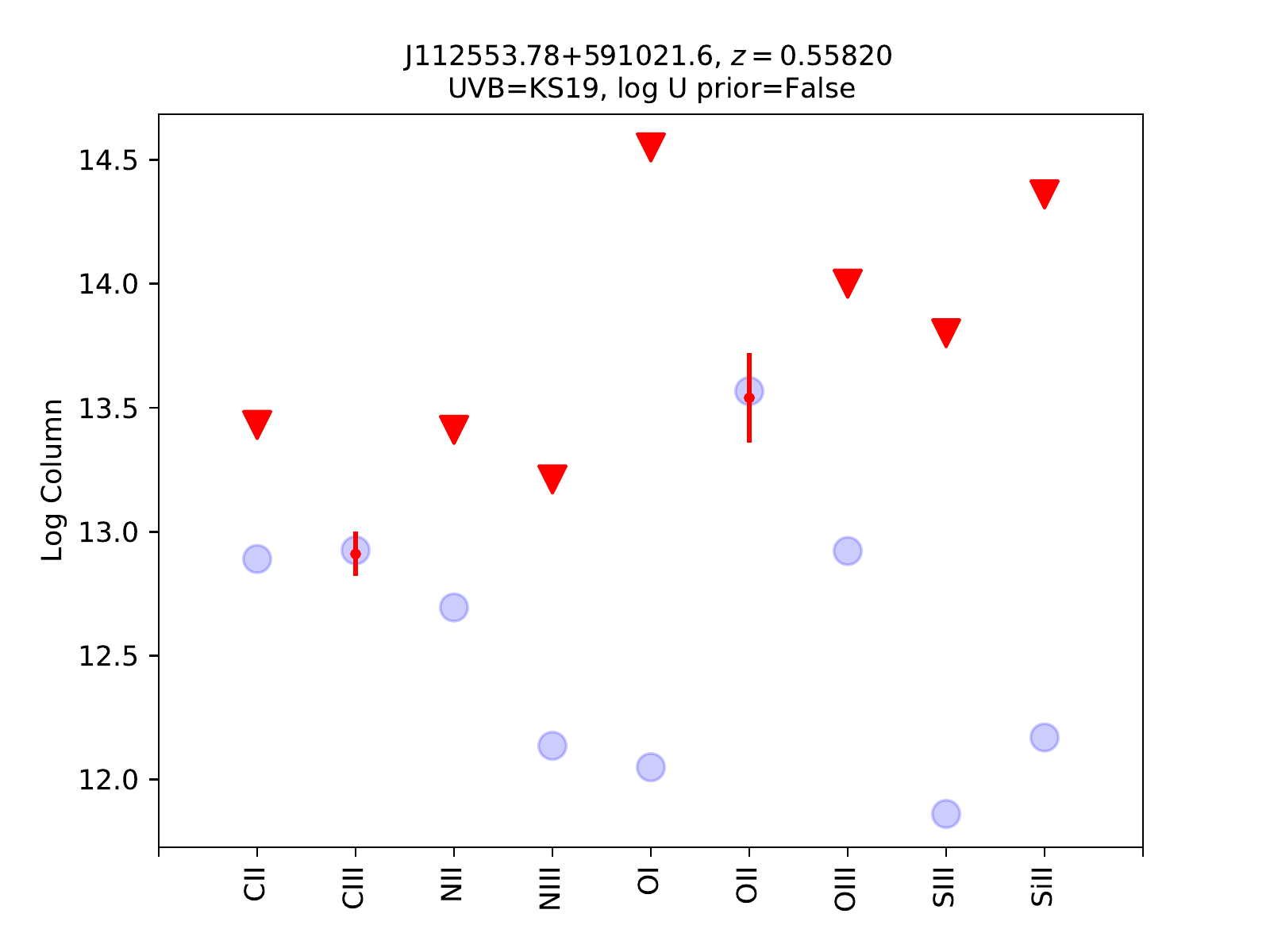}
\caption{Same as Figure~\ref{fig:A1}, but for absorber 16.
\label{fig:A16}
}
\end{figure}

\begin{figure}[tbp]
\epsscale{0.5}
\plotone{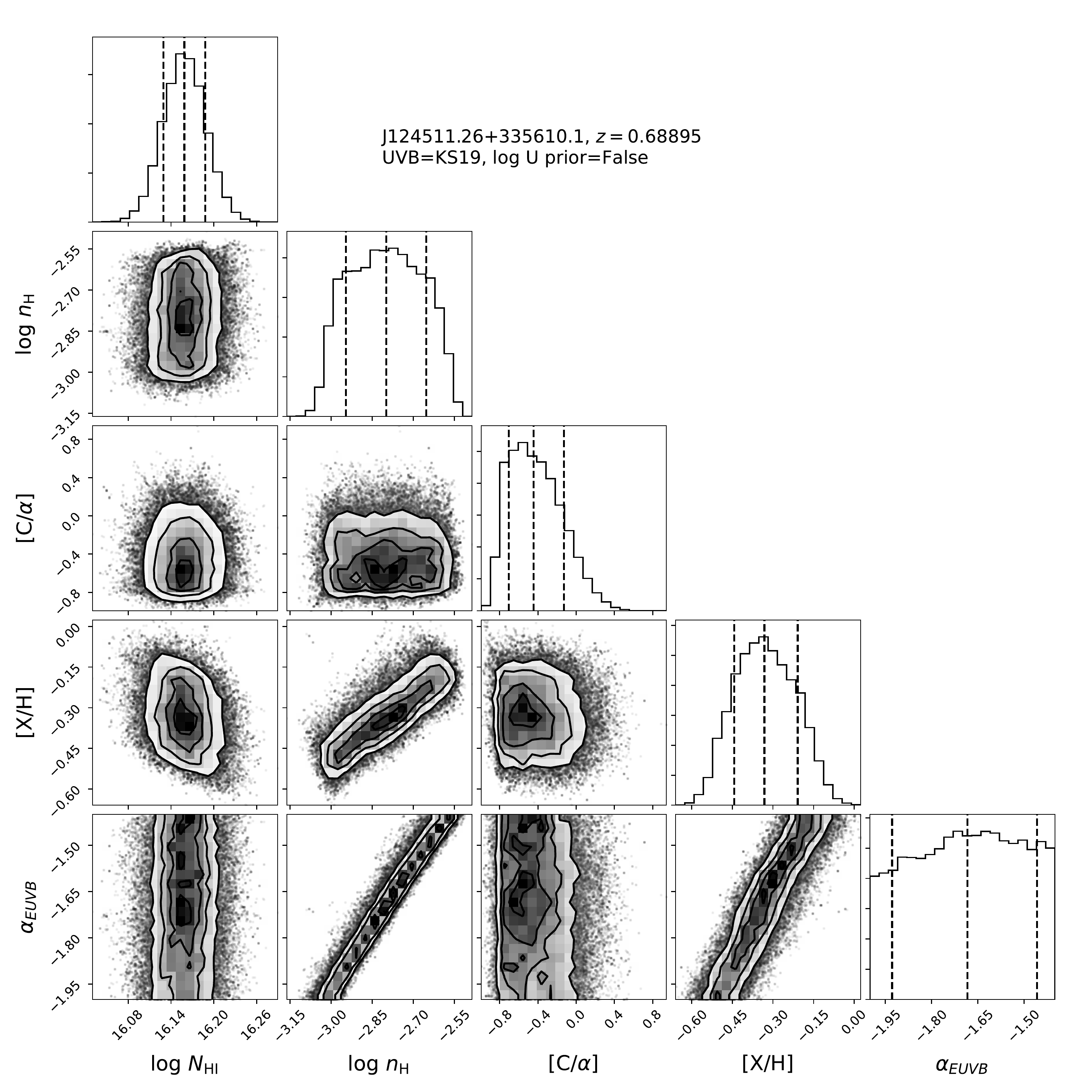}
\plotone{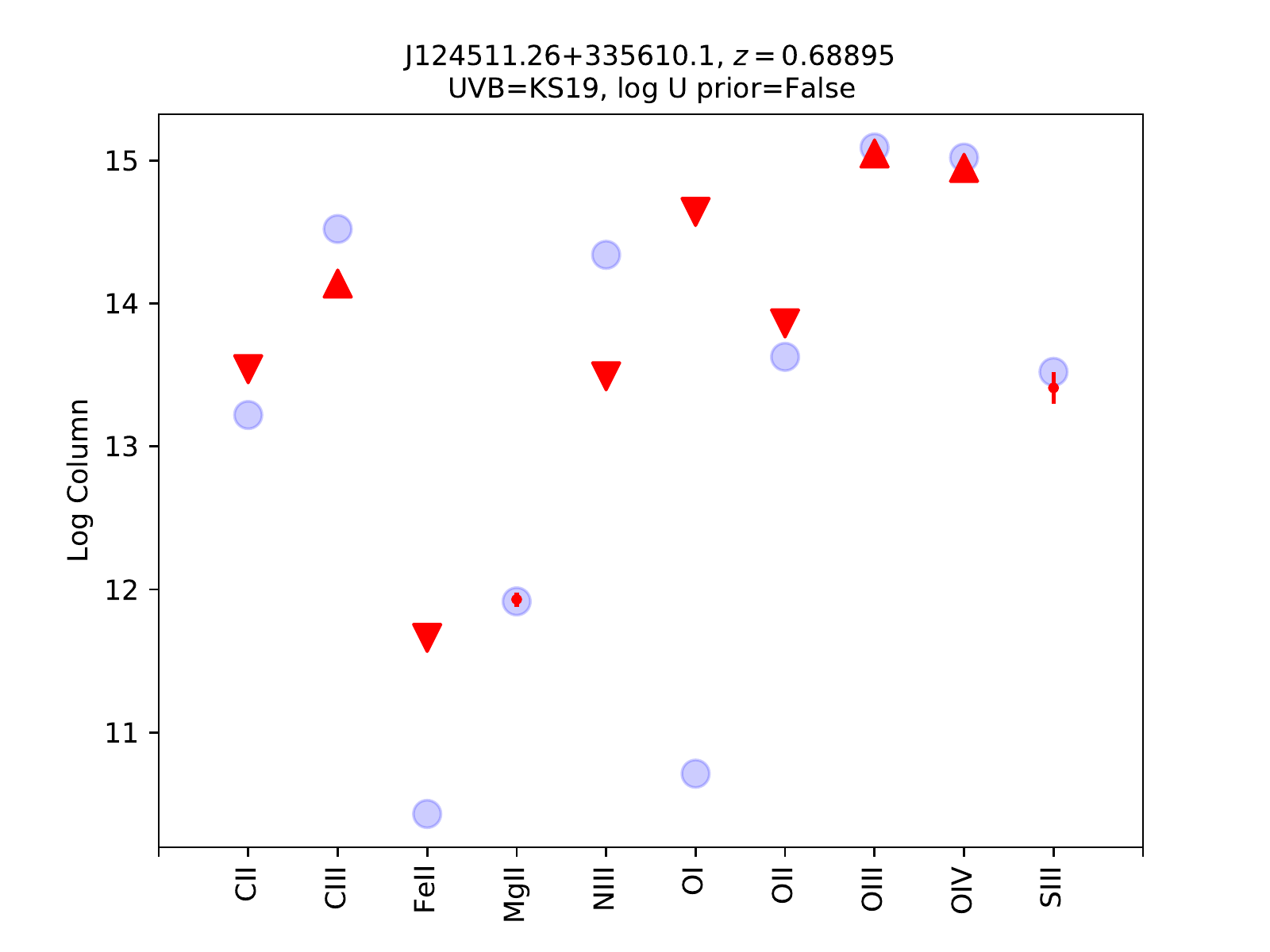}
\caption{Same as Figure~\ref{fig:A1}, but for absorber 17.
\label{fig:A17}
}
\end{figure}

\begin{figure}[tbp]
\epsscale{0.5}
\plotone{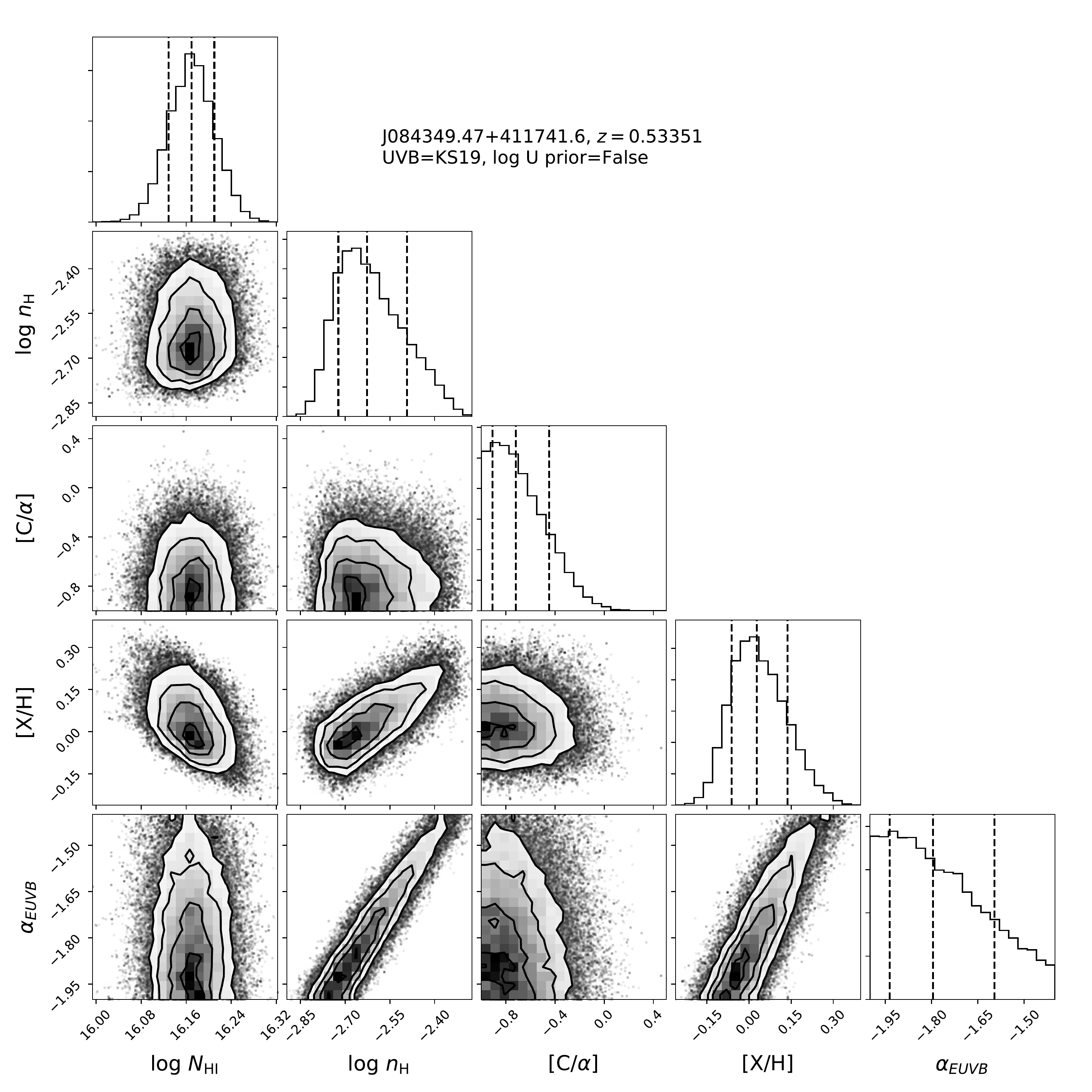}
\plotone{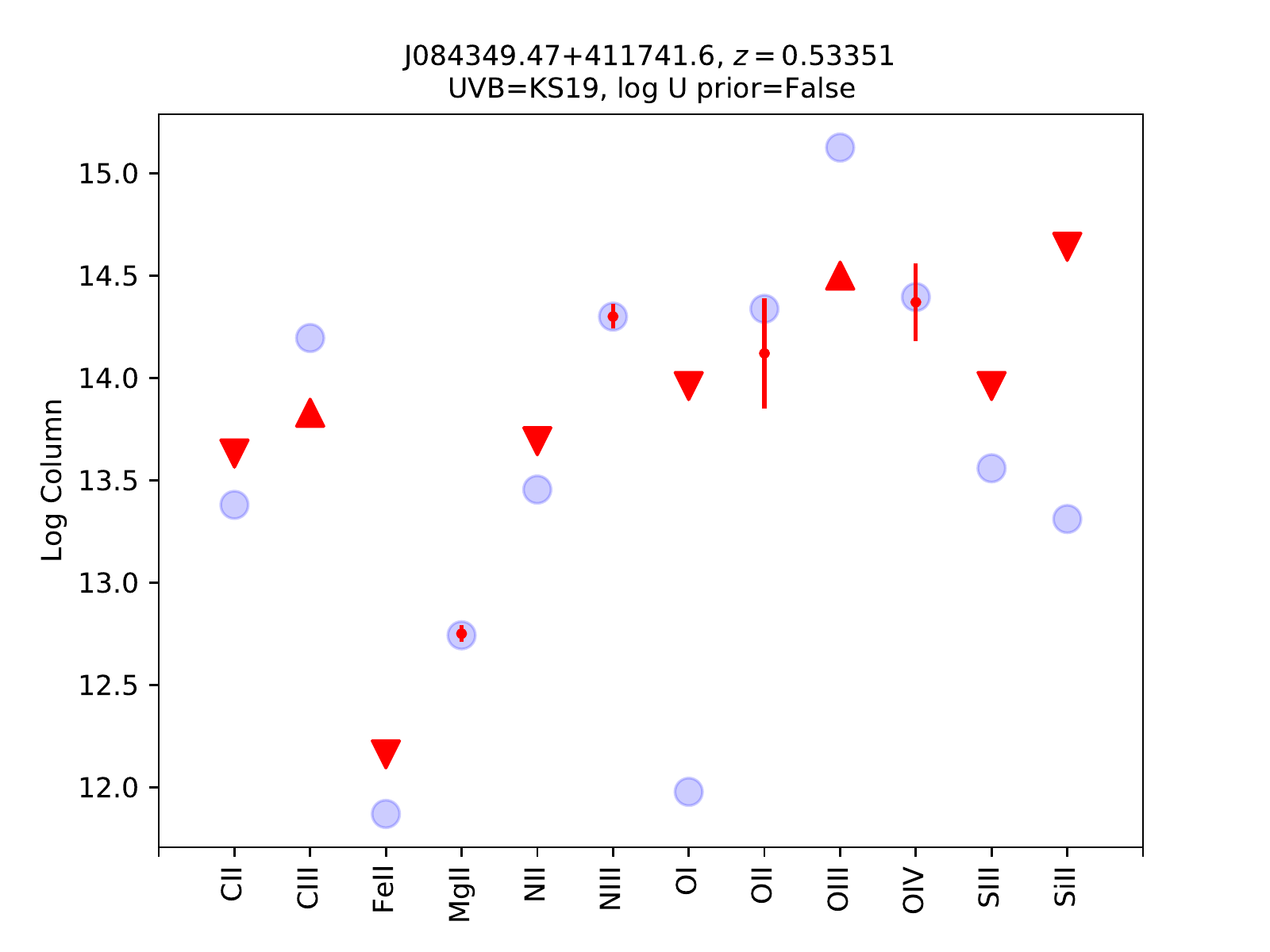}
\caption{Same as Figure~\ref{fig:A1}, but for absorber 18.
\label{fig:A18}
}
\end{figure}

\begin{figure}[tbp]
\epsscale{0.5}
\plotone{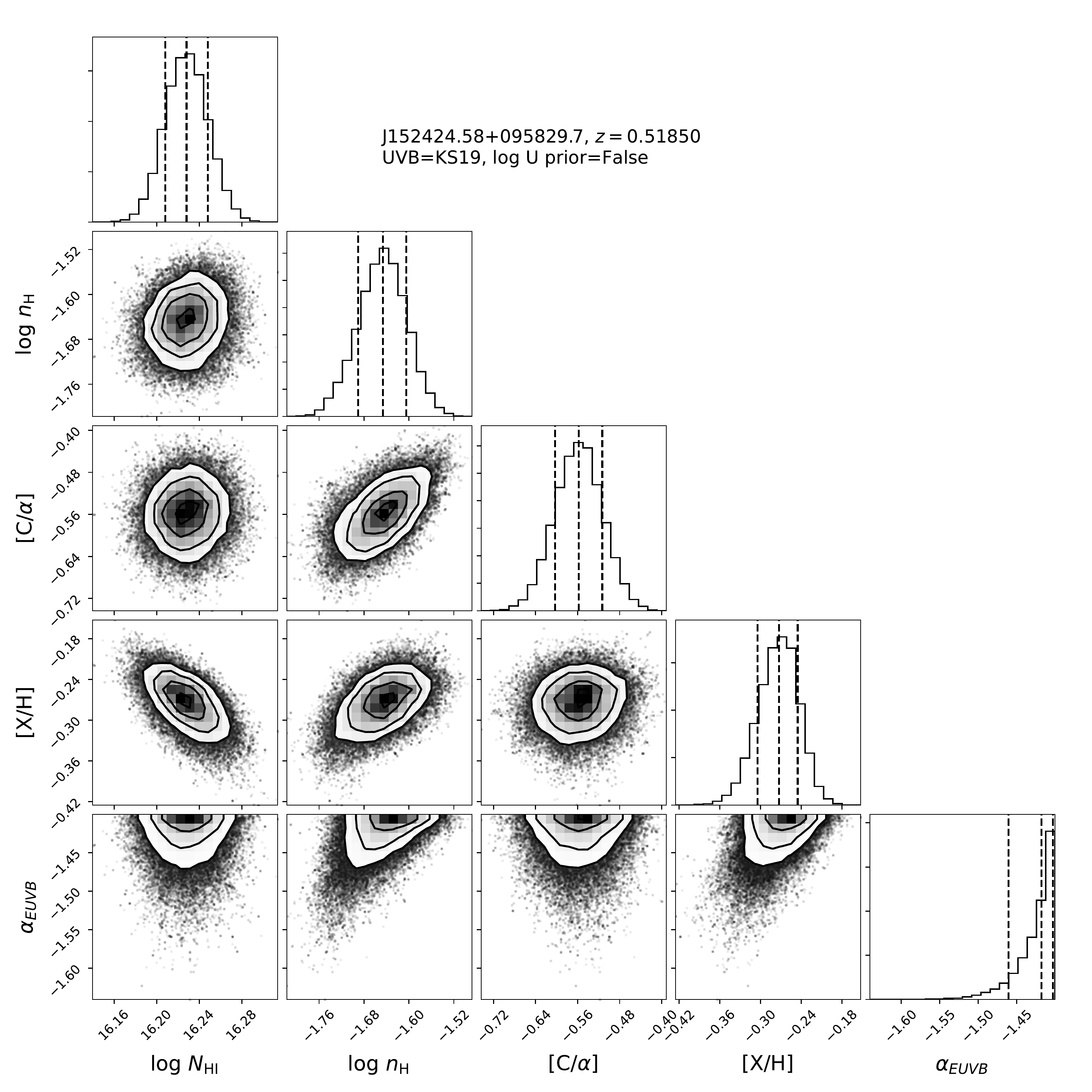}
\plotone{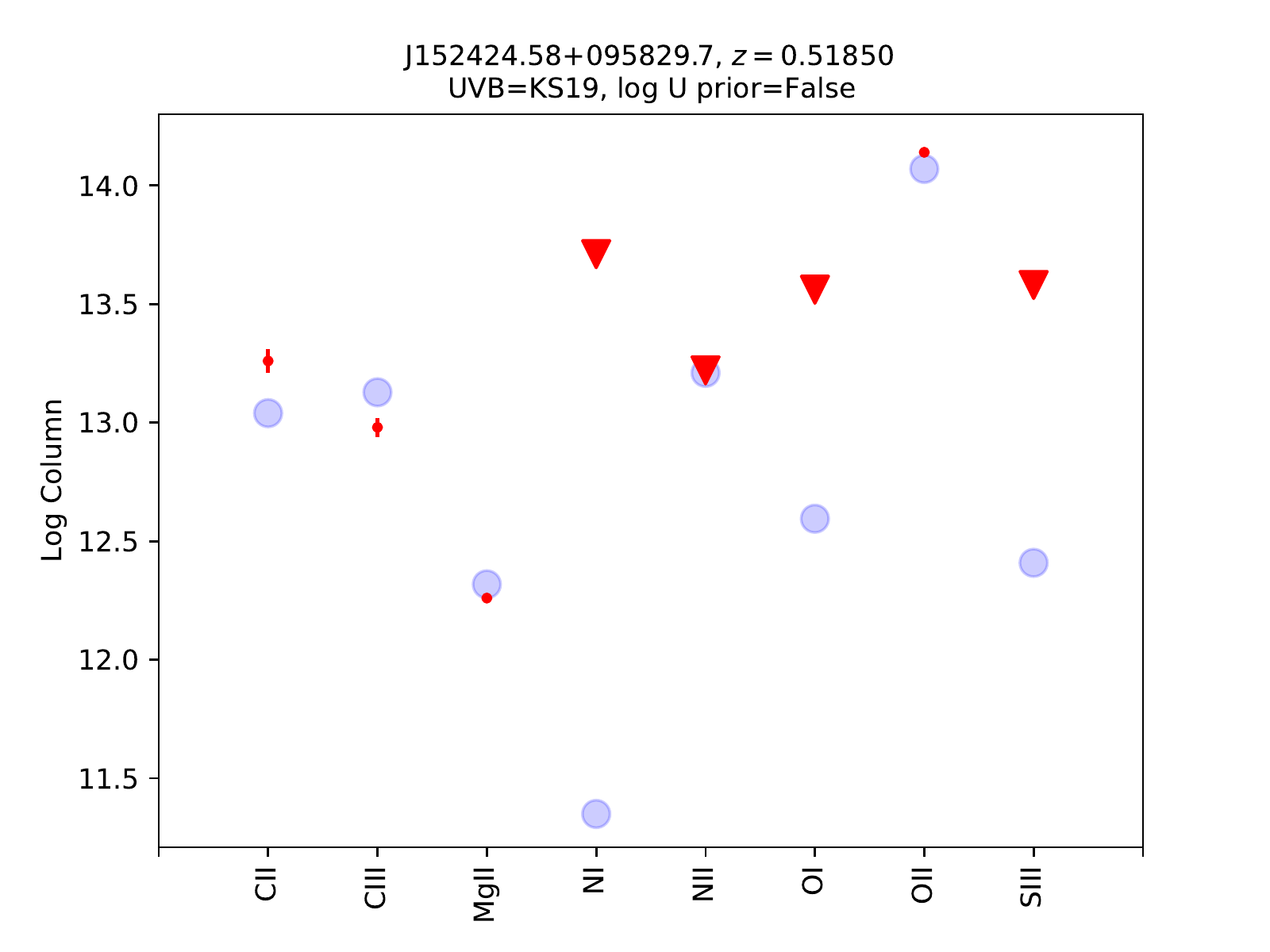}
\caption{Same as Figure~\ref{fig:A1}, but for absorber 19.
\label{fig:A19}
}
\end{figure}

\begin{figure}[tbp]
\epsscale{0.5}
\plotone{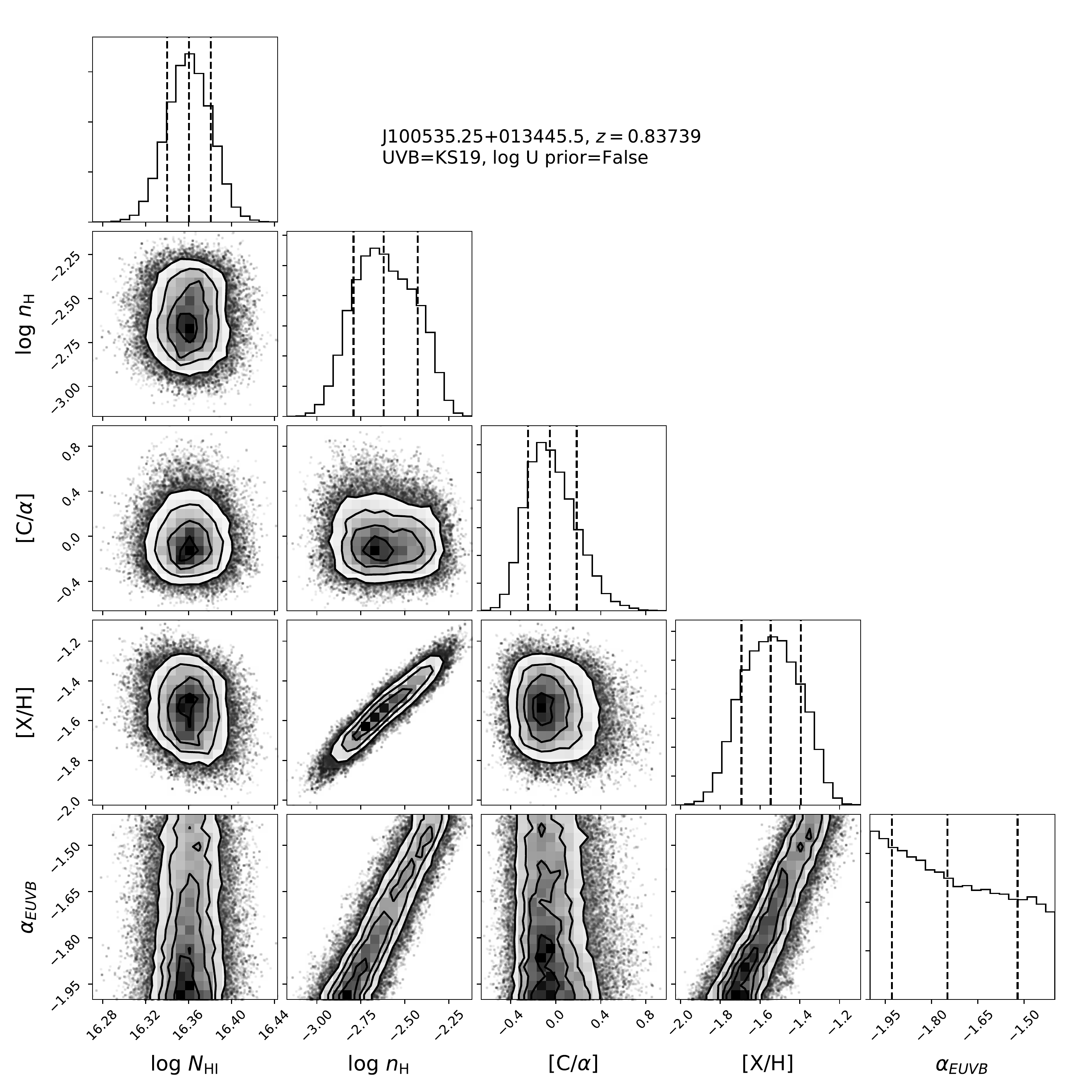}
\plotone{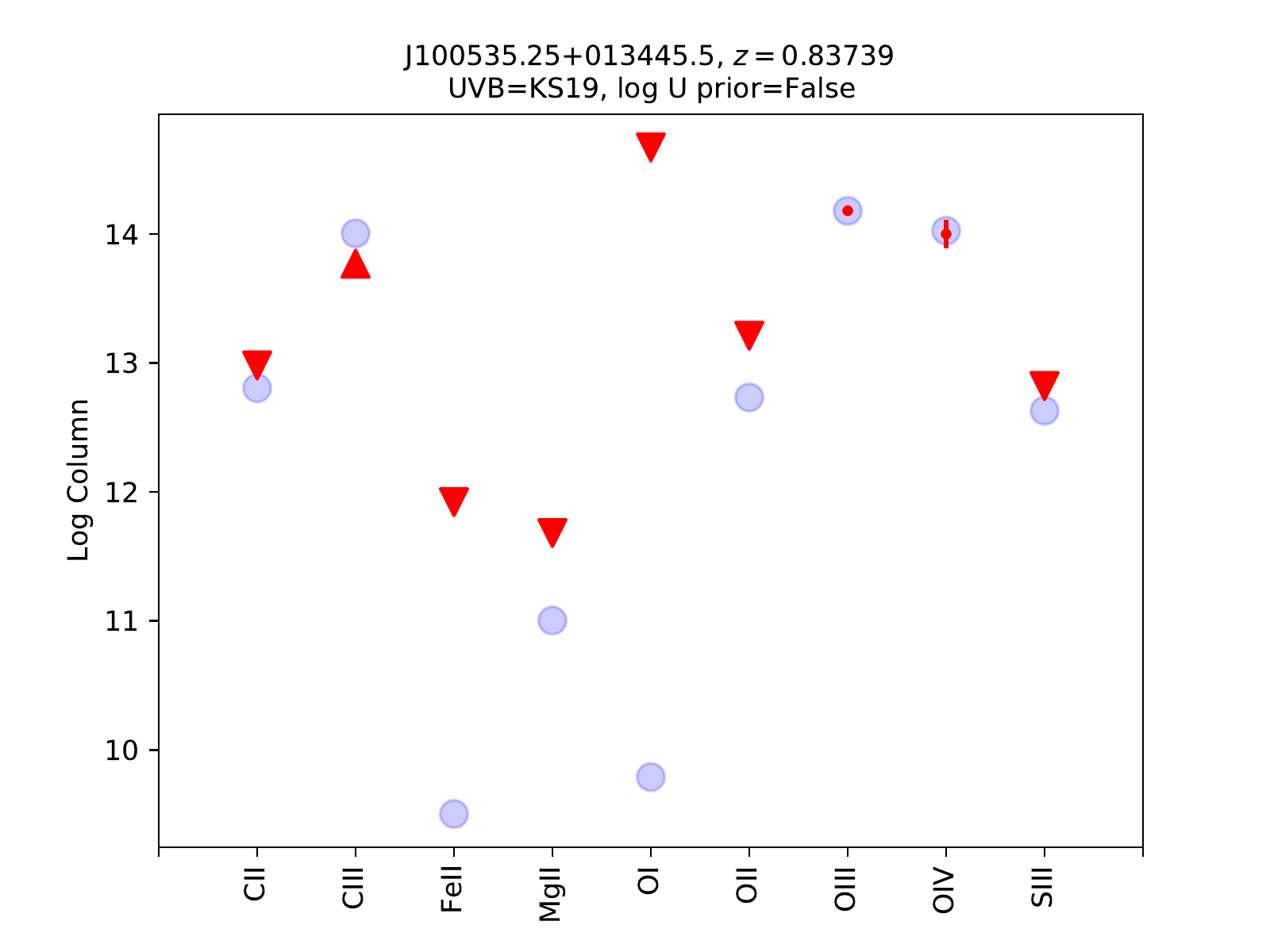}
\caption{Same as Figure~\ref{fig:A1}, but for absorber 20.
\label{fig:A20}
}
\end{figure}

\begin{figure}[tbp]
\epsscale{0.5}
\plotone{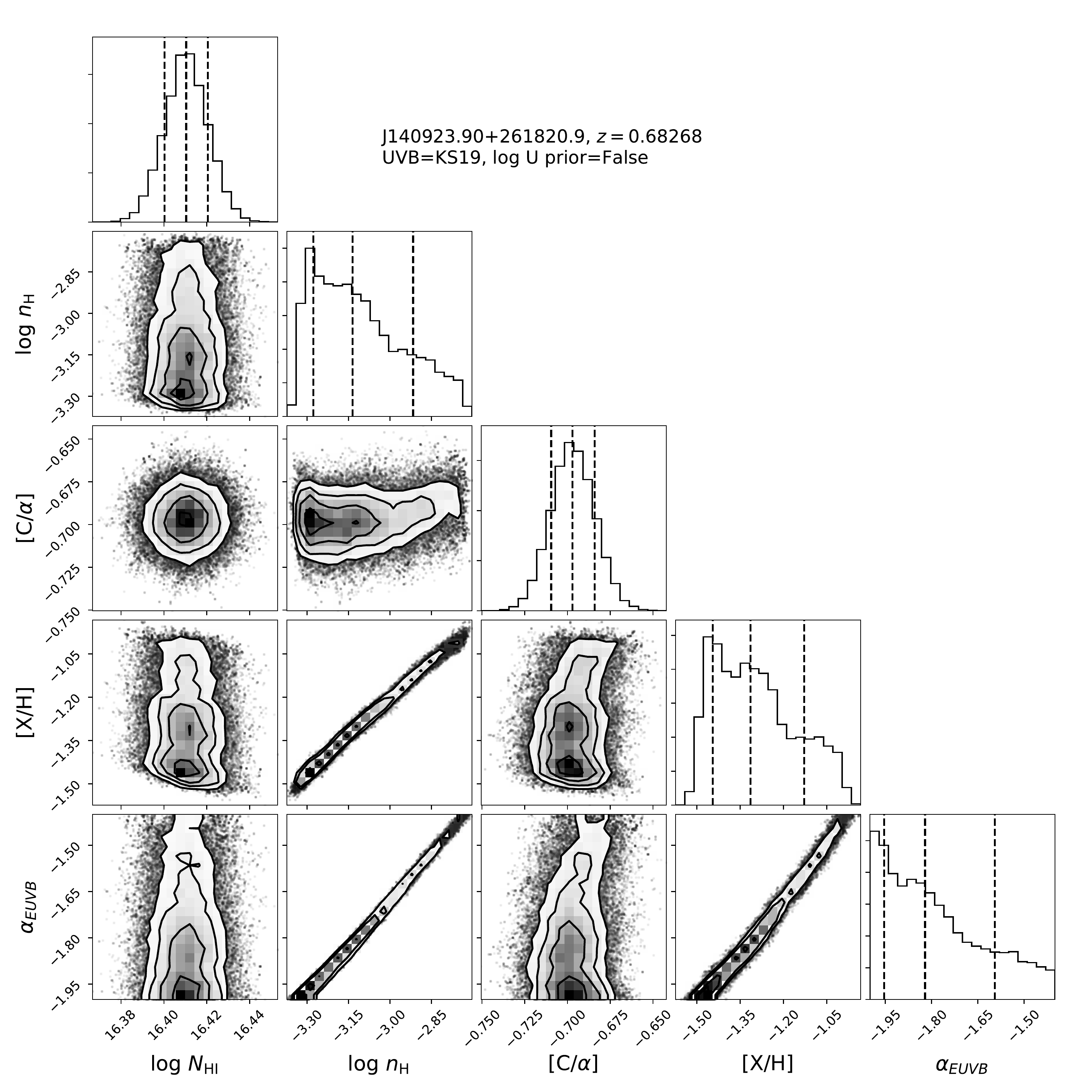}
\plotone{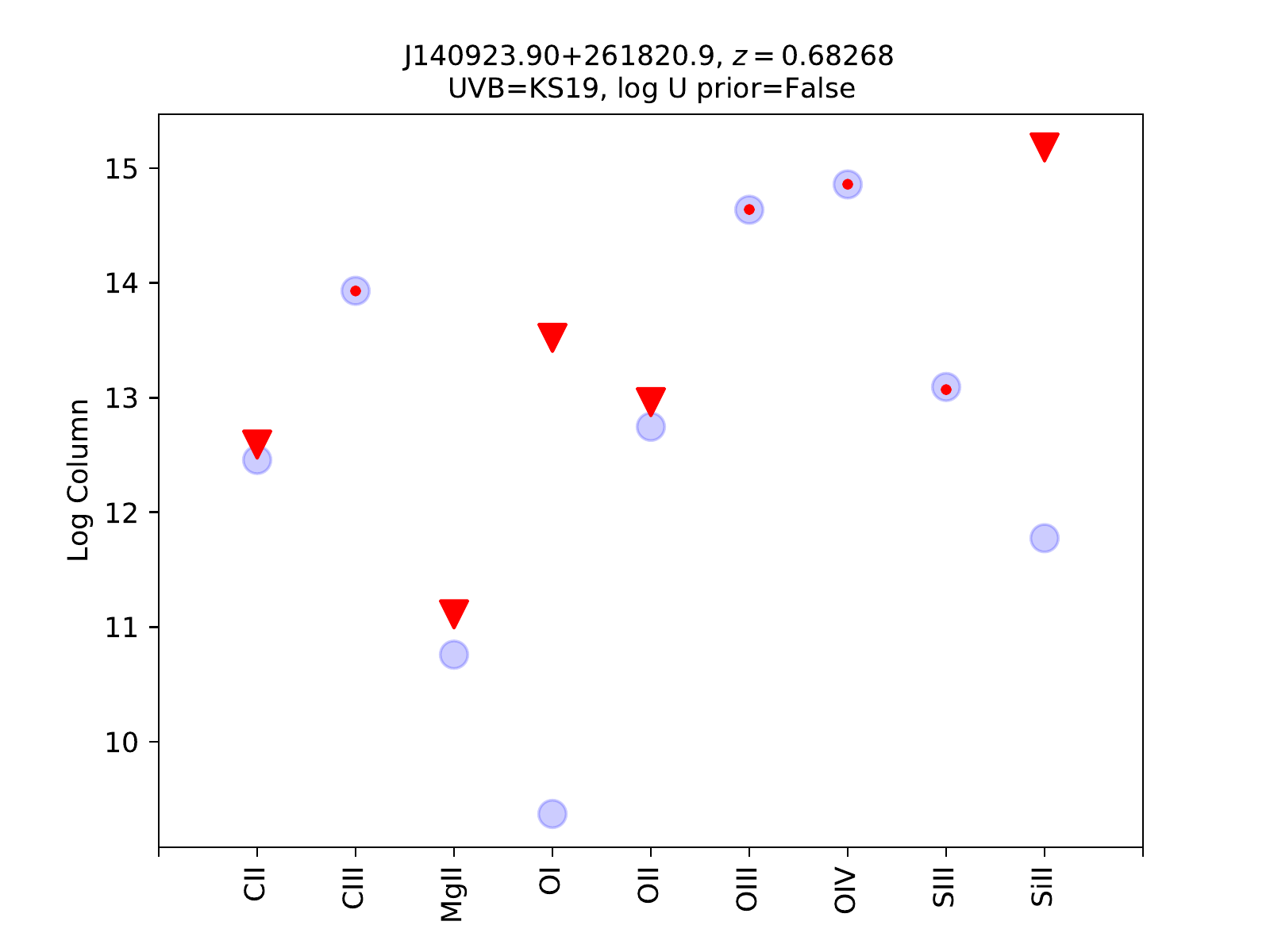}
\caption{Same as Figure~\ref{fig:A1}, but for absorber 21.
\label{fig:A21}
}
\end{figure}

\begin{figure}[tbp]
\epsscale{0.5}
\plotone{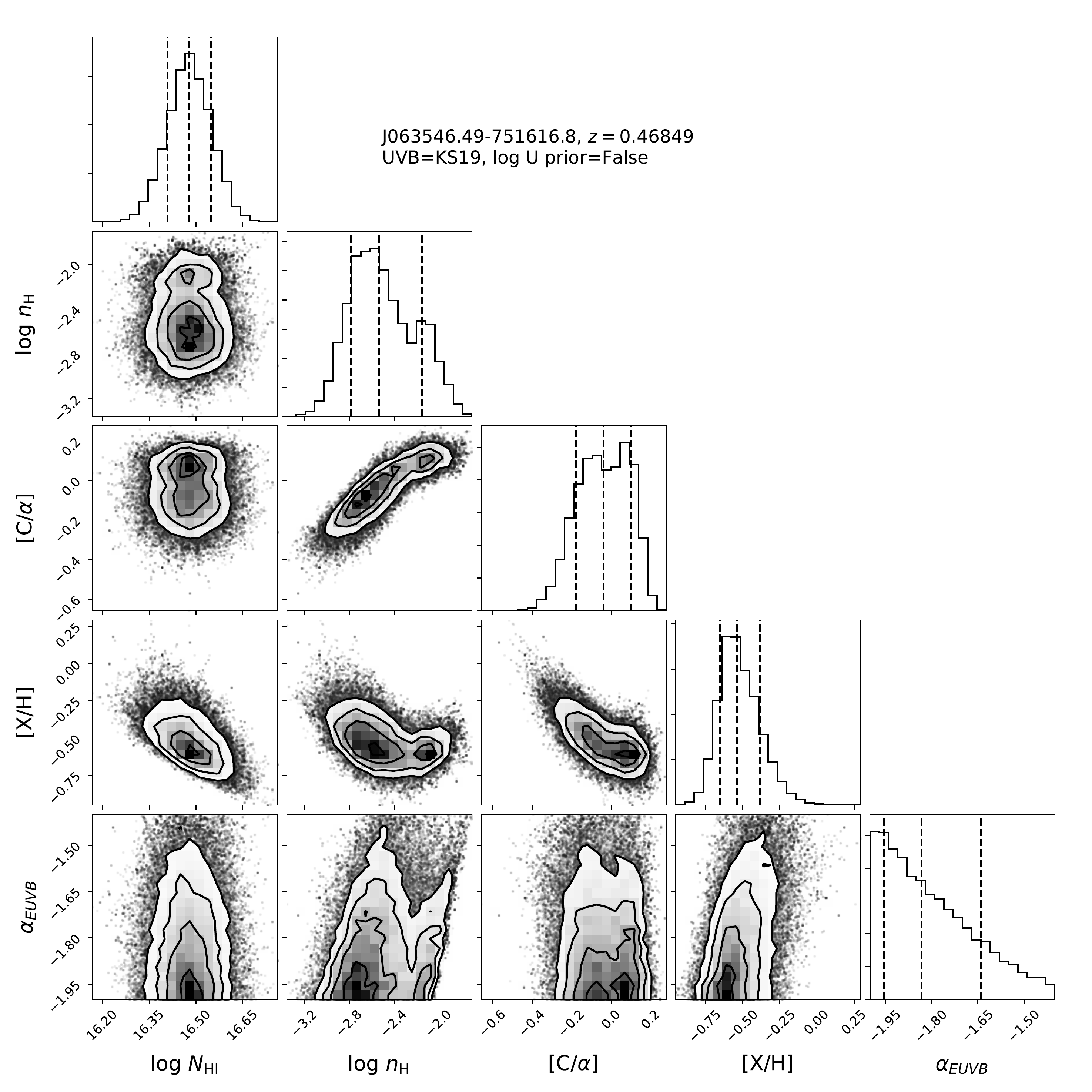}
\plotone{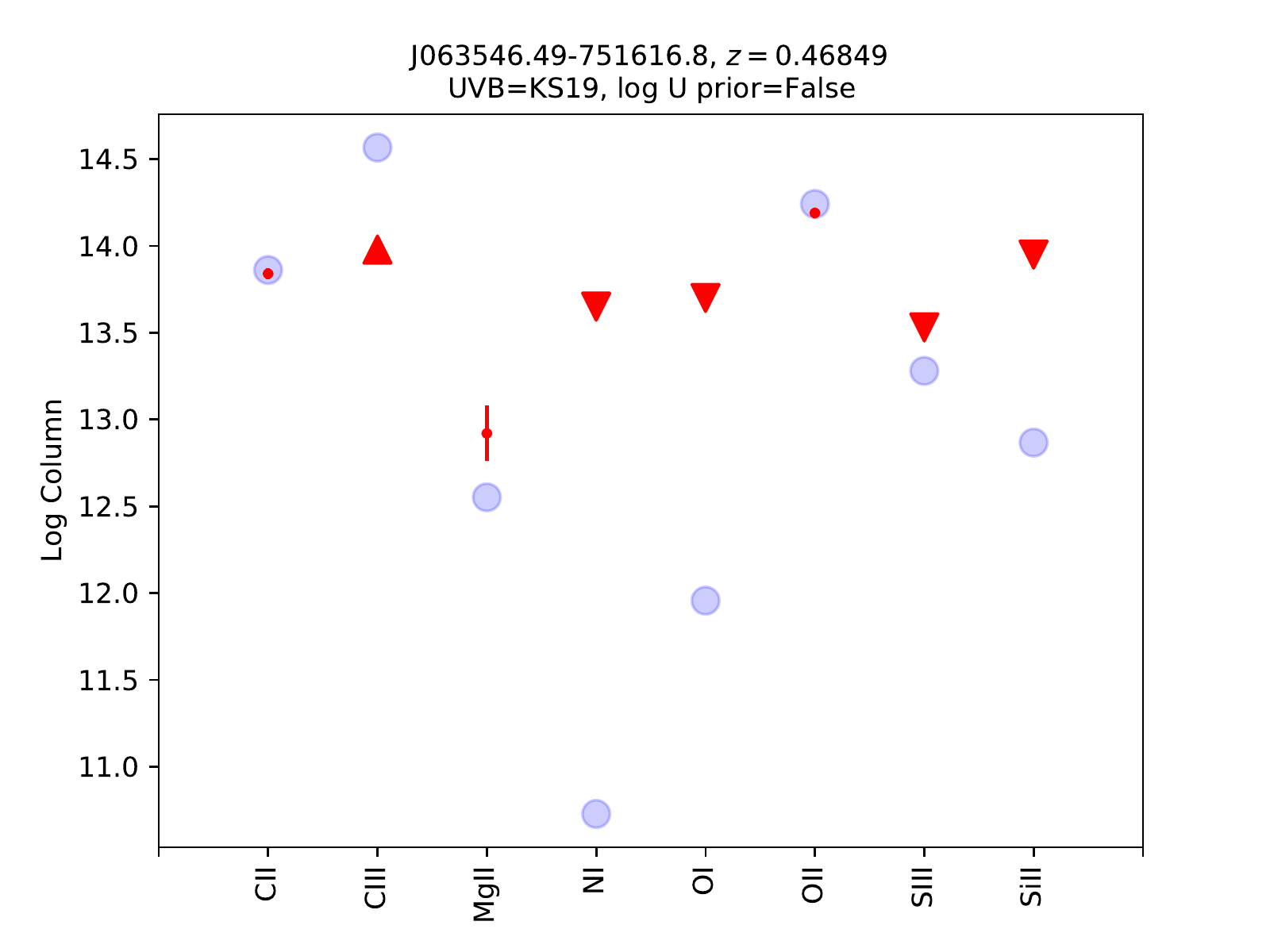}
\caption{Same as Figure~\ref{fig:A1}, but for absorber 22.
\label{fig:A22}
}
\end{figure}

\begin{figure}[tbp]
\epsscale{0.5}
\plotone{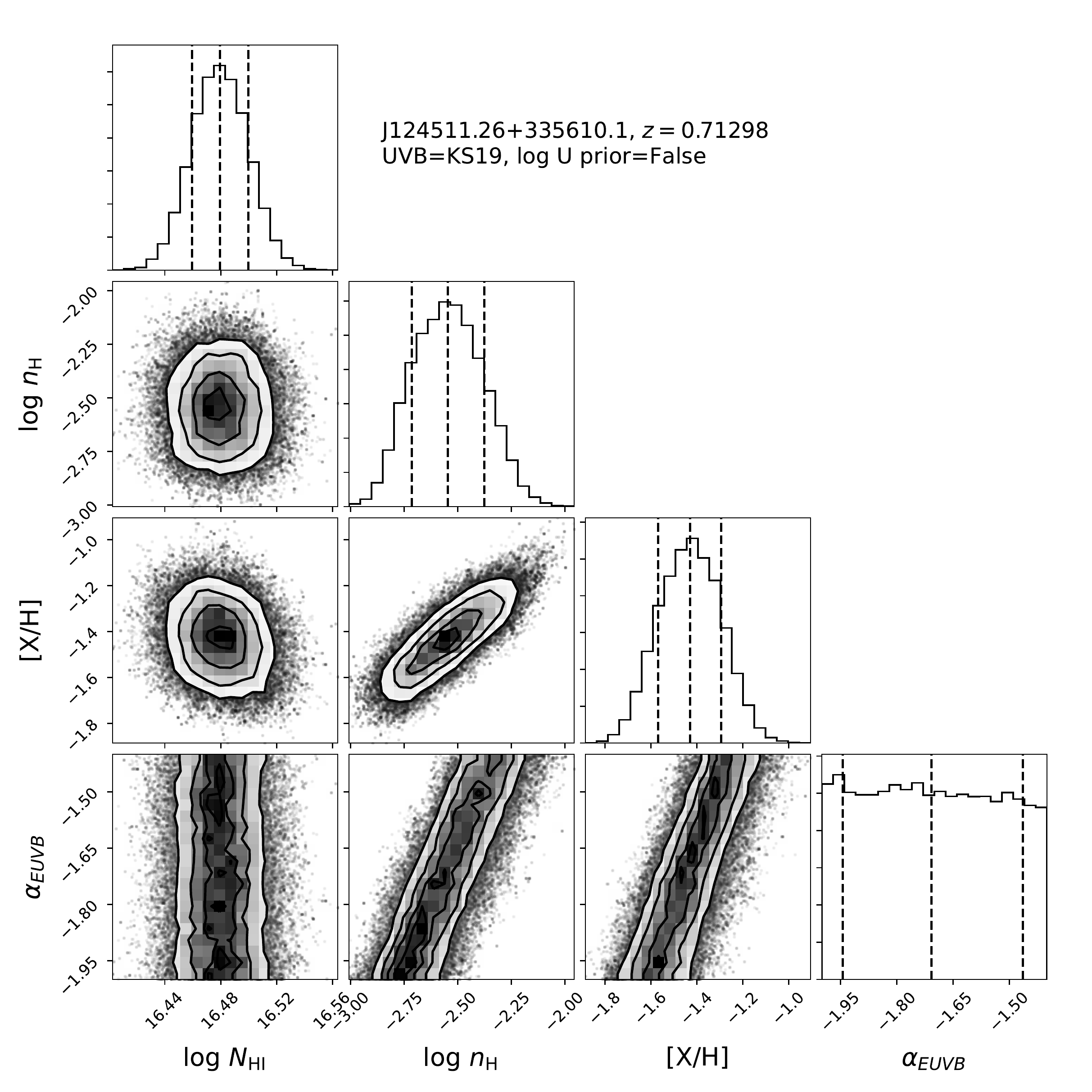}
\plotone{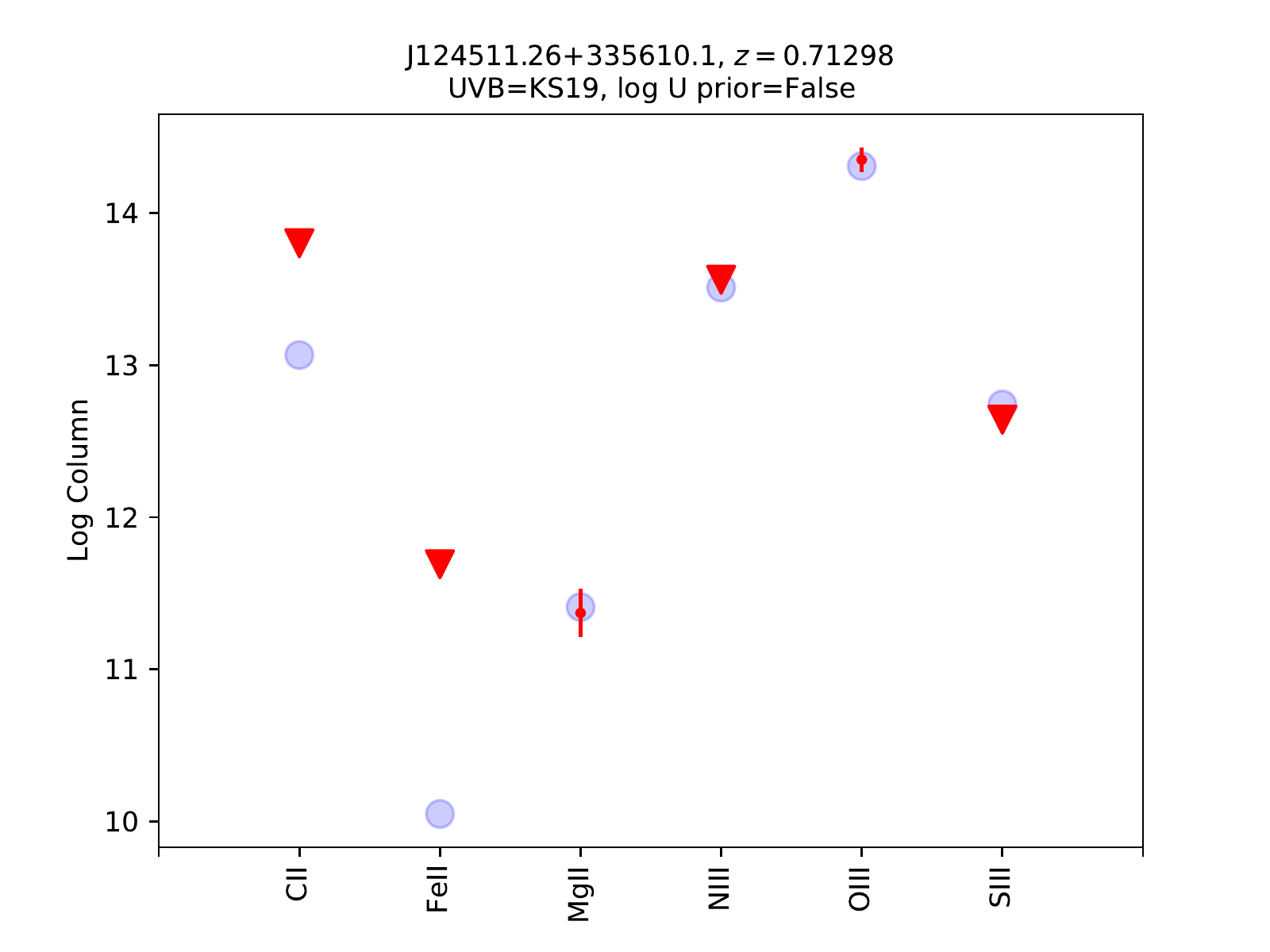}
\caption{Same as Figure~\ref{fig:A1}, but for absorber 23.
\label{fig:A23}
}
\end{figure}

\begin{figure}[tbp]
\epsscale{0.5}
\plotone{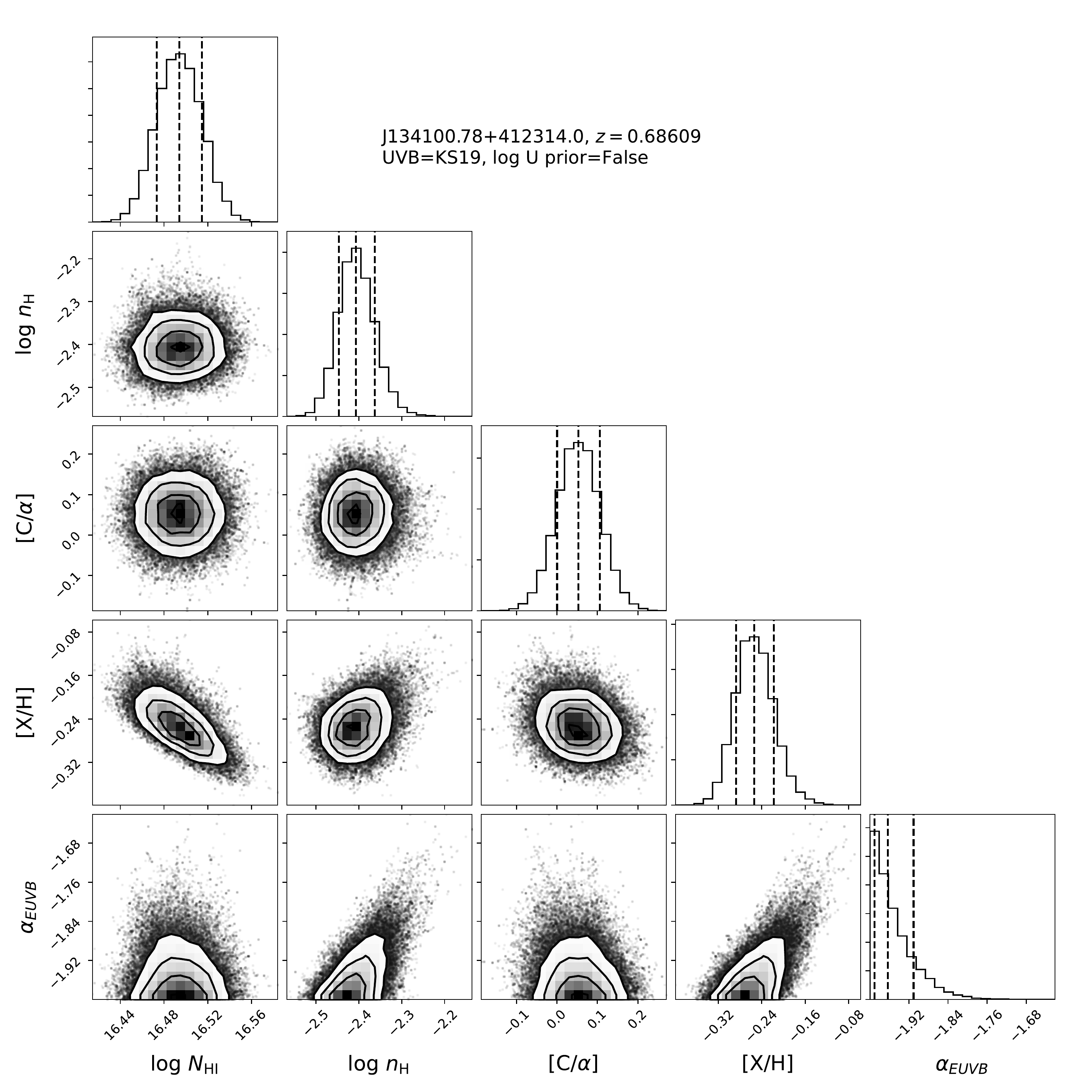}
\plotone{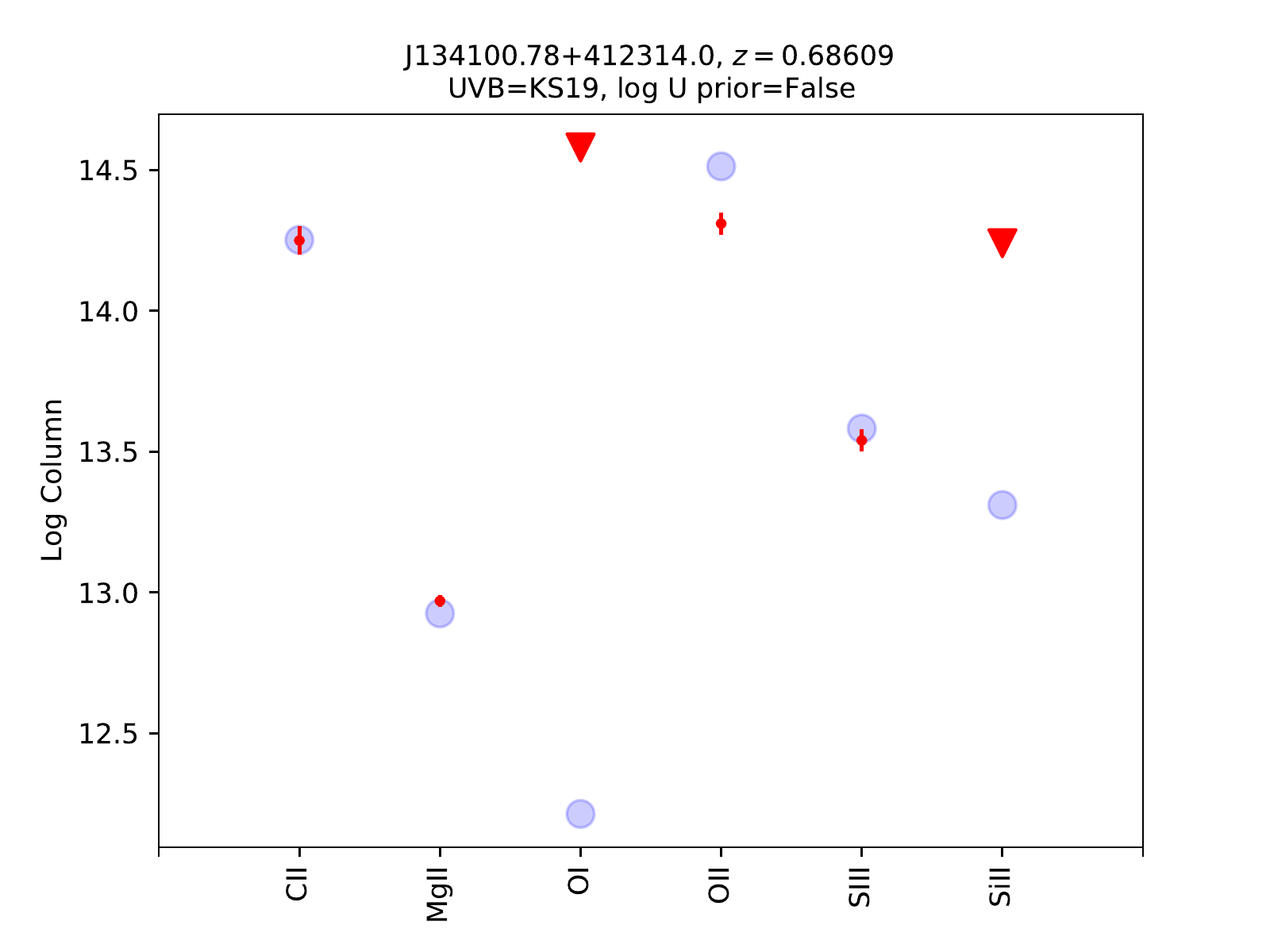}
\caption{Same as Figure~\ref{fig:A1}, but for absorber 24.
\label{fig:A24}
}
\end{figure}

\begin{figure}[tbp]
\epsscale{0.5}
\plotone{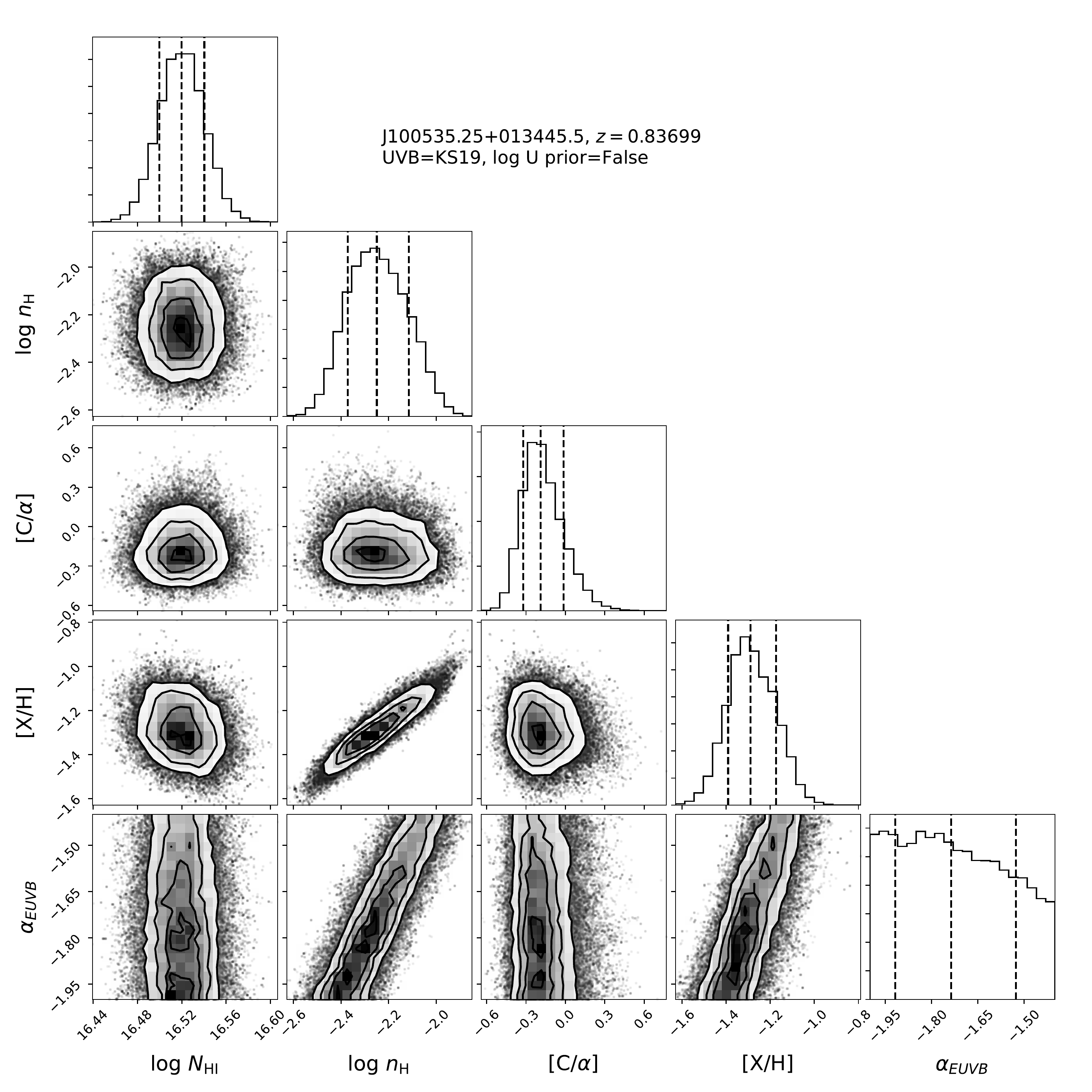}
\plotone{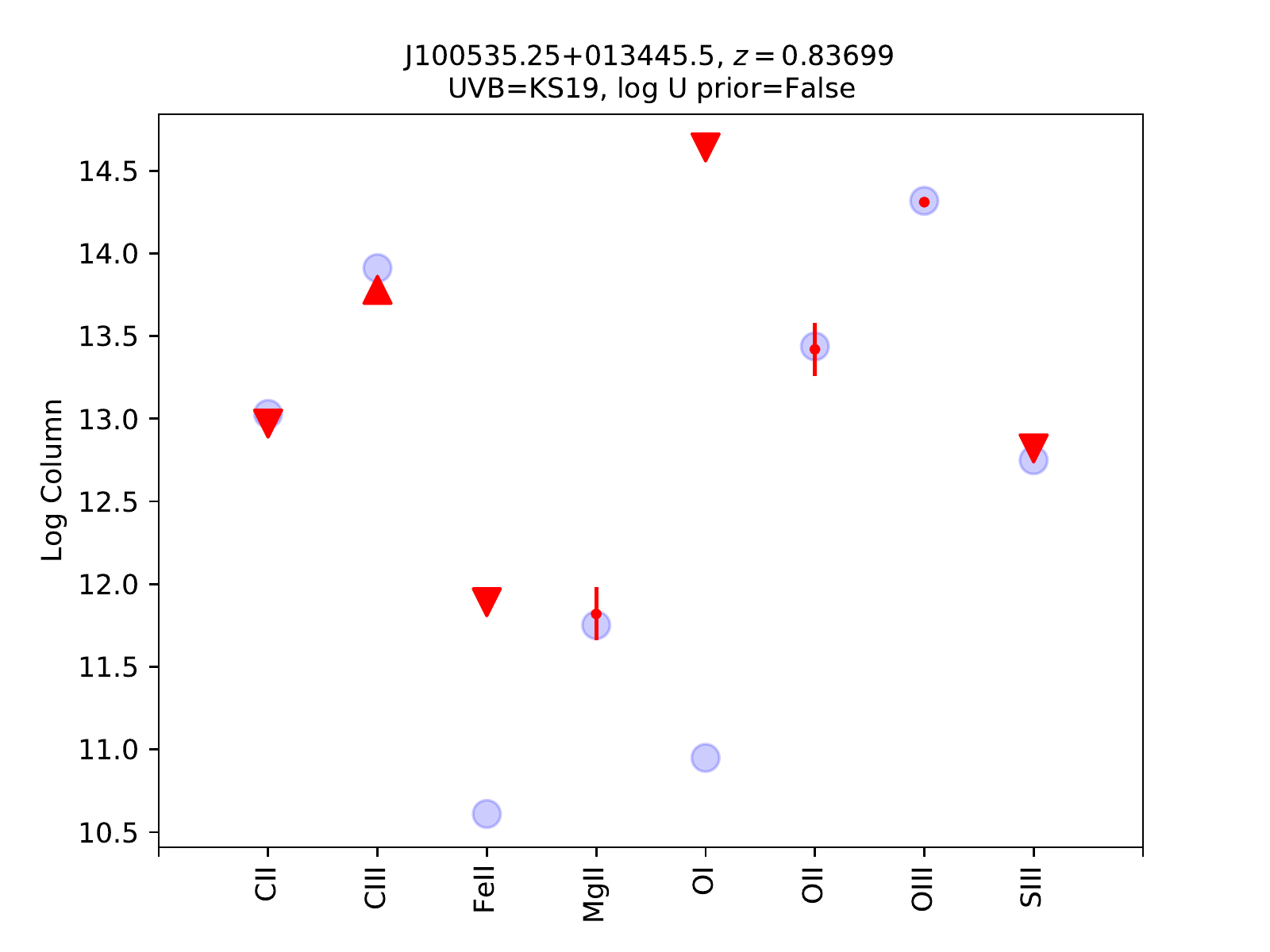}
\caption{Same as Figure~\ref{fig:A1}, but for absorber 25.
\label{fig:A25}
}
\end{figure}

\begin{figure}[tbp]
\epsscale{0.5}
\plotone{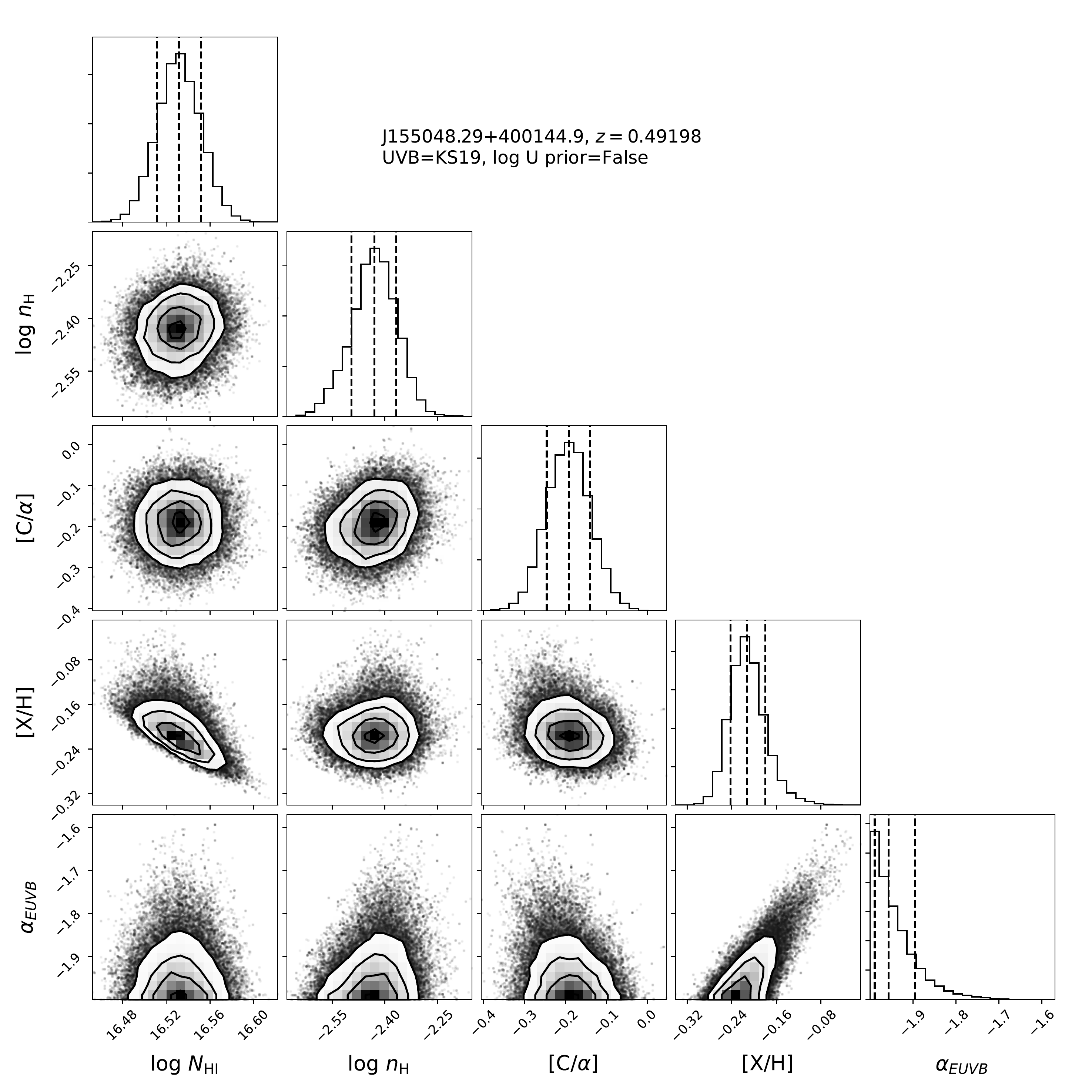}
\plotone{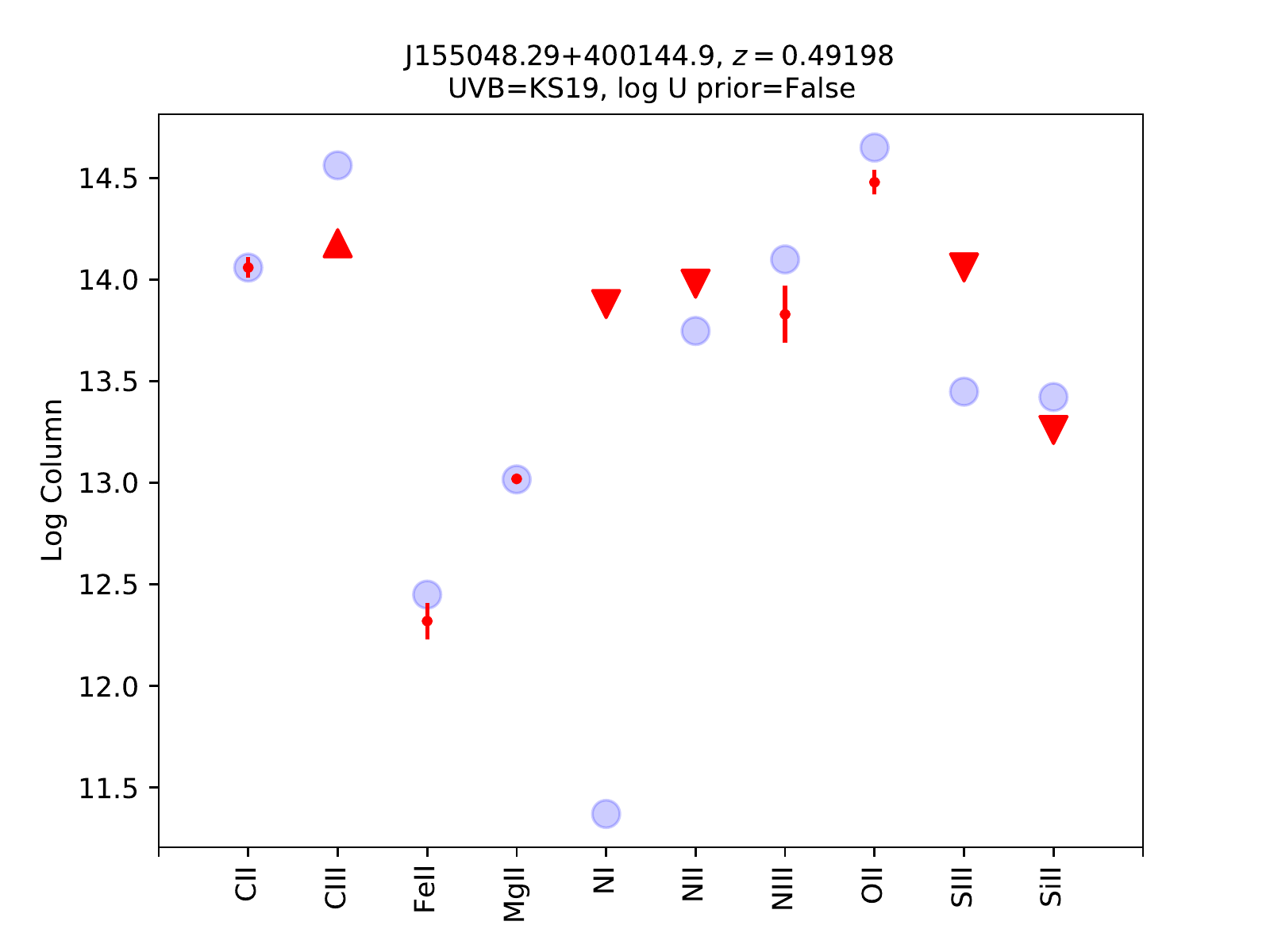}
\caption{Same as Figure~\ref{fig:A1}, but for absorber 26.
\label{fig:A26}
}
\end{figure}

\begin{figure}[tbp]
\epsscale{0.5}
\plotone{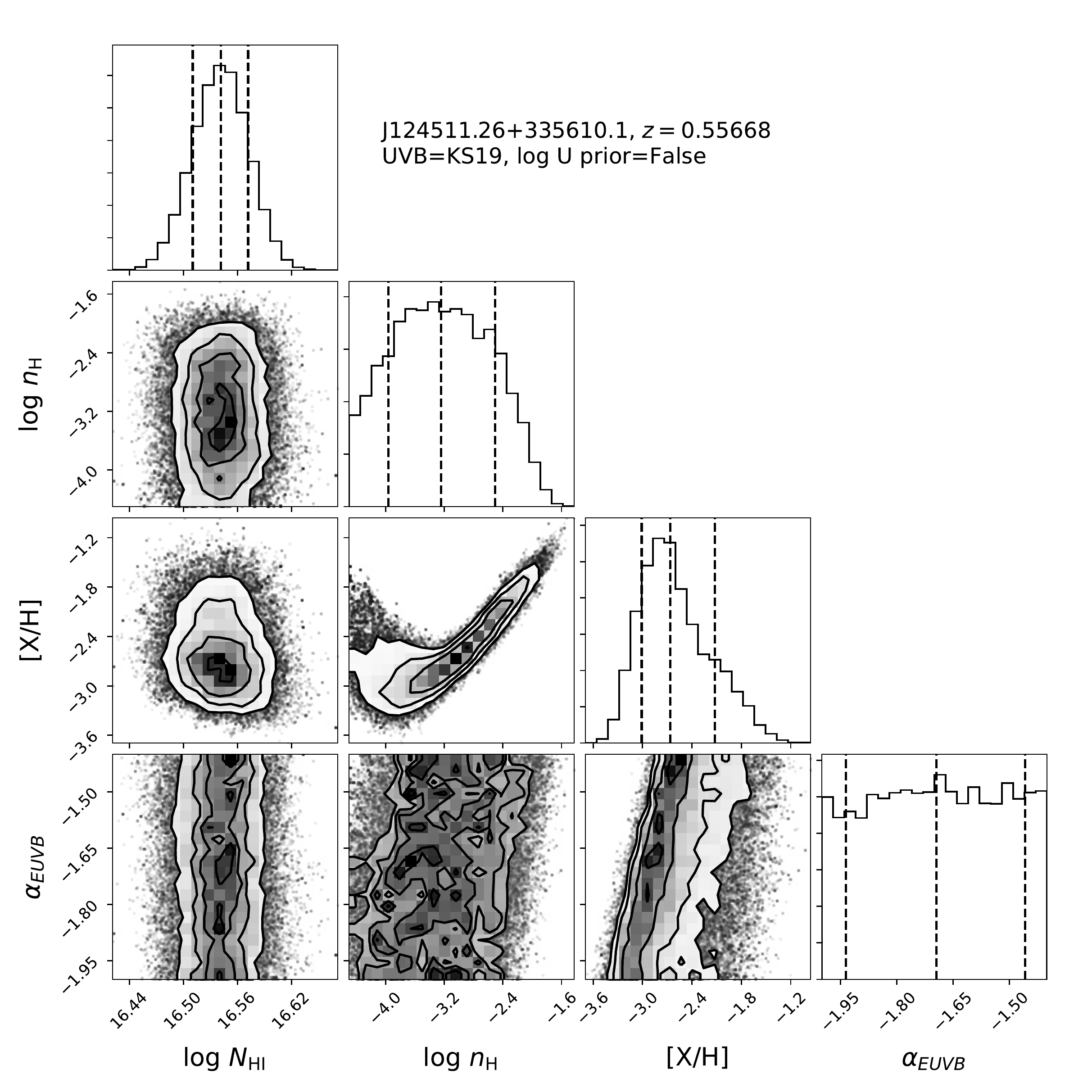}
\plotone{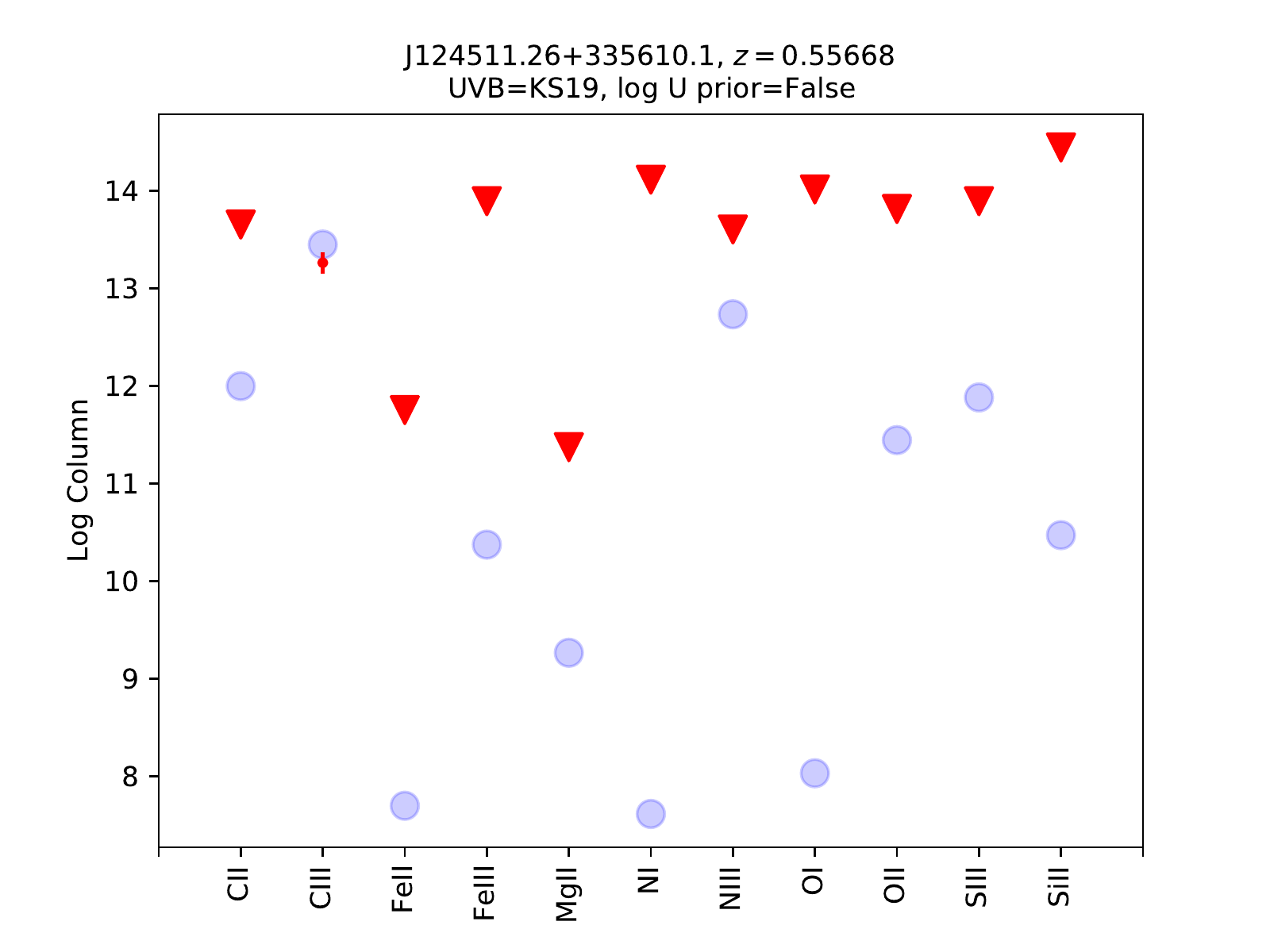}
\caption{Same as Figure~\ref{fig:A1}, but for absorber 27.
\label{fig:A27}
}
\end{figure}

\begin{figure}[tbp]
\epsscale{0.5}
\plotone{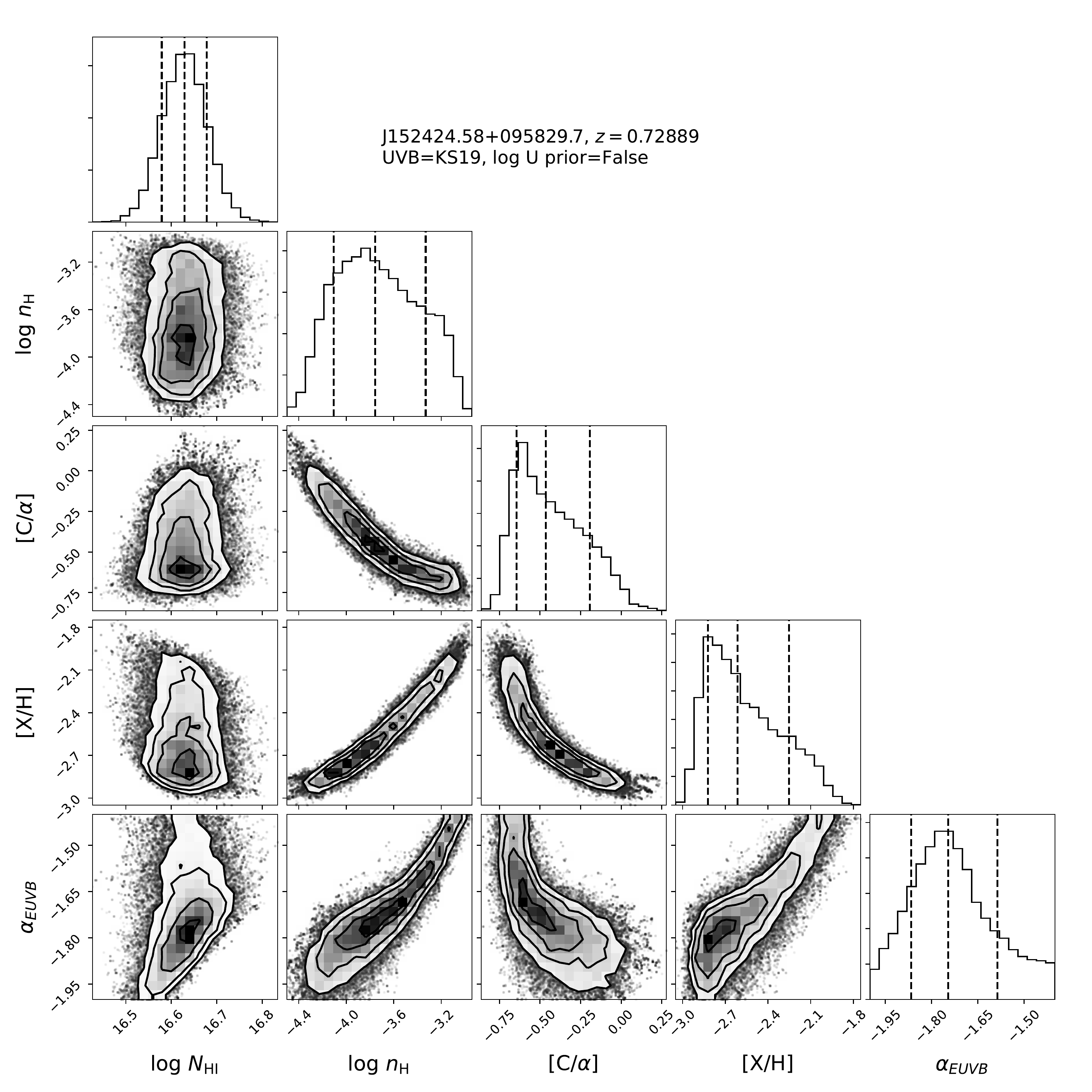}
\plotone{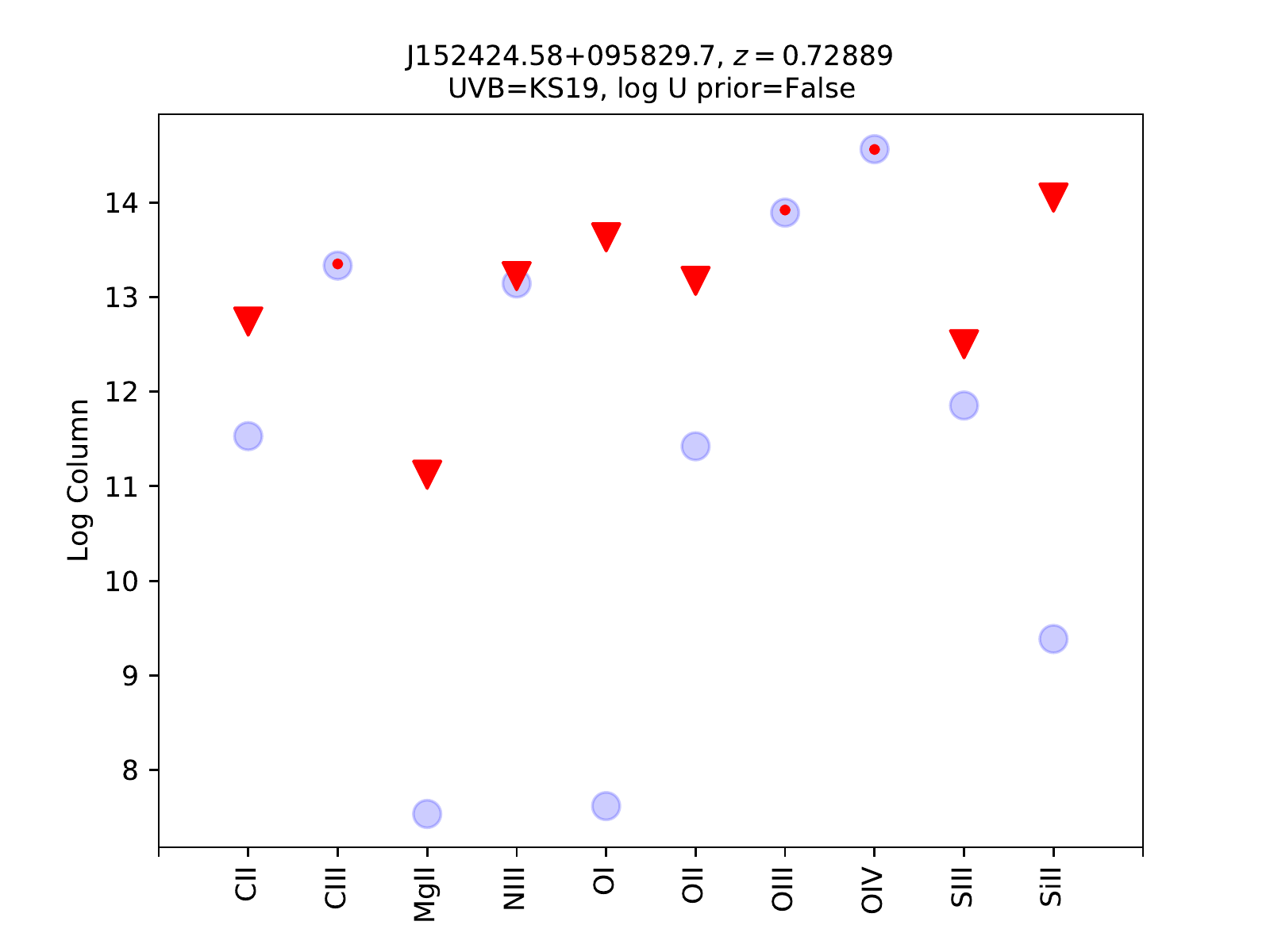}
\caption{Same as Figure~\ref{fig:A1}, but for absorber 28.
\label{fig:A28}
}
\end{figure}

\begin{figure}[tbp]
\epsscale{0.5}
\plotone{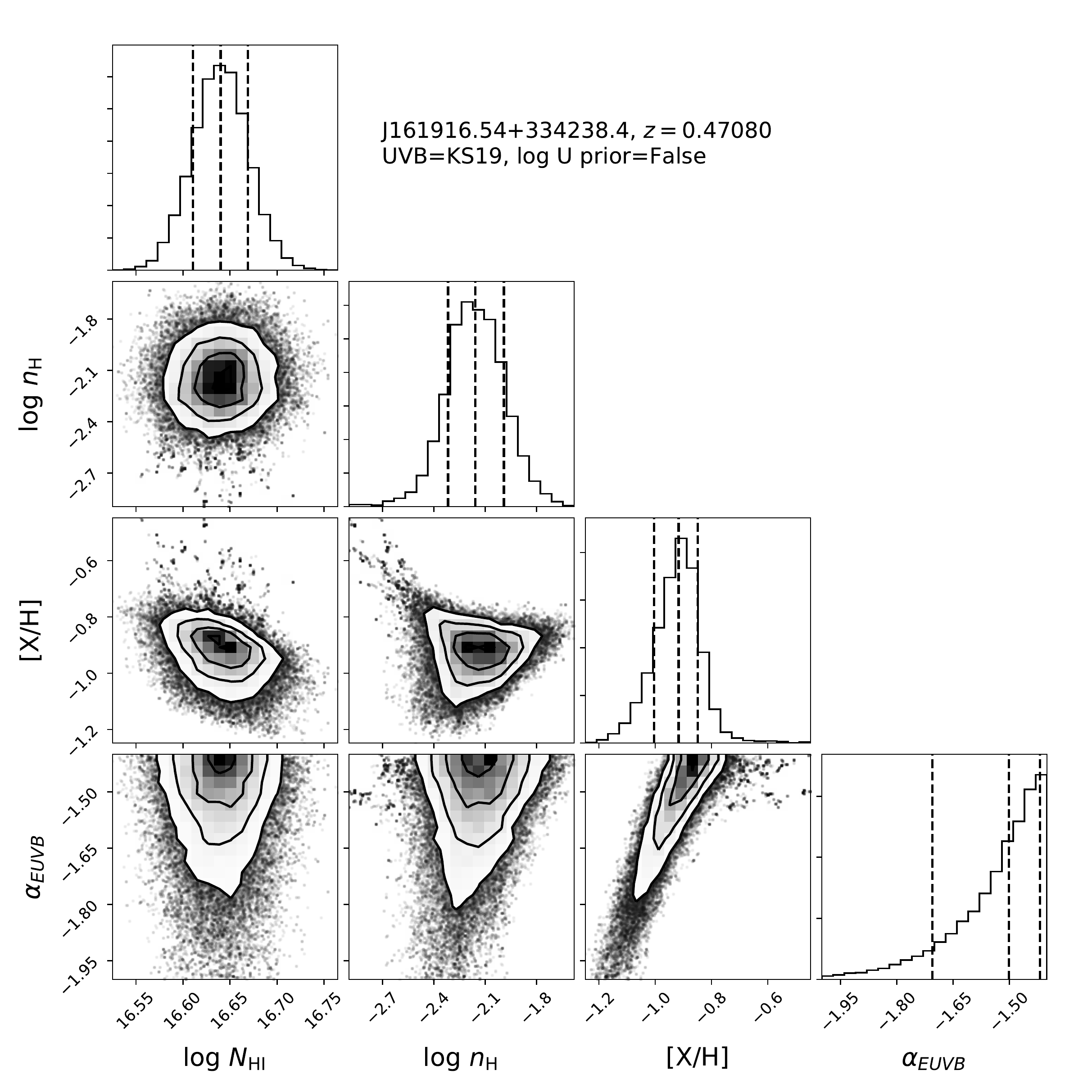}
\plotone{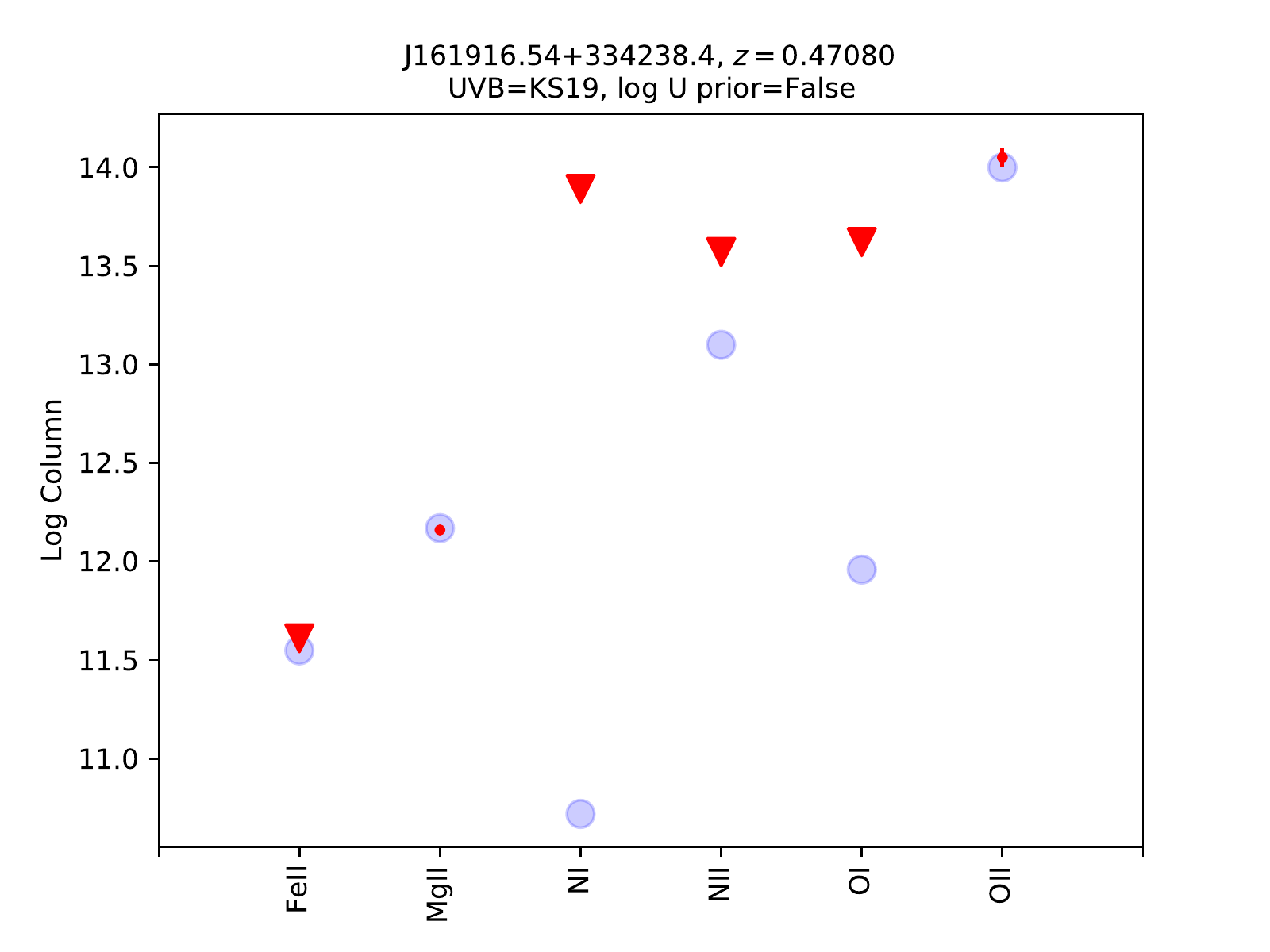}
\caption{Same as Figure~\ref{fig:A1}, but for absorber 29.
\label{fig:A29}
}
\end{figure}

\begin{figure}[tbp]
\epsscale{0.5}
\plotone{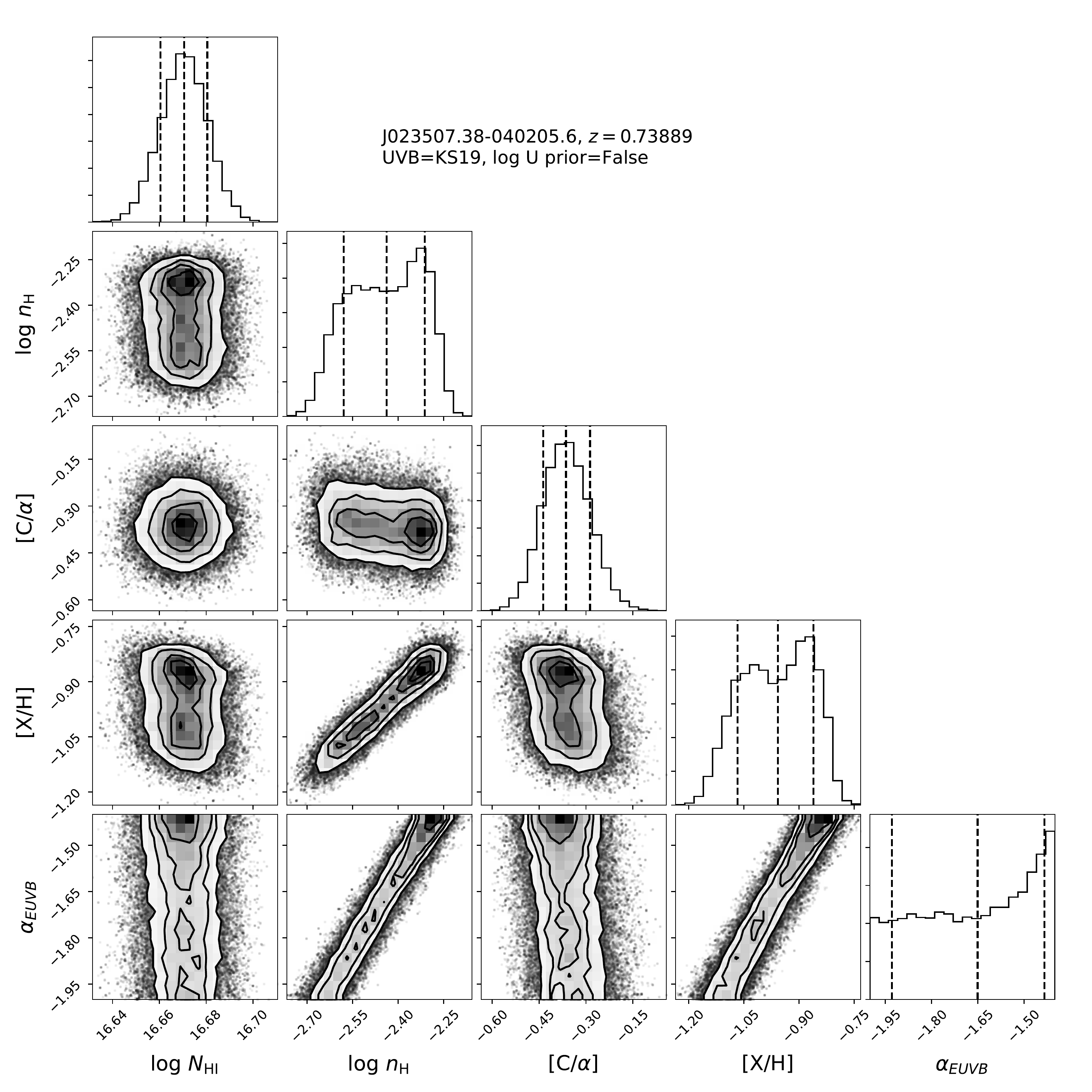}
\plotone{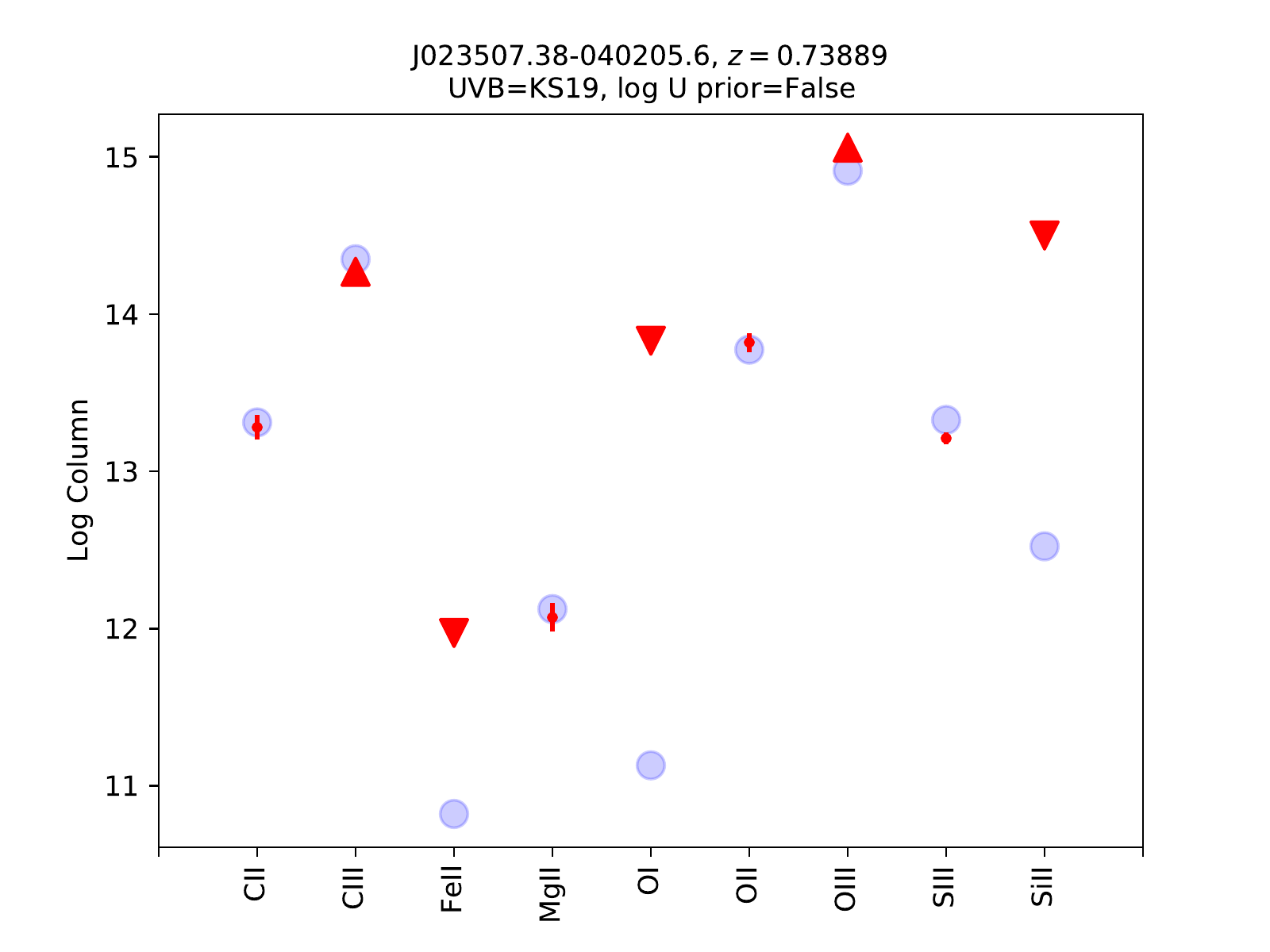}
\caption{Same as Figure~\ref{fig:A1}, but for absorber 30.
\label{fig:A30}
}
\end{figure}

\begin{figure}[tbp]
\epsscale{0.5}
\plotone{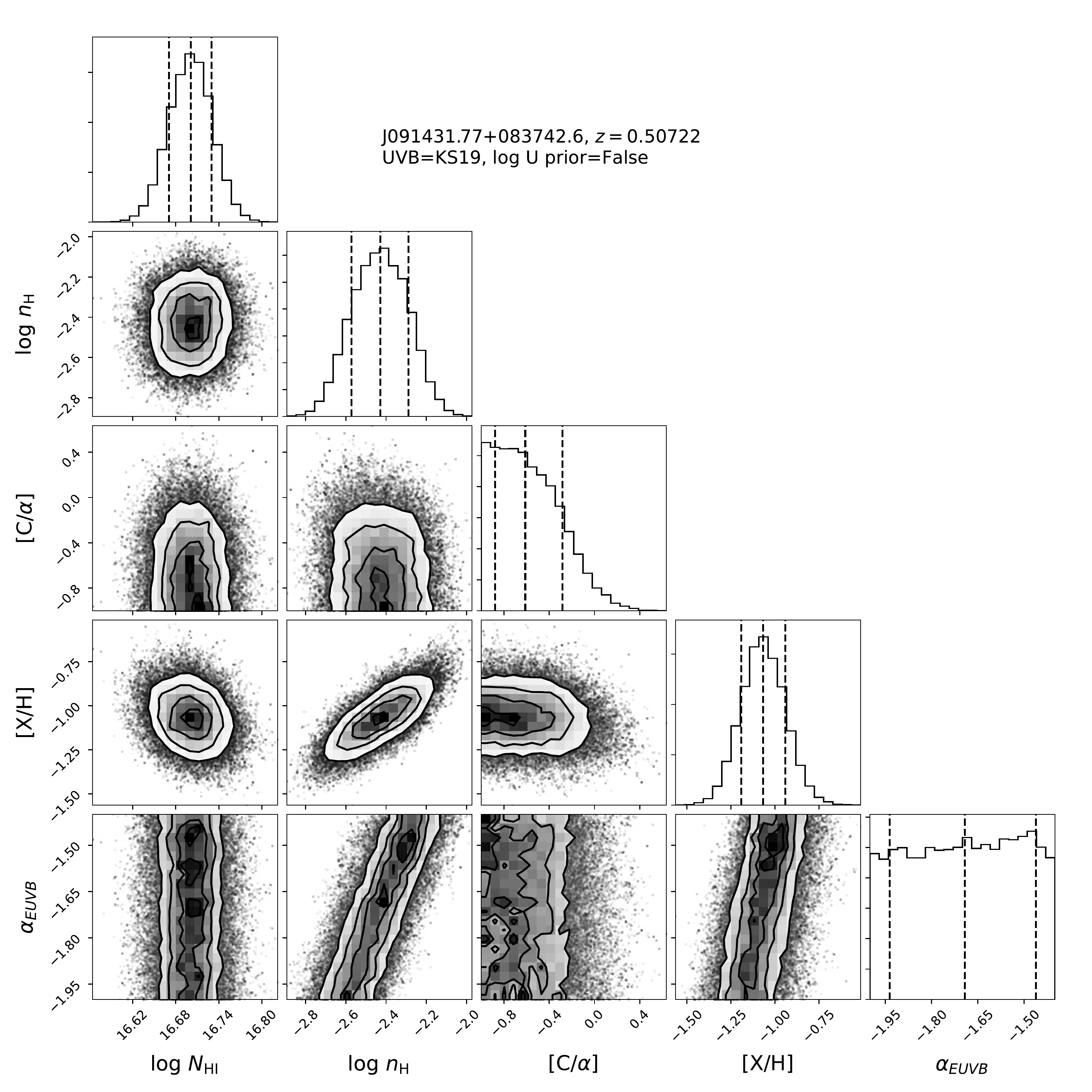}
\plotone{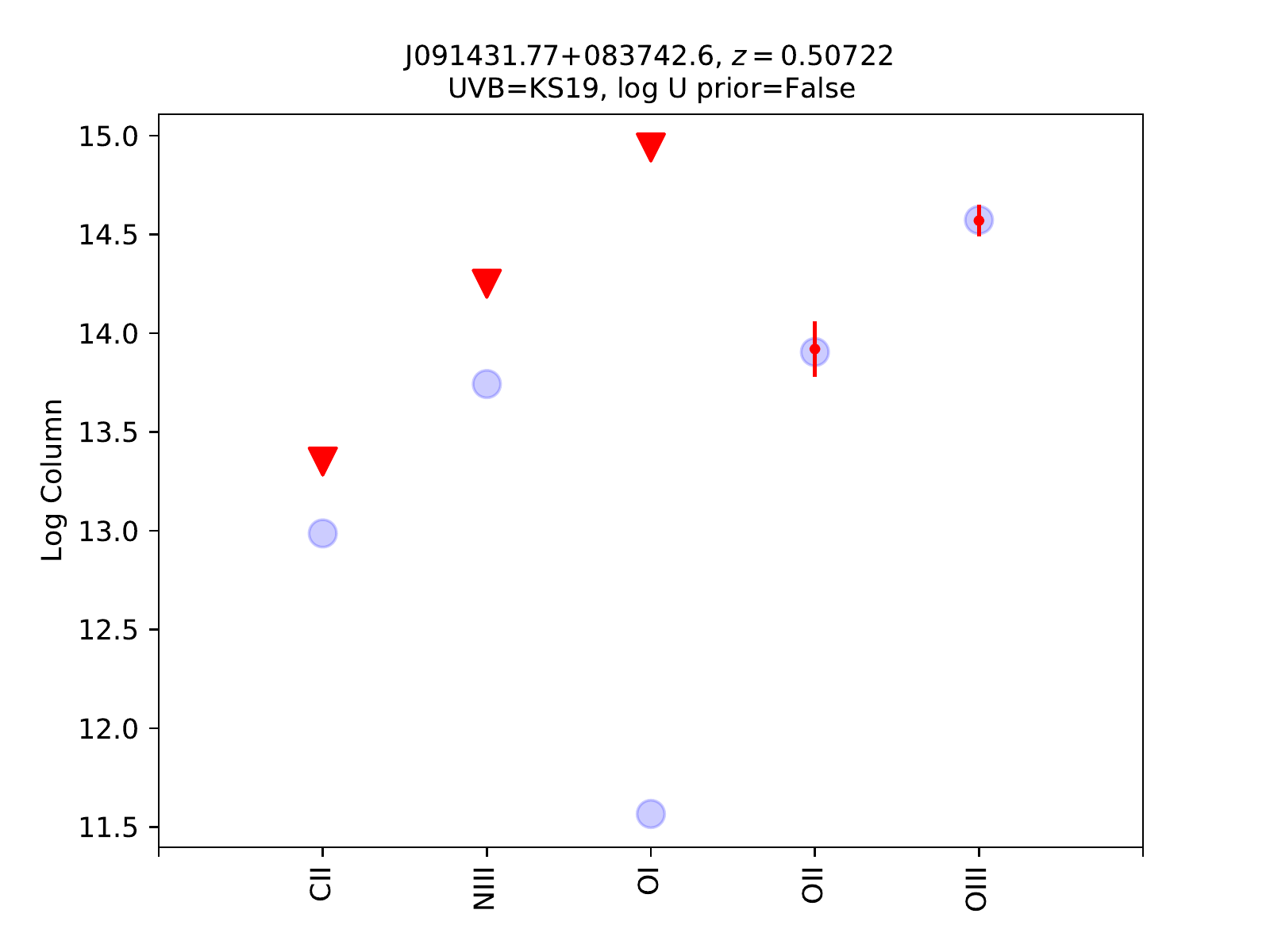}
\caption{Same as Figure~\ref{fig:A1}, but for absorber 31.
\label{fig:A31}
}
\end{figure}

\begin{figure}[tbp]
\epsscale{0.5}
\plotone{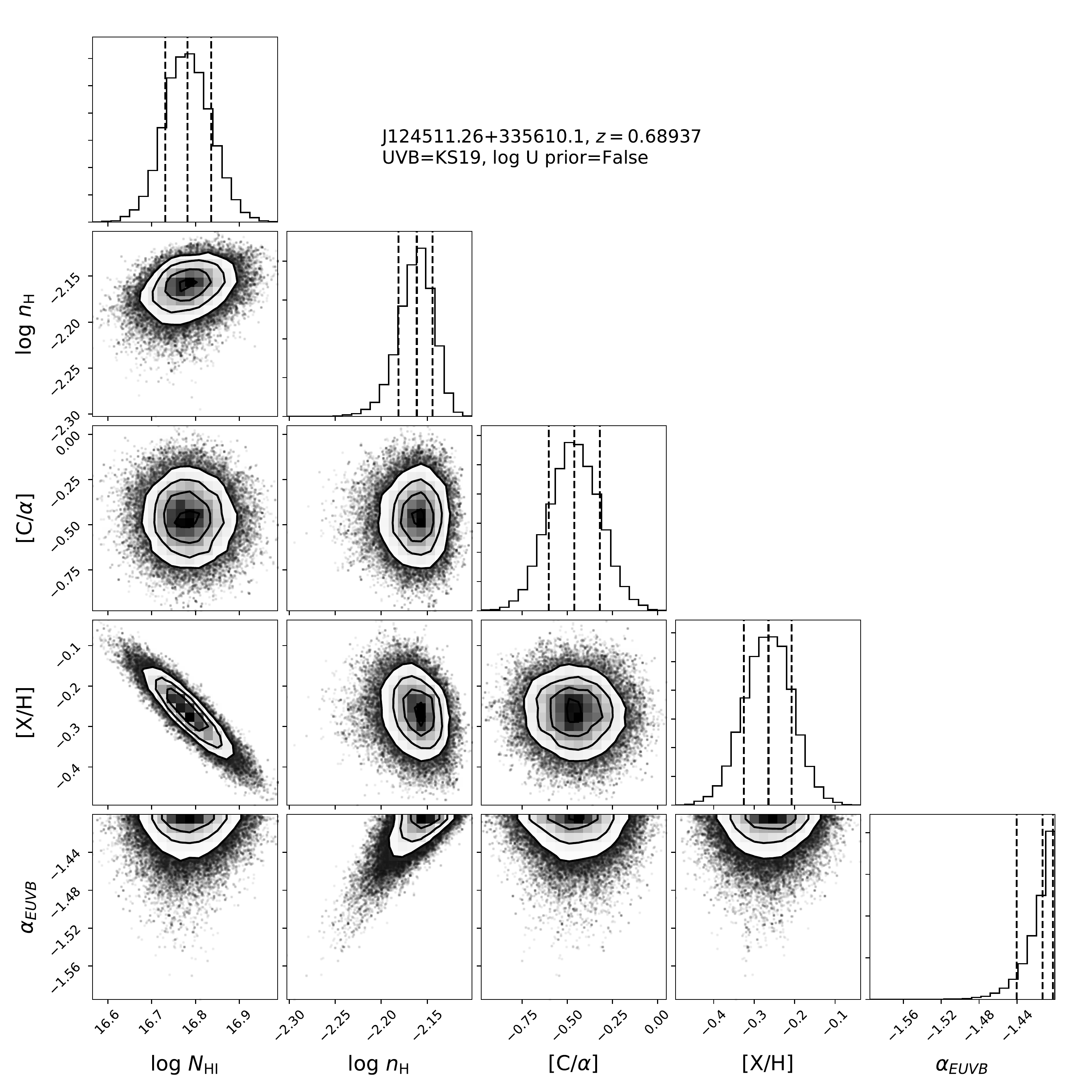}
\plotone{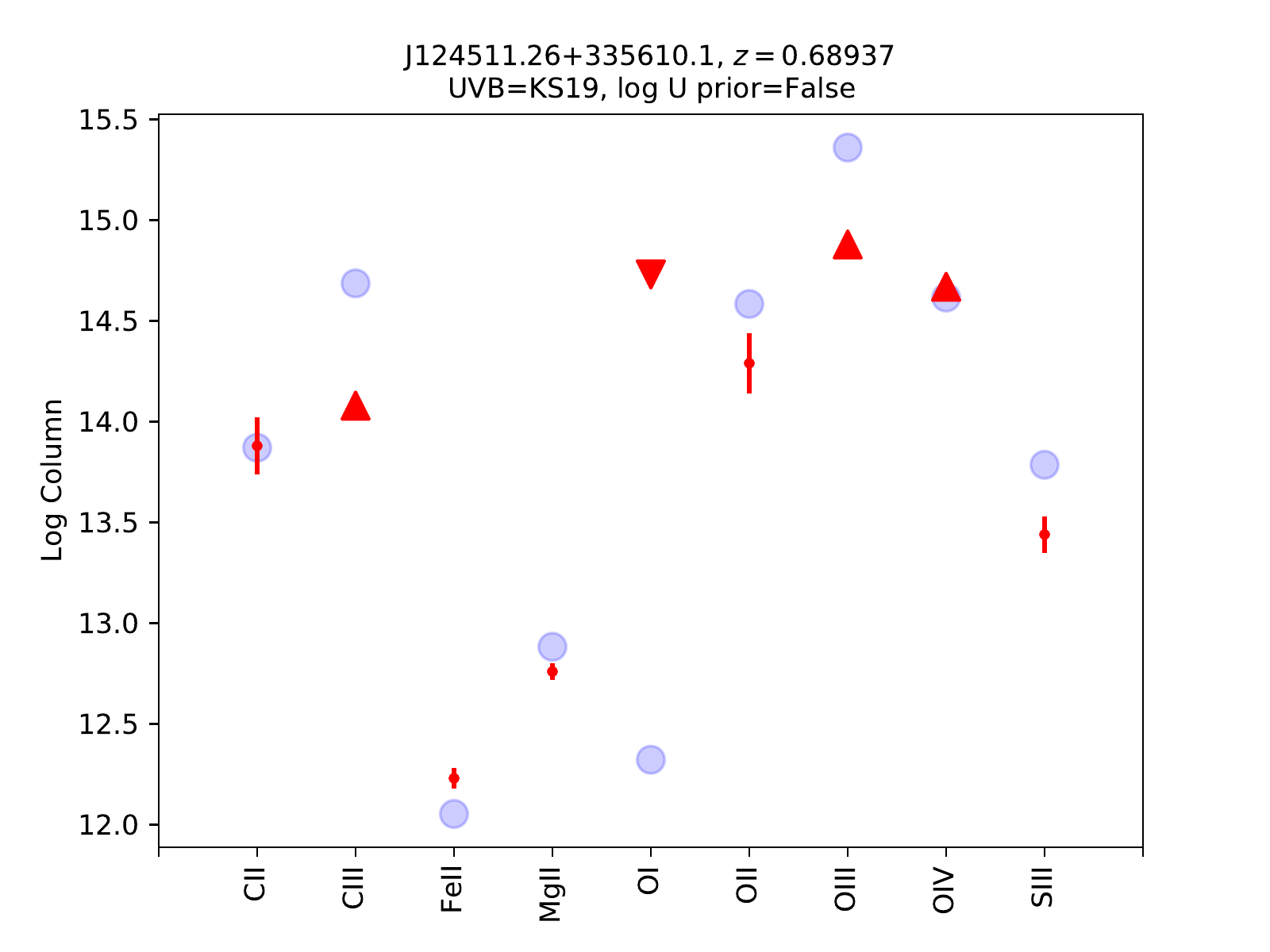}
\caption{Same as Figure~\ref{fig:A1}, but for absorber 32.
\label{fig:A32}
}
\end{figure}

\begin{figure}[tbp]
\epsscale{0.5}
\plotone{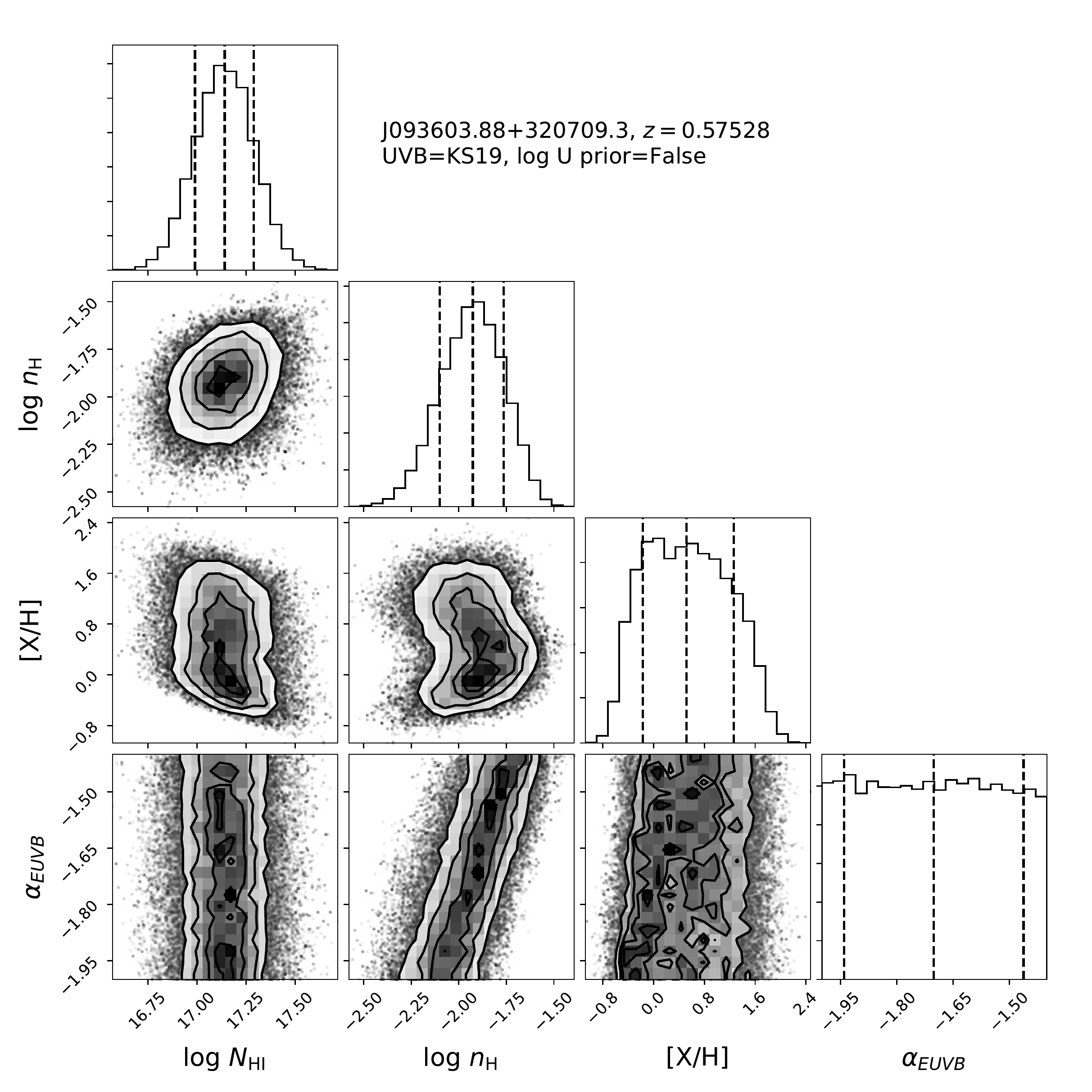}
\plotone{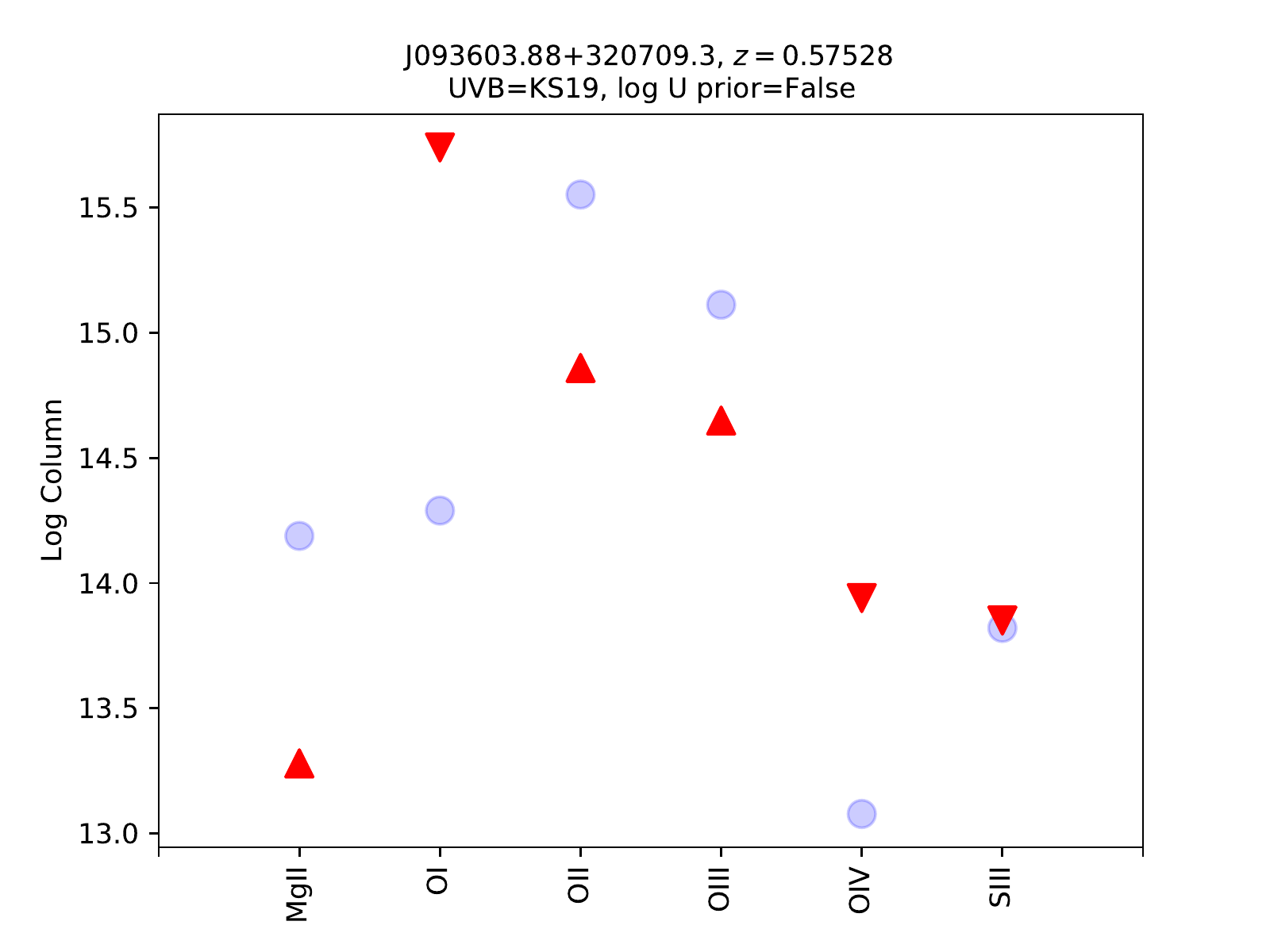}
\caption{Same as Figure~\ref{fig:A1}, but for absorber 33.
\label{fig:A33}
}
\end{figure}

\begin{figure}[tbp]
\epsscale{0.5}
\plotone{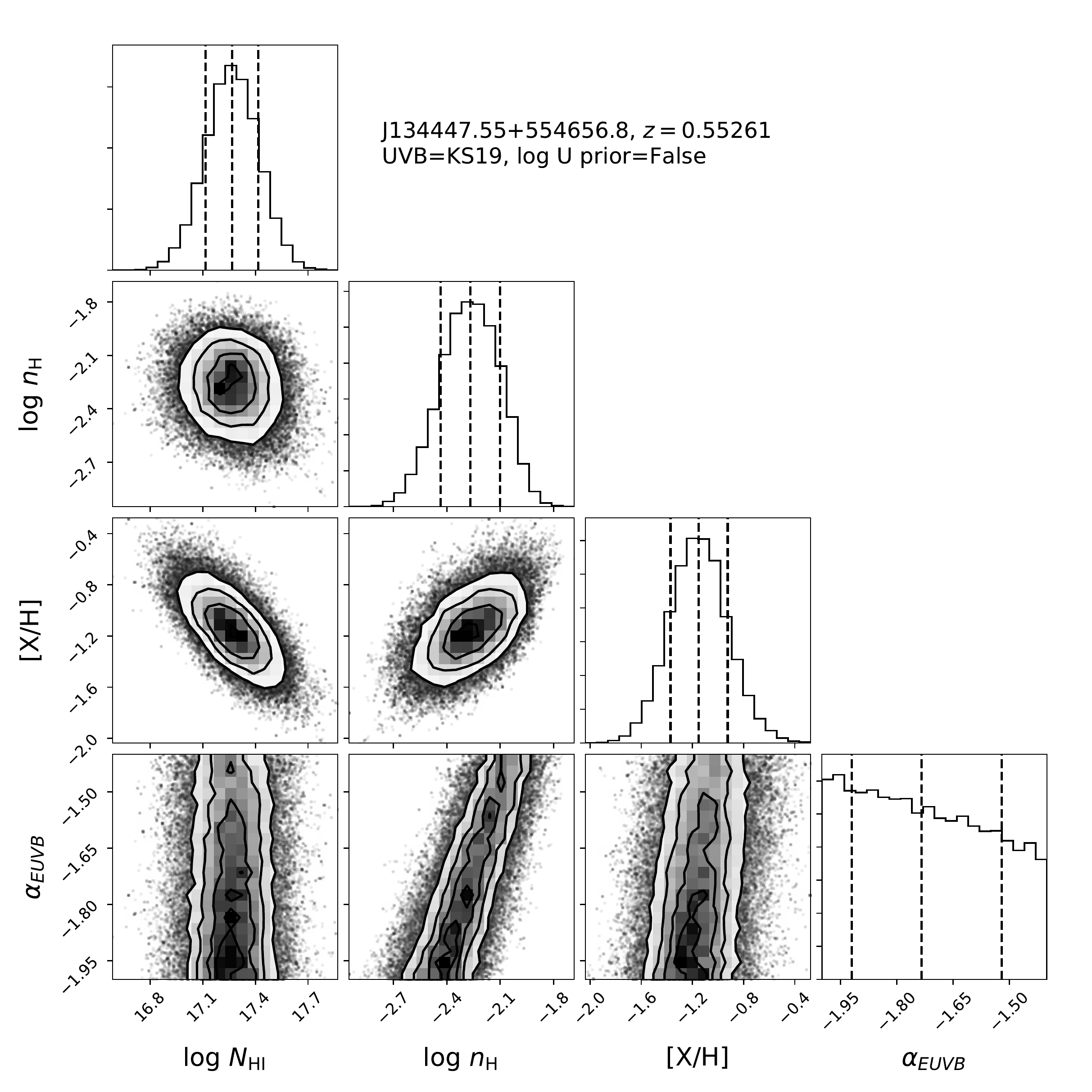}
\plotone{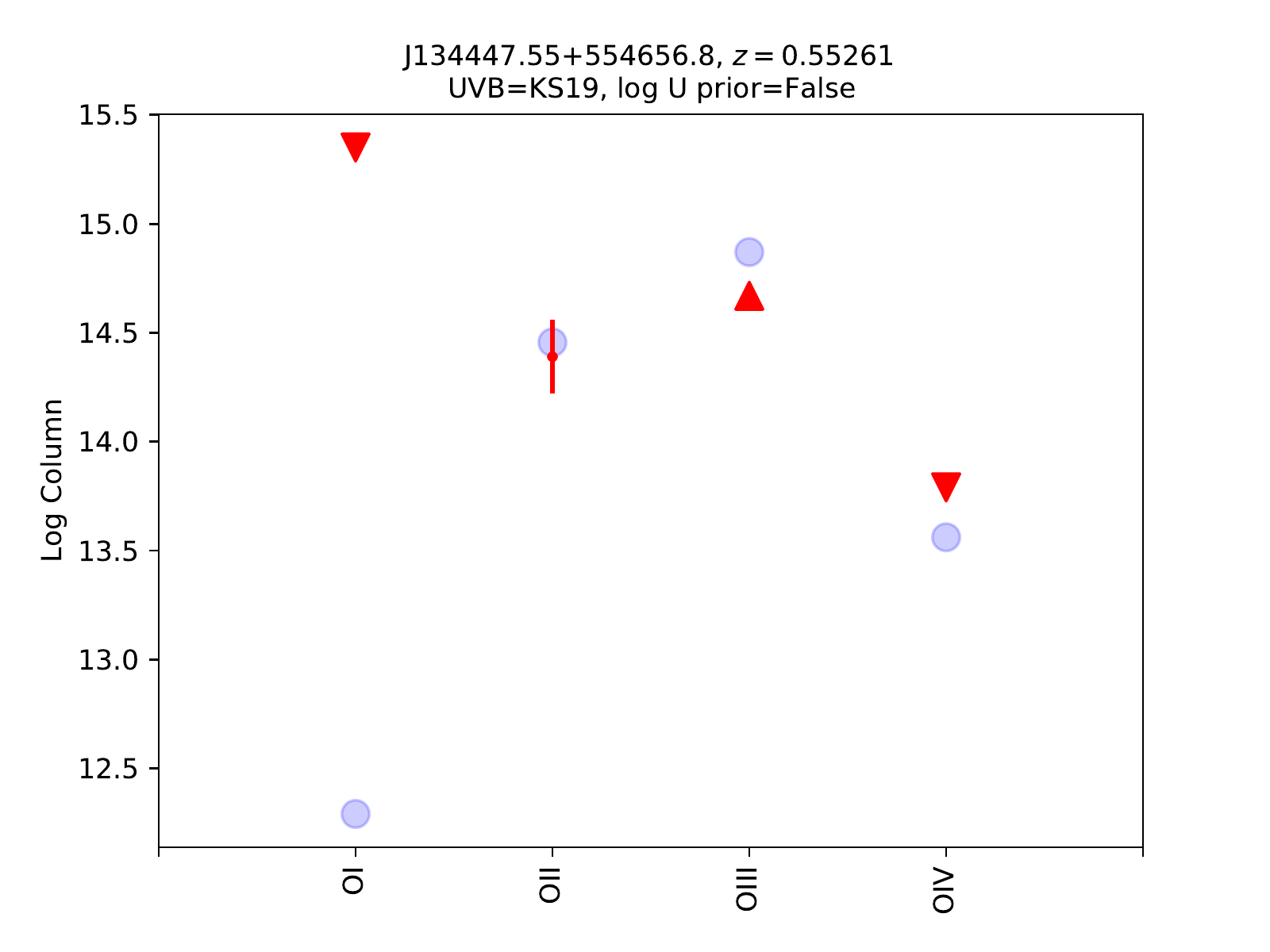}
\caption{Same as Figure~\ref{fig:A1}, but for absorber 34.
\label{fig:A34}
}
\end{figure}

\end{document}